\documentclass[11pt,a4paper]{article}
\usepackage{jheppub}
\usepackage{amscd}
\usepackage{amsfonts}
\usepackage{amsthm}
\usepackage{rotating}
\usepackage{stmaryrd}
\usepackage[normalem]{ulem}
\usepackage{tikz}
\usepackage{subcaption}
\usetikzlibrary{arrows,shapes}
\usepackage{pdflscape}

\usepackage{empheq}
\usepackage[most]{tcolorbox}
\usepackage{fancybox}

\newcommand{\into}{\hookrightarrow}
\newcommand{\onto}{\twoheadrightarrow}

\definecolor{cambridgeblue}{rgb}{0.64, 0.76, 0.68}
	\definecolor{lapislazuli}{rgb}{0.15, 0.38, 0.61}
\definecolor{awesome}{rgb}{1.0, 0.13, 0.32}
\definecolor{aureolin}{rgb}{0.99, 0.93, 0.0}
\definecolor{almond}{rgb}{0.94, 0.87, 0.8}
\definecolor{antiquewhite}{rgb}{0.98, 0.92, 0.84}

\newcommand{\Tr}{\operatorname{Tr}}

\newcommand\fverb{\setbox\pippobox=\hbox\bgroup\verb}
\newcommand\fverbdo{\egroup\medskip\noindent%
            \fbox{\unhbox\pippobox}\ }
\newcommand\fverbit{\egroup\item[\fbox{\unhbox\pippobox}]}
\newbox\pippobox

\def    \bea    {\begin{eqnarray}}
\def    \eea    {\end{eqnarray}}
\def \ba{\begin{align}}
\def \ea{\end{align}}
\def \pl {\partial}
\def\non{\nonumber}
\def\be{\begin{equation}}
\def\ee{\end{equation}}

\newcommand{\X}{\mathbb{X}}

\newcommand{\s}{\sigma}

\renewcommand{\H}{\mathcal{H}}

\def\V{\mathcal V}

\title{The Duality Symmetric String at Two-loops}

\author[a]{Neil B. Copland,}
\author[b]{Giacomo Piccinini,}
\author[a,b]{ and Daniel C. Thompson}
\affiliation[a]{
Theoretische Natuurkunde, Vrije Universiteit Brussel\\  \& The International Solvay Institutes\\
Pleinlaan 2, B-1050 Brussels, Belgium }
\affiliation[b]{Physics Department, Swansea University, Singleton Campus, SA2 8PP Swansea, United Kingdom}

\emailAdd{g.piccinini.987589@swansea.ac.uk, d.c.thompson@swansea.ac.uk}

\abstract{The Tseytlin duality symmetric string makes manifest the $O(n,n)$ T-duality symmetry on the worldsheet at the expense of manifest Lorentz invariance.  Here we consider the two-loop renormalisation of this model in the context of  ``cosmological'' spacetimes consisting of an internal $n$-dimensional torus   fibred over a one-dimensional base manifold.   The lack of manifest Lorentz symmetry introduces a range of complexities in momenta loop integrals which we approach using different methods.  Whilst the results do satisfy a number of key consistency criteria, we find however that the two-loop counter-terms are incompatible with $O(n,n)$ symmetry and obstruct the renormalisability of the duality symmetric string.     }


\begin{document}
 
\maketitle

\setcounter{equation}{0}

\section{Introduction}
  The   renormalisation group flow of the two-dimensional non-linear $\sigma$-model provides a linkage between the worldsheet and spacetime points of view of String Theory. Demanding the vanishing of the Weyl anomaly associated to a closed string propagating in a spacetime geometry requires that the target space background data (i.e.  metric and NS two-form in the bosonic sector) obey a set of equations \cite{Friedan:1980jf,Callan:1989nz,Callan:1985ia}.  These equations can be interpreted as field equations of an appropriate target space gravitational theory.\footnote{Though somewhat tricky to source, we found the contemporary lectures notes    \cite{Hull:1986hn,Callan:1989nz} to be a particularly useful resource.}     At one-loop on the worldsheet, this target space theory has a common bosonic sector shared with the type II Supergravity theories.   Higher loop quantum calculations  \cite{Hull:1987pc,Metsaev:1987zx,Jack:1989vp} lead in turn to modifications to the target space effective theory, organised in an expansion in derivatives.\footnote{The full description of String Theory is richer still, since this two-dimensional quantum expansion needs to be supplemented with a $g_s$ expansion in the worldsheet genus.} 

This picture largely retains the conventional geometric notions associated to point particles and does not capture all the features we have come to associate with String Theory.  In particular, it is now well understood that there is a rich duality symmetry structure that heavily constrains the form of the target space effective description.   The $O(n,n)$ T-duality symmetry of strings on a $n$-dimensional torus 
\cite{Buscher:1987sk,Buscher:1987qj} indicates that fields should be repackaged into appropriate representations of this symmetry group.  The target metric and two-form become unified into a combined object, a representative of the coset space $O(n,n)/ (O(n)\times O(n))$, known as the {\em generalised metric}, which we denote as $\H$.   The target space formulation of the dynamics of this generalised metric has been well expounded \cite{Tseytlin:1990nb,Tseytlin:1990va,Siegel:1993th,Siegel:1993xq,Hull:2009mi,Hohm:2010pp} and has become known as Double Field Theory.\footnote{There is a similar approach to make the exceptional U-duality groups of eleven-dimension supergravity manifest \cite{Berman:2010is,Hohm:2013pua}, though this is beyond the scope of this paper. }   Starting with the seminal work of Meissner \cite{Meissner:1996sa}, there have been significant efforts \cite{Hohm:2013jaa,Coimbra:2014qaa,Lee:2015kba,Marques:2015vua,Hohm:2015mka,Baron:2017dvb} in developing the theory of higher derivative corrections in a T-duality covariant fashion (see  \cite{Lescano:2021lup} for  recent lecture notes on this topic).   

This progress poses a sharp question:   {\em can one exploit  worldsheet renormalisation to obtain higher derivative corrections of the target space theory in a way that maintains T-duality covariance throughout? }

Key to this is to adopt a reformulation of the string worldsheet theory in which T-duality is promoted to a manifest symmetry  \cite{Duff:1989tf,Tseytlin:1990nb,Tseytlin:1990va,Maharana:1992my,Hull:2004in,Hull:2006va,  Hull:2007jy,Hull:2009sg}.  A conventional non-linear $\sigma$-model is described by $D$ bosons that map the worldsheet  into the target space $M$.  For the target to exhibit $O(n,n)$ T-duality  we impose a  $U(1)^n$ isometry group, such that the target is a torus fibration $T^n \into  M \onto  B$ over a base $B$. On a patch, we let $x^i$, $i=1\dots n$, be coordinates adapted to the isometry that parametrise the fibre, and $y^a$  those on the base.  To allow T-duality to be exhibited manifestly we consider a larger {\em doubled} space $\mathbb{M}$ with fibration $T^{2n}\into  \mathbb{M} \onto  B $ in which the original coordinates $x^i$ are supplemented with   additional $\tilde{x}_i$, $i=1,\dots, n$, that can be thought of as describing the T-dual space, $\widetilde{T}^n$, to the torus such that locally $T^{2n}= T^{n} \times \widetilde{T}^n$.    
  
This doubling of coordinates allows us to introduce an $O(n,n)$ ``vector'' of coordinates $\X^I  = (\tilde{x}_i, x^i)$, and to combine the internal toroidal components of the metric and $B$-field into a generalised metric, 
\begin{equation}
    \H_{IJ}(y) =  \begin{pmatrix}  g^{-1}  & g^{-1}b \\ -b g^{-1} & g - bg^{-1}b
    \end{pmatrix} \, . 
\end{equation}
which   preserves the $O(n,n)$ pairing 
\begin{equation}
\eta_{IJ} = \begin{pmatrix}
    \, 0 &  &{\bf 1} \,  \\
 \,   {\bf 1} & &0 \,  
\end{pmatrix} \,  .
\end{equation}

A crucial aspect of $\H$, with one index raised with $\eta$,  is that it is idempotent. As a consequence, we can define projectors $\mathcal{P}_\pm = \frac{1}{2}(1 \pm \mathcal{H})\eta$, obeying the  rules $\mathcal{P}_\pm^2 = \mathcal{P}_\pm$, $\mathcal{P}_+ \mathcal{P}_-=0$. 
These objects are particularly relevant for us since the doubling trick comes at a cost.  The   T-dual coordinates $\tilde{x}_i$ are not completely independent variables:  indeed, they should be constrained in terms of the initial $x^i$ by the T-duality rules 
\begin{equation} \label{eq:chirality}
    {\cal P}_\pm \partial_\mp \X = 0\, ,
\end{equation}
where lightcone coordinates $\sigma^\pm = \sigma^0 \pm \sigma^1$ have been introduced. 
These equations are essentially chirality constraints which are notoriously tricky to handle.   In this work, we  will follow the approach  pioneered by Tseytlin \cite{Tseytlin:1990nb,Tseytlin:1990va} in which the Floreanini and Jackiw  treatment \cite{Floreanini:1987as} of chiral bosons is employed to provide a  Lagrangian description for the eq.~\eqref{eq:chirality}.  We can then hope to apply conventional Quantum Field Theory techniques to such a Lagrangian so as to investigate its quantum properties.  The price we pay for doing so is that we loose manifest Lorentz invariance of the worldsheet theory, now specified by\footnote{We choose Lorentzian $(+,-)$ signature on the worldsheet with Lorentz indices $\mu,\nu, \dots= 0,1$, so that e.g. $p^2 = p^\mu p_\mu = p_0^2 - p_1^2$ in two dimensions.} 
\begin{equation} \label{eq:TAction}
S =  \int \mathrm{d}^2 \s \left(- \frac{1}{2}\H_{IJ} \pl_1 \X^I \pl_1 \X^J + \frac{1}{2}\eta_{IJ} \pl_0 \X^I \pl_1 \X^J +\frac{1}{2} \lambda \, \pl_\mu y \pl^\mu y  \right)\ .
\end{equation}

 On the base manifold, where Lorentz invariance is unbroken, we have accordingly retained the use of a vector $\sigma^{\mu} = (\sigma^0, \sigma^1)$. 
A number of simplifications have been assumed: only a single coordinate $y$ on $B$ is considered, such that the base geometric data is just encoded in the coupling $\lambda$ which can be thought of as a partial fixing of the target space lapse variable (the reason for not further fixing this variable to unity will be discussed momentarily).  Furthermore, we have set to zero off-diagonal components of the metric and two-form which together provide a connection in the doubled bundle.  On a general worldsheet, this action should be supplemented with a Fradkin-Tseytlin term that induces a coupling to the duality-invariant dilaton $\Phi = \phi - \frac{1}{4} \ln \det g$ in which $\phi$ is the conventional dilaton field.   In the present context, we will consider the theory on two-dimensional Minkowski space such that this dilaton coupling is absent.\footnote{In general, the dilaton does make a contribution to the Weyl anomaly which can be established by calculating on a curved worldsheet.  However a pragmatic approach is to determine the dilaton contributions  through consistency requirements as was done for the Tseytlin string at one-loop in \cite{Berman:2007yf}.}  Though this is a heavily simplified set-up, even here we will find sufficient complexities.   

The one-loop  quantum effective action resulting from the Lagrangian in eq.~\eqref{eq:TAction} was first considered in \cite{Berman:2007xn,Berman:2007yf}.   This was later expanded to a more democratic approach in which both the fibre {\em and} base space are doubled allowing direct contact to be made with Double Field Theory \cite{Copland:2011yh,Copland:2011wx}.   The doubled action of eq.~\eqref{eq:TAction} can be generalised  to accommodate non-Abelian and Poisson-Lie  generalisations  of T-duality \cite{Klimcik:1995ux}, and its renormalisation was considered in \cite{Avramis:2009xi,Sfetsos:2009vt}.  More recently the one-loop calculation of  \cite{Berman:2007xn,
  Berman:2007yf} has been revisited and refined in \cite{Bonezzi:2021mub} with the inclusion of a connection in the doubled bundle $\mathbb{M}$\footnote{The Tseytlin string can also be viewed as a gauge-fixed version of a covariant theory whose one-loop renormalisation was considered in \cite{GrootNibbelink:2013ncv}; though  puzzlingly,  the resultant background field equations do not appear to agree with those Double Field Theory (and hence those of the standard string).}.

  In these notes we will push the techniques  of \cite{Berman:2007xn,
  Berman:2007yf} further and apply them at two-loop order.       There are several critical aspects that make this calculation extremely challenging:
  \begin{itemize}
      \item The lack of manifest worldsheet Lorentz invariance.  This has two consequences, the first is that   it requires treating non-covariant loop momenta integrals and the second is that we can not disregard contributions arising from the connection $\Omega= {\cal V}^{-1} \mathrm{d} {\cal V}$ associated to the vielbein $ {\cal V}$ that diagonalises $\H$.
 \item  To treat the loop integrals we consider two different methods.  In Method 1 we employ dimensional regularisation and  move  immediately  to $d=2 + \epsilon$ and evaluate the wide variety of tensor  integrals encountered by reducing them in a basis of standard scalar integrals.  In Method 2 we work in $d=2$ to perform the maximal number of simplifications of loop momenta (by e.g. replacing  $p^2_0 - p_1^2= p^2$) expressing the result implicitly in terms of a set of only five independent integrals whose regularisation and evaluation is left initially implicit.  We are then able to extract conclusions that are largely independent of the choice of regularisation method\footnote{For instance one could invoke prescriptions that do \textit{not} require analytic continuation of the worldsheet dimensions (Pauli-Villars and cut-off regularisations among the others) which are potentially more compatible with the chiral nature of the duality-symmetric string.}.  For concreteness, we then complete Method 2 by employing the dimensional regularisation approach of Method 1 to the few remaining loop integrals. 
 
      \item The chirality constraint \eqref{eq:chirality} which obstructs a straightforward regularisation of IR divergences. Whilst this may prove challenging in general, we are able to address this potential troublesome issue  in a rather naive fashion which we find to be satisfactory.    By shifting momenta  $p^2 \rightarrow p^2- m^2$ in the denominator of loop integrals one regulates the divergence, and all features (such as mixed IR and UV divergences at two-loops) associated to the introduction of $m^2$ are removed once we include a ``mass'' term for the background fields. 
      \item The departure from conventional Riemannian geometry invoked by the generalised metric $\H$ prevents the use of known (target space) covariant background field methods \cite{Alvarez-Gaume:1981exa,Mukhi:1985vy}  to simplify the calculation. 
      Instead, we resort to a non-covariant background field expansion with a linear split between the classical background and quantum fluctuation.  To tackle the abundance of diagrams produced in this expansion, we complement a  pen-and-paper calculation of the counter-terms for the generalised metric $\mathcal{H}$ with a  \texttt{Mathematica} implementation to determine the renormalisation of the coupling $\lambda$. 
  \end{itemize}

If the renormalisation of the action \eqref{eq:TAction} could be successfully performed, regardless of the detailed choice of method for evaluating the diagrams, within dimensional regularisation the bare and renormalised generalised metric and base coupling are to be related by 
\be
\H_B = \mu^\epsilon \left( \H + \sum_{n=1}^\infty \epsilon^{-n} T_n(\H, \lambda )  \right) \, , \quad \lambda_B = \mu^\epsilon \left( \lambda + \sum_{n=1}^\infty \epsilon^{-n} \widetilde{T}_n(\H, \lambda )  \right) \, ,
\ee 
with $\mu$ some mass scale and   $T_n(\H, \lambda), \widetilde{T}_n(\H, \lambda )$   the, $\epsilon$-independent, counter-terms that render the effective action finite when phrased in terms of $\H, \lambda$.     The coupling $\lambda$  also provides a loop counting parameter such that we can further express $T_n(\H, \lambda ) = \sum_{L} T^{(L)}_n(\H) \lambda^L$ and $\widetilde{T}_n(\H, \lambda )  =\sum_{L} \widetilde{T}^{(L)}_n(\H) \lambda^{L-1}  $.  The $O(n,n)$ invariant metric on the other hand will not be renormalised $\eta_B = \mu^\epsilon \eta$.

Because at the classical level the lapse variable can be fixed to  $\lambda=1$ by  rescaling $y$, one anticipates it should be   irrelevant to the quantum theory.  However, some care is required since the one-loop result \cite{Berman:2007xn} indicates that a non-constant divergent counter-term proportional to $\partial_\mu y \partial^\mu y$ is produced even when $\lambda =1$.  Such divergences at one-loop can be removed by the addition of a counter-term proportional to the equation of motion for the background fields (equivalently via a field re-definition of $y$). The required term depends rather implicitly on $\widetilde{T}$, and can be established exactly for particular models as in \cite{Tseytlin:1991ht}.  Doing so will induce a modification of the fibre counter-term.  We prefer to keep $\lambda$ explicit as it serves as a loop counting parameter and mirrors the way the lapse function enters in the target space description \cite{Meissner:1996sa}.\footnote{In Appendix \ref{app:ex} we illustrate this point with an explicit example and further more show that such field redefinitions do not ameliorate the difficulties encountered at two-loop order.    }

As the bare couplings are independent of the scale we can take the derivative of $\H_B, \lambda_B$  to determine the $\beta$-functions in terms of the  $\epsilon^{-1}$ poles\footnote{The derivation of this result slightly deviates from that in the conventional non-linear $\sigma$-model, as the counter-terms are not homogeneous in $\H$ alone but are homogeneous when viewed as a function of both $\H$ and $\eta$.  In particular $T^{(L)}_n(\H ,\eta) $ is of homogeneity degree $1$.  One must then keep track of the $\eta$ dependence, even though all its counter-terms vanish.  }:
\be \label{eq:bdef}
\beta^{\H} \equiv  \mu \frac{\partial \H}{\partial \mu}  +\epsilon \H    =  - \sum_L L T_{1}^{(L)}(\H, \eta) \lambda^{-L} \ , \quad \beta^{\lambda} \equiv  \mu \frac{\partial \lambda}{\partial \mu}  +\epsilon \lambda    =  - \sum_L L \widetilde{T}_{1}^{(L)}(\H, \eta) \lambda^{-L+1} \, . 
\ee

  The higher order poles are not independent and instead provide a set of consistency relations known as the pole equations\footnote{As we are employing a simple linear splitting of a quantum fluctuation on top of a classical background field it is sufficient to use the pole equations as presented here rather than the more general ones \cite{Howe:1986vm,Hull:1985rc} required of a geometric non-linear background field method.}.   
  Specialising to the two-loop case we have in particular that on the fibre\footnote{The symbol $\circ$ is used here to indicate both index contraction and integration, e.g. $X \circ \frac{\delta}{\delta Y} \equiv \int \mathrm{d}y  X_{IJ}(y) \frac{\delta}{\delta Y_{IJ}(y)}$.} 
\be \label{eq:polefibre}
  0 =  2  T^{(2)}_2  -  T^{(1)}_1  \circ  \frac{\delta   ~  }{\delta \H }T^{(1)}_1  +   \widetilde{T}^{(1)}_1  T^{(1)}_1 \, . 
  \ee

In addition to this we have a further consistency requirement that comes from demanding that $\H_B$ preserves $\eta_B$, which invokes not only $\eta = \H \eta^{-1} \H$ but also 
\color{black}
\be
0 = T_n \eta^{-1}\H + \H \eta^{-1} T_n + \sum_{m=1}^{n-1} T_m \eta^{-1} T_{n-m} \, . 
\ee
\color{black}
Inserting the loop expansion of the counter-term one obtains the requirements relevant for two loops:  
\be
\begin{aligned}
0 =& T_1^{(1) }\eta^{-1}\H + \H \eta^{-1} T_1^{(1) } \, ,  \\
0 =& T_1^{(2) }\eta^{-1}\H + \H \eta^{-1} T_1^{(2) } \, , \\
0 =& T_2^{(2) }\eta^{-1}\H + \H \eta^{-1} T_2^{(2) } + T_1^{(1)} \eta^{-1} T_1^{(1)} \, .  \label{eq:consistencycond}
\end{aligned}
\ee

Our findings however show that this renormalisation programme, whilst valid at one-loop order, is unsuccessful at two-loop order.  

Certain features are valid and provide strong checks on the calculation; the $\epsilon^{-2}$ pole of fibre obeys both the pole equation of eq.~\eqref{eq:polefibre} and the consistency relation of the last of eq.~\eqref{eq:consistencycond}.     The $\epsilon^{-2}$-pole on the base $\widetilde{T}_2^{(2)}$ reflects the unbroken Lorentz covariance of the base part of the Lagrangian in eq.~\eqref{eq:TAction} - this is despite that fact that the constituent diagrams that produce $\partial_0 y \partial_0 y$ and $\partial_1 y \partial_1 y$ are vastly different.   Moreover, $\widetilde{T}_2^{(2)}$ is expressible only in terms of $\H$ and $\eta$ which is a deeply non-trivial fact since the individual diagrams depend, in addition, on the connection $\Omega = \V^{-1} \mathrm{d} \V$.   

However, the simple pole on the fibre $T^{(2)}_1$ does not, regardless of the method used to evaluate momentum integrals, satisfy the $O(n,n)$ compatibility requirement  eq.~\eqref{eq:consistencycond}.   This alone is enough to raise serious questions as to the quantum validity of the Lagrangian in eq.~\eqref{eq:TAction} since it would seem to destroy the possibility of integrating out the T-dual coordinates and hence prevent the matching of degrees of freedom with the standard string.   Moreover,   this conclusion is robust and can be reached regardless of the precise prescription for evaluating loop integrals. 

The situation on the base is less pleasant still.  We find that the simple poles do not respect the unbroken Lorentz covariance of the base part of the Lagrangian in eq.~\eqref{eq:TAction}.  That is, $\partial_0 y \partial_0 y$ and $\partial_1 y \partial_1 y$  come with different counter-terms and, moreover, a counter-term is produced for $\partial_0 y \partial_1 y$.   Furthermore, unlike the one-loop result or the two-loop $\epsilon^{-2}$ pole, these counter-terms are not evidently expressible in terms of just $\eta$ and $\H$.

The outline of the paper is as follows. In Section \ref{sec:exp}, we recall the basics of the background field method and furnish an all-order expansion of the action. In Section \ref{sec:oneloop}, we summarise one-loop results as first obtained in \cite{Berman:2007xn, Berman:2007yf}. In Section \ref{sec:twoloop} we first detail the relevant Wick contractions that produce the contributing diagrams to the two-loop calculation.   
We consolidate these in an appendix by then summing all contributing diagrams, reorganised in a basis of tensors of $\H$ and its derivatives, expressing the result in terms of un-evaluated momentum integrals\footnote{This provides an intermediate result that is largely independent of any details of the regularisation procedure. It could be harnessed in further studies which might employ different regularisation schemes.}.   Then, in Section \ref{sec:evaluation} we tackle the evaluation of the resultant loop integrals using the two alternative methodologies described above.    Multiple  technical appendices detailing the calculation supplement the presentation.

\section{Expansion}\label{sec:exp}

To calculate the renormalisation of the action \eqref{eq:TAction} we adopt a background field expansion method, (Taylor) expanding around some classical saddle for the fibre and base coordinates
\begin{equation}\label{eq:expansion}
\X^I = \X_{\mathrm{cl}}^I + \xi^I \ , \quad y = y_{\mathrm{cl}} + \zeta \ . 
\end{equation}

The fluctuations $\xi$ and $\zeta$ are dynamical, whilst the classical background fields $\X_{\mathrm{cl}}$ and $y_{\mathrm{cl}}$ are frozen (i.e. we will not integrate over them in the path-integral approach).
The quantum effective action $\Gamma$ is then obtained via Wick contraction of the exponential of the interacting Lagrangian $\mathcal{L}_\mathrm{I}$
\begin{equation}
e^{i\Gamma}=   e^{i S_{\mathrm{cl}}} \, \langle \exp \left(i \int \mathrm{d}^{2} \sigma {\cal L}_{\mathrm{I}}   \right) \rangle_{1\mathrm{PI}} \, .
\end{equation} 

Some comments are in order. On the right-hand side, only one-particle-irreducible (1PI) diagrams need to be considered and the average is taken with respect to the quantum fluctuations $\xi, \zeta$. $S_{\mathrm{cl}}$ denotes the classical action, i.e. the one comprising of the classical fields $\mathbb{X}_{\mathrm{cl}}$ and $y_{\mathrm{cl}}$ only.\footnote{To ease the notational burden we henceforward  omit the subscript $\X_{\mathrm{cl}}\rightarrow \X$ on the classical background and, where useful, indicate the worldsheet point with a subscript, e.g. $\zeta(\sigma)\equiv \zeta_\sigma$. }  
 We choose\footnote{This is not mandatory, and one could decide to work off-shell for the classical background.} the background fields to be on-shell: terms in $\mathcal{L}_{\mathrm{I}}$ linear in the fluctuations are necessarily proportional to the equations of motion and will be dropped.  
The effective action is effectively recovered taking logarithms, thus removing disconnected diagrams. 

As this is a Taylor expansion of actual coordinate values, it is evidently not geometrically covariant.  In the context of the conventional non-linear $\sigma$-model it is much more preferable to maintain geometric covariance in the calculation, and this can be achieved by means of a covariant background field expansion \cite{Mukhi:1985vy}.   Here, however, the departure from conventional Riemannian geometry entailed by the introduction of the generalised metric ${\cal H}$ means such notion of ``covariant'' background field expansion is currently (at least to our knowledge) lacking.  Instead, we will proceed non-covariantly adopting \eqref{eq:expansion}.

Our first task is to single out, for each fluctuation type, a kinetic term. While for $\zeta$ the latter has a standard form, $\frac{1}{2} \lambda \partial_\mu\zeta \partial^\mu \zeta$, fluctuations along the fibre coordinates require the introduction of a (generalised) vielbein $\V_I{}^A$ to remedy for the non-constant  $\H_{IJ}(y)$ factor. The vielbein and its derivatives are defined by
\begin{equation}
    \H_{IJ} = \V_I{}^A \H_{AB} \V_J{}^B \, , \qquad \eta_{IJ} = \V_{I}{}^A \eta_{AB} \V_{J}{}^B \, , \qquad \Omega_\mu{}^{IJ} = \V^{I}{}_A \partial_\mu \V^{JA} \,  ,
\end{equation}
where $\Omega$ is known as the Weitzenb\"ock connection. Early alphabet capital Latin indices $A, B, \dots$ are used to indicate the flat generalised tangent bundle.  Accordingly, the fluctuations on the torus can be ``flattened'' by defining $\xi^I = \mathcal{V}^I{}_A \xi^A$.

Both $\xi^A$'s and $\zeta$'s have now a canonical kinetic term
\begin{equation} \label{eq:Kinetic} \begin{gathered}
    \mathcal{L}_{\mathrm{K}} = - \frac{1}{2} \H_{AB} \partial_1 \xi^A \partial_1 \xi^B + \frac{1}{2}\eta_{AB} \partial_0 \xi^A \partial_1 \xi^B + \frac{1}{2}\lambda  \pl_\mu  \zeta \pl^\mu  \zeta \, ,
\end{gathered}\end{equation} 
from which two-point functions are easily extracted\footnote{ There is one subtlety if we were to consider dimensional regularisation; when computing the Green function for   $\xi$   one ends up with an expression such as $p_0^2 - p_1^2$.  It is not clear that one should directly trade this combination  for $p^2$ in $d$-dimensions. However our general approach will be to remain strictly in $d=2$ where  $p_0^2 - p_1^2\equiv p^2 $ and only continuing to  $d=2+\epsilon$ when evaluating integrals.} as
\begin{align}
    \langle \xi^A(\sigma) \xi^B(\sigma') \rangle =   \H^{AB}   \Delta(\sigma - \sigma') + \eta^{AB} \theta(\sigma - \sigma')   \,  , \qquad 
    \langle \zeta(\sigma) \zeta(\sigma')\rangle = \lambda^{-1} \Delta(\sigma - \sigma') \, ,
\end{align}
where 
\begin{equation} \label{eq:dt}
    \Delta(\sigma) = \int \frac{\mathrm{d}^{2} p }{(2 \pi)^2} e^{- i p \cdot \sigma} \frac{i}{p^2} \, , \qquad    \theta(\sigma) =  \int \frac{\mathrm{d}^{2} p }{(2 \pi)^2} e^{- i p \cdot \sigma} \frac{i}{p^2} \frac{p_0}{p_1} \, . 
\end{equation}  
Performing the Taylor series one finds the entire interaction Lagrangian, at any order, is given by
\begin{align} \label{eq:GenExp}
    -2\mathcal{L}_{\mathrm{I}} &= -\xi^{I} \xi^{J}  \Omega_{1I}{}^{K} \Omega_{0JK} -\xi^I \partial_1 \xi^A \Omega_{0IA}-  \xi^I \partial_0 \xi^A \Omega_{1IA} \nonumber \\
    &+\sum_{n \geq 2} \biggl( \frac{1}{n!} \mathcal{H}^{(n)}_{IJ} \zeta^n \partial_1 \mathbb{X}^I \partial_1\mathbb{X}^J + \frac{2}{(n-1)!} \H_{IA}^{(n-1)} \zeta^{n-1} \partial_1 \xi^A \partial_1 \mathbb{X}^I \nonumber\\
    &-  \frac{2}{(n-1)!} \H_{IJ}^{(n-1)} \Omega_1{}^J{}_K \zeta^{n-1} \xi^K \partial_1 \mathbb{X}^I + \frac{1}{(n-1)!} \H_{AB}^{(n-1)} \zeta^{n-1} \partial_1 \xi^A \partial_1 \xi^B  \nonumber \\
    &- \frac{2}{(n-2)!} \mathcal{H}^{(n-2)}_{AJ} \Omega_1{}^J{}_I \zeta^{n-2} \xi^I \partial_1 \xi^A + \frac{1}{(n-2)!} \H_{IJ}^{(n-2)} \Omega_1{}^I{}_K \Omega_1{}^J{}_L \zeta^{n-2} \xi^K \xi^L 
    \biggr)  \,  .
\end{align}  

To simplify the presentation of the tensorial structure we have introduced some notation.   We indicate derivatives with respect to the base coordinate  with a dot, e.g. $\dot{\mathcal{H}} \equiv \partial_y \mathcal{H}$, or in general for the $n$-th derivative we use $\mathcal{H}^{(n)}$ (so that   $\mathcal{H}^{(0)} \equiv \mathcal{H}$).  
Concatenations of  (matrix) products of $\mathcal{H}$'s and their derivatives (assuming $\eta$ is used to raise indices whenever needed) will be indicated with  $\cdot$, e.g.  $\mathcal{H}^{(i,j,k)} = \mathcal{H}^{(i)} \cdot \mathcal{H}^{(j)} \cdot \mathcal{H}^{(k)}$. 
Contractions of $\mathcal{H}$  with an external torus leg $\partial_1 \mathbb{X}$ will be shown with a $\bullet$, i.e. $\H_{I\bullet} \equiv \H_{IJ} \partial_1 \mathbb{X}^J$.  We also use as shorthand  $\H^{(n)}_{AB} \equiv \V_A{}^I \H^{(n)}_{IJ} \V_B{}^J$.

 By inspection of this expansion we can immediately make two general statements that are true to all orders perturbatively.  First is that   the $O(n,n)$ pairing $\eta$ does not receive quantum corrections at any order in perturbation theory; as no $\partial_0 \mathbb{X}^I$ legs appear in \eqref{eq:GenExp}\footnote{The reason why this is the case is straightforward:  any $\partial_0 \mathbb{X}^I$ leg must necessarily come from the background field expansion of $\eta_{IJ} \partial_0 \mathbb{X}^I \partial_1 \mathbb{X}^J$; however, $\eta$ is constant and terms linear in the fluctuations are discarded on-shell.}, it is impossible to generate terms proportional to $\partial_0 \mathbb{X}^I \partial_1 \mathbb{X}^J$ upon expanding $\exp(i S_{\mathrm{I}})$.   Secondly, we cannot generate any mixed base-fibre terms of the form $\partial_1   \mathbb{X} \partial_\mu   y$ in the effective action.  This follows as any such term would necessarily involve an odd number of $\xi$ fields, and thus vanish upon Wick contraction. 

\section{One-loop Recap} \label{sec:oneloop}
Before moving on to the two-loop calculation, let us recapitulate the situation at one-loop.  At this order it is sufficient to consider only quadratic terms in the fluctuations and the effective action is given by   (assuming only 1PI and connected diagrams are considered)
\be
\Gamma = S_{\mathrm{cl}} + \langle S_{\mathrm{I}} \rangle + \frac{i}{2} \langle S_{\mathrm{I}}^2 \rangle \, ,
\ee
where $S_{\mathrm{I}} = \int \mathrm{d}^2 \sigma \mathcal{L}_{\mathrm{I}}$. To calculate the quantum correction to $\H_{IJ}\pl_1 \X^I \pl_1 \X^J$ we can ignore the Weitzenb\"ock connection pieces all together, as any such contribution implicitly contains derivatives of the base coordinates $y$ such that the relevant interaction term reads
\begin{equation} \label{eq:SI1}
    {\cal L} _\mathrm{I} \supset    - \frac{1}{4} \ddot{\mathcal{H}}_{IJ} \zeta^2_\sigma \partial_1 \mathbb{X}^I \partial_1 \mathbb{X}^J - \dot{\mathcal{H}}_{AI} \zeta_\sigma \partial_1 \xi^A_\sigma \partial_1 \mathbb{X}^I \,  .
\end{equation}
There are two contributions to consider here.  The first requires only a single copy of the worldsheet and is given by 
\begin{equation}
    - \frac{1}{4} \ddot{\mathcal{H}}_{IJ}  \partial_1 \mathbb{X}^I \partial_1 \mathbb{X}^J \langle \zeta_\sigma \zeta_\sigma \rangle  =  - \frac{1}{4} \H^{(2)}_{\bullet\bullet}   \Delta(0) = - \frac{i}{4} \H^{(2)}_{\bullet\bullet}{\bf I}     ,
\end{equation}
in which $\H^{(2)}_{\bullet\bullet} \equiv \ddot{\H}_{IJ}\pl_1 \X^I \pl_1 \X^J$. 
We will denote this contraction as a {\em bubble} and introduce the divergent integral 
\be
{\bf I} = - i \Delta(0) =  \int \frac{\mathrm{d}^{2} p }{(2 \pi)^2}   \frac{1}{p^2} \, . 
\ee 

The second contribution arises from $\langle S_{\mathrm{I}}^2 \rangle$ and requires two copies of the worldsheet: 
\begin{equation}
     \int \mathrm{d}^2 \sigma_2   \left(\frac{i}{2} \dot{\H}_{AI}\dot{\H}_{BJ} \partial_1 \mathbb{X}^I \partial_1 \mathbb{X}^J\langle\zeta_{\sigma_1} \zeta_{\sigma_2} \rangle \langle \partial_1 \xi_{\sigma_1}^A \partial_1 \xi_{\sigma_2}^B\rangle \right) =   \frac{ i}{2 \lambda} {\bf L}  \H^{(1,1,0)}_{\bullet \bullet} -\frac{ i}{2 \lambda} \widetilde{{\bf L}}  \H^{(1,1)}_{\bullet \bullet}  \,  .
\end{equation}

We will denote such contraction as a {\em loop} and introduce the   integrals 
\begin{equation}
{\bf L}  = \int \frac{\mathrm{d}^2 p }{(2 \pi)^2} \frac{p_1^2}{(p^2)^2}  \, ,  \quad \widetilde{{\bf L}}  = \int \frac{\mathrm{d}^2 p }{(2 \pi)^2} \frac{p_1 p_0 }{(p^2)^2}  \, . 
  \end{equation}
  
On general grounds (e.g. integration over an odd integrand) we may assume $\widetilde{{\bf L}} = 0$, however the integral ${\bf L}$ is expected to result in a UV divergence. 

At one-loop order we can simply regulate IR divergences by the replacement of $p^2 \rightarrow p^2- m^2$ in integrals, and UV divergences can be unambiguously regulated in $d=2+\epsilon$.  The fundamental divergent integral ${\bf I}$ evaluates, in dimensional regularisation, to 
 \begin{equation}
   {\bf I}=    {\bf P}  + \frac{  i \bar{\gamma} }{4\pi} \, , \quad   {\bf P}\equiv \frac{i}{2 \pi \epsilon} \, ,
\end{equation}
 where $O(\epsilon)$ contributions have been dropped and the combination $\bar{\gamma} = \gamma_E + \log(m^2/4 \pi)$ introduced. 
In $d=2+\epsilon$ we can invoke  Lorentz invariance to relate  
\be \label{eq:LtoI}
 \int \frac{\mathrm{d}^d  p }{(2 \pi)^d} \frac{p_\mu p_\nu}{(p^2-m^2 )^2} = \frac{\eta_{\mu \nu} }{d} {\bf I} \, . 
 \ee 

 A naive prescription to compute ${\bf L}$ is to simply set the free Lorentz indices $\mu = \nu = 1$ to  give 
 \be 
{\bf L} \sim  -\frac{1}{2} {\bf I}\, .
 \ee 
 
 This is sufficient at one-loop, and in general allows an unambiguous determination of the leading divergence of any integral we encounter.  However, strictly speaking  the relation \eqref{eq:LtoI} is only valid in $d=2+\epsilon$ and the process of specifying the component $p_1$ of a $d=2 + \epsilon$ dimensional vector is rather ambiguous.  Different prescriptions for doing so will lead to different finite parts.  At two-loop this ambiguity becomes more acute since whilst the $\epsilon^{-2}$ pole will be well determined, a prescription needs to be given to find the $\epsilon^{-1}$ pole.   
 
 The minimal subtraction procedure (i.e. removal of $\frac{1}{\epsilon}$ divergences only) then gives a counter-term   to $\H_{\bullet \bullet}$ of 
\begin{equation} \label{eq:CTt1}
    T_1^{(1)} = \frac{1}{4 \pi \lambda} \left( \H^{(2)} + \mathcal{H}^{(1,1,0)}\right) \,  .
\end{equation}

Calculating the correction to the base term is a little more complicated since there are a number of possible diagrams involving the vielbein pieces.  After some work, and invoking identities such as
\begin{equation}
    \Tr(\H \Omega_0 \H \Omega_0) = \Tr(\Omega_0 \Omega_0) + \frac{1}{2}\Tr(\H^{(1,1)})\partial_0 y\partial_0 y  \,  ,
\end{equation} 
one finds a counter-term to $\lambda$ of 
\begin{equation} \label{eq:CTb1}
    \widetilde{T}_1^{(1)}=- \frac{1}{16 \pi } \Tr\left(\mathcal{H}^{(1,1)}\right) \,  .
\end{equation}
Though far from obvious from the intermediate stages of the calculation, one finds that the counter-term for $\lambda (\partial_0 y)^2$ matches that of $\lambda (\partial_1 y)^2$ and that no terms  in the form of $\partial_0 y \partial_1 y$ are generated, as expected.   According to the general results as per \eqref{eq:bdef}, the $\beta$-functions are extracted as
 \begin{align} \label{eq:betaf1l}
     \beta^{\H}_{(1)} = - \frac{1}{4 \pi \lambda} (\H^{(2)} + \H^{(1,1,0)}) \, , \qquad 
     \beta^\lambda_{(1)} = \frac{1}{16 \pi} \Tr(\H^{(1,1)}) \,  ,
 \end{align}
where the subscript is used to emphasise the loop order we are working at.  
 
In the above we regulated IR divergences in an ad-hoc fashion by replacing  $p^2 \rightarrow p^2 - m^2$.    However, the inclusion of a mass term is a delicate matter as it is in general disruptive to the chiral nature of the fields $\mathbb{X}$.   Indeed, it seems hard to find a local term that precisely recreates this prescription. We come closer by introducing
 \begin{equation}\label{eq:massLag}
      \mathcal{L}_{\mathrm{mF}} = - \frac{m^2}{4} \H_{IJ} \mathbb{X}^I \mathbb{X}^J \, , \qquad \mathcal{L}_{\mathrm{mB}} = - \frac{\lambda}{2} m^2 y^2 \, , 
 \end{equation}
  as mass terms on the fibre and base, respectively.  For two-loop calculations both the background field expansion of $\mathcal{L}_{\mathrm{mF}}$  mass term, and its one loop renormalisation are  important.   A straightforward calculation shows that the counter-term for the mass is 
\begin{equation}\label{eq:massCT}
    T_m = \frac{1}{4 \pi \lambda} \H^{(2)} \, .
\end{equation}

Together we end up with a one-loop counter-term Lagrangian
\be\label{eq:Lagcounter} 
  \mathcal{L}_{\mathrm{CT}} = - \frac{1}{2\epsilon} \left(T^{(1)}_{1}\right)_{IJ} \partial_1 \X^I \partial_1 \X^J +\frac{1}{2\epsilon} \widetilde{T}^{(1)}_{1} \partial_\mu y \partial^\mu y - \frac{m^2}{4\epsilon} \left(T_m\right)_{IJ}\mathbb{X}^I \mathbb{X}^J \,  .
\ee

In addition, we have a freedom to add any terms that vanish as a consequence of the classical equations of motion satisfied by the background
\be
\Box y + \frac{1}{2} \dot{\H}_{IJ}  \partial_1 \X^I \partial_1 \X^J = 0 \,  .
\ee 
Multiplying this equation by a function $f(y)$ assumed to be first order in derivatives, and integrating by parts allows us to consider the addition of 
\be
 \mathcal{L}_{\mathrm{on-shell}} =  - \frac{1}{\epsilon} \dot{f}(y)\partial_\mu y \partial^\mu y +\frac{1}{2 \epsilon} f(y) \dot{\H}_{IJ}  \partial_1 \X^I \partial_1 \X^J \, . 
\ee  
Choosing $\dot{f}(y) =\frac{1}{2 } \widetilde{T} $  eliminates the base divergence in 
\be 
 \mathcal{L}_{\mathrm{CT}} +  \mathcal{L}_{\mathrm{on-shell}}=  - \frac{1}{2\epsilon} \left(T^{(1)}_{1}-f \dot{\H}\right)_{IJ} \partial_1 \X^I \partial_1 \X^J  - \frac{m^2}{4\epsilon} \left(T_m\right)_{IJ}\mathbb{X}^I \mathbb{X}^J \, .
\ee 
The same result can be obtained by performing a field redefinition $y \rightarrow y - \frac{1}{  \lambda \epsilon} f(y) $.
In what follows choose we will not perform this redefinition (though examine its consequence for the two-loop result in appendix \ref{app:ex}).  
\section{Two-loop Expansion and Wick Contractions}   \label{sec:twoloop}

The two-loop effective action is evaluated to
\begin{equation}
\Gamma  = S_{\mathrm{cl}} + \langle S_{\mathrm{All}} \rangle  + \frac{i}{2}\langle S_{\mathrm{All}}^2 \rangle -  \frac{1}{6}\langle S_{\mathrm{All}}^3 \rangle -  \frac{i}{24}\langle S_{\mathrm{All}}^4 \rangle  
\end{equation}
where the restriction to 1PI connected diagrams is understood.  Here 
$S_{\mathrm{All}} = \int \mathrm{d}^2 \sigma \mathcal{L}_{\mathrm{All}} $ contains the interaction Lagrangian expansion to quartic order in fluctuations {\em and} the background field expansion of the one-loop counter-term Lagrangian to quadratic order\footnote{This approach of expanding the counter term Lagrangian is something of a shortcut and one of the great virtues of the background field method; however when  employing covariant background field expansions in which the quantum-classical splitting is non-linear this approach is not complete \cite{Howe:1986vm} and instead one should renormalise each and every vertex in $\mathcal{L}_{\mathrm{I}}$.   
Here however we {\em are} employing a linear splitting and so anticipate that the resultant Ward identity ensures we can complete the renormalisation by considering only diagrams with external classical background fields.  For completeness we have calculated the full one-loop renormalisation of   $\mathcal{L}_{\mathrm{I}}$ by splitting the fluctuations into a quantum-background (indicated with a tilde) and a dynamical part $\xi \rightarrow \xi + \tilde{\xi}$, $\zeta \rightarrow \zeta+ \tilde{\zeta}$ performing the path-integral over the latter. As expected, doing so recovers the expansion of the one-loop counter-term Lagrangian to quadratic order.}.
  Because this is quite involved we will treat the fibre and the various contributions to Lorentz components  $\partial_0 y \partial_0 y$, $\partial_1 y \partial_1 y$ and $\partial_0 y \partial_1 y$   separately.

  \subsection{Fibre Contributions}
 
 To renormalise the term $\H_{IJ}\pl_1 \X^I \pl_1 \X^J $  we can discard all terms in which the classical background $\Omega$ is involved.  This leaves only a few contributors    
 \bea\non 
\mathcal{L}_{\mathrm{All}} &\supset&\phantom{+} \left(-\frac{1}{4}\H^{(2)}\zeta^2 -\frac{1}{12}\H^{(3)}\zeta^3  -\frac{1}{48}\H^{(4)} \zeta^4   - \frac{1}{2} X^{(2)} \zeta^2 \right)_{IJ} \pl_1 \X^I \pl_1 \X^J\\ \non 
&&+  \left(- \H^{(1)}\zeta -\frac{1}{2} \H^{(2)}\zeta^2 -\frac{1}{6} \H^{(3)}\zeta^3 - X^{(1)} \zeta \right)_{IJ} \partial_1 \xi^{I} \partial_1 \X^J \\\non 
&& +\left(-\frac{1}{2} \H^{(1)} \zeta  - \frac{1}{4}\H^{(2)}\zeta^2  -X \right)_{AB}\partial_1 \xi^{A} \partial_1 \xi^{B}    \\ 
&& + \frac{1}{2}Y \partial_\mu \zeta \partial^\mu \zeta \,  .
\eea

Here   $X_{IJ} = \frac{1}{\epsilon}   \left(T^{(1)}_{1}\right)_{IJ}$ and $Y   =    \frac{1}{\epsilon}  \tilde{T}^{(1)}_{1}$ arise from the expansion of the counter-term Lagrangian in the MS scheme.  
Schematically we group the terms here in the number of classical background fields as
 \bea
\mathcal{L}_{\mathrm{All}} &\supset&  \underbrace{A^{[2]}_{IJ} \pl_1 \X^I \pl_1 \X^J}_{{\cal A}^{[2]}}  +   \underbrace{A^{[1]}_{A } \partial_1 \X^A}_{{\cal A}^{[1]}}    + {\cal A}^{[0]} \, , 
\eea
where we further denote by  ${\cal A}^{[n]}_i$ the term in ${\cal A}^{[n]}$ that contains $i$ derivatives of $\H$.    
As the loop expansion is organised into a derivative expansion of the generalised metric, two-loop contributions occur with fourth order in derivatives.   Since ${\cal A}^{[0]} $ carries at least one derivative, the expansion of $\Gamma$ truncates to this order with the term $\langle \mathcal{L}_{\mathrm{All}}^4 \rangle$.  We   only require terms with exactly two occurrences of the background field $\partial_1 \X$.  This will give the following contributions to deal with 
 \bea
&& a_1= \langle  {\cal A}^{[2]}   \rangle \,  ,\quad  a_2 = i \langle  {\cal A}^{[0]}   {\cal A}^{[2]}    \rangle  \,  ,\quad 
a_3  = \frac{i}{2} \langle  {\cal A}^{[1]}   {\cal A}^{[1]}   \rangle  \, , \non \\
&& a_4= -\frac{1}{2} \langle  {\cal A}^{[0]}   {\cal A}^{[0]} {\cal A}^{[2]}  \rangle\, , \quad 
a_5  =  -\frac{1}{2} \langle {\cal A}^{[0]}   {\cal A}^{[1]} {\cal A}^{[1]}    \rangle \, , \quad  
a_6 = -\frac{i}{4} \langle {\cal A}^{[0]}   {\cal A}^{[0]}  {\cal A}^{[1]} {\cal A}^{[1]}   \rangle  \,  .
\eea
 
The first step is to evaluate the Wick contractions to obtain expressions containing  an un-evaluated   momentum integral of the form 
\be
[[T(p_0,p_1,k_0,k_1)]]_{i, j, k} =  \int \frac{\mathrm{d}^{2}k}{(2\pi)^{2}} \frac{\mathrm{d}^{2}p}{(2\pi)^2}  \frac{T(p_0,p_1,k_0,k_1) }{(p^{2})^{i}(k^{2})^{j}[(k+p)^{2}]^{k}  } \, , 
\ee
where the $T(p_0,p_1,k_0,k_1)$ will be some specific components of momenta $k$ and $p$, arising predominantly from the fibre propagator terms.  The Wick contraction is standard though tedious, here we only present $a_6$ in detail.  In $a_6$ there are three distinct contractions to consider. The first is
\bea
a_{6_a} &=& -\frac{i}{4}\dot{\H}_{A B} \dot{\H}_{C D} \dot{\H}_{\bullet E} \dot{\H}_{\bullet F} \langle\zeta_{\s_1} \zeta_{\s_3}    \rangle \langle\zeta_{\s_2} \zeta_{\s_4}    \rangle  \langle \partial_1\xi^{A}_{\s_1} \partial_1 \xi^{C}_{\s_2}    \rangle   \langle \partial_1\xi^{B}_{\s_1} \partial_1 \xi^{D}_{\s_2}    \rangle    \langle\partial_1\xi^{E}_{\s_3} \partial_1 \xi^{F}_{\s_4}    \rangle \nonumber  \\ 
&=&- \frac{1}{4 \lambda^2} \Tr(\H^{(1,1)})\H^{(1,1,0)}_{\bullet \bullet} \times \left( [[(p_1+k_1)k_1 p_1^2 ((p_0 +k_0) k_0 - (p_1+k_1)k_1)]]_{3,1,1}  \right) \,  . 
\eea. In deriving this expression we have made use of cyclicity of trace to discard terms involving $\Tr(\H^{(1,1,0)}) = - \Tr(\H^{(1,1,0)})$.
The diagram that gives rise to a $[[\dots ]]_{3,1,1}$ we call {\em square envelope} topology (see Appendix \ref{app:Integrals} for more details).    The remaining contractions  here yield
\bea
a_{6_b} &=& -\frac{i}{4}\dot{\H}_{A B} \dot{\H}_{C D } \dot{\H}_{\bullet E} \dot{\H}_{\bullet F } \langle\zeta_{\s_1} \zeta_{\s_2}    \rangle \langle\zeta_{\s_3} \zeta_{\s_4}    \rangle  \langle \partial_1\xi^{A}_{\s_1}  \partial_1\xi^{C}_{\s_2}    \rangle   \langle \partial_1\xi^{B}_{\s_1}  \partial_1\xi^{E}_{\s_3}    \rangle   \langle \partial_1\xi^{D}_{\s_2}  \partial_1\xi^{F}_{\s_4}    \rangle \nonumber \\ 
&=&  \frac{1}{2 \lambda^2} \H^{(1,1,1,1,0)}_{\bullet \bullet} \left( [[k_1^2 p_0^2 p_1^2]]_{3,1,1} - 2[[k_0 k_1 p_0 p_1^3]]_{3,1,1} + [[k_1^2 p_1^4]]_{3,1,1}  \right)  \, , 
\eea 
which is another square envelope and 
\begin{align}
a_{6_c} &= -\frac{i}{4}\dot{\H}_{AB} \dot{\H}_{C D } \dot{\H}_{\bullet E} \dot{\H}_{\bullet F } \langle\zeta_{\s_1} \zeta_{\s_3}    \rangle \langle\zeta_{\s_2} \zeta_{\s_4}    \rangle  \langle \partial_1\xi^{A}_{\s_1}  \partial_1\xi^{C}_{\s_2}    \rangle   \langle \partial_1\xi^{B}_{\s_1} \partial_1 \xi^{F}_{\s_4}    \rangle   \langle \partial_1\xi^{D}_{\s_2}  \partial_1\xi^{E}_{\s_3}    \rangle \nonumber \\ 
&= \frac{ 1}{2 \lambda^2} \H^{(1,1,1,1,0)}_{\bullet \bullet} \biggl( [[(p_1+k_1)^2 p_1^2 k_1^2]]_{2,2,1} - 2[[(p_1+k_1) (p_0+k_0)  p_1 p_0 k_1^2]]_{2,2,1} \nonumber \\
&+ [[(p_1+k_1)^2 p_1  p_0 k_1 k_0 ]]_{2,2,1}  \biggr)   \, ,  
\end{align}
 where the $[[\dots ]]_{2,2,1}$ integral arises from what we call {\em diamond sunset} topology. 
 
The remaining contributions $a_1, \dots, a_5$ are dealt with, in an similar fashion, in Appendix \ref{app:Fibre}. 
Combining these contributions results in the following tensor structures with coefficients given by un-evaluated integrals:

\begin{empheq}[box=\ovalbox]{align} 
    \H^{(4)}_{\bullet\bullet}:\quad&  \frac{1}{16} {\bf I}^2 - \frac{1}{8} {\bf I} {\bf P} \nonumber \\  \H^{(3,1,0)}_{\bullet\bullet}:\quad&  \frac{1}{2} ( {\bf P} - {\bf I}) [[p_1^2]]_{2,0,0}  - \frac{1}{4} {\bf I}{\bf P}\nonumber \\  \H^{(2,0,2)}_{\bullet\bullet}:\quad& \frac{1}{4} {\bf I }{\bf P} + \frac{1}{4} [[p_1^2]]_{1,1,1}\nonumber  \\  \H^{(2,1,1)}_{\bullet\bullet}:\quad& \frac{1}{2}{\bf I} {\bf P} - \frac{1}{2} {\bf P} [[p_1^2]]_{2,0,0} + [[p_1 k_1 p \cdot k]]_{2,1,1}\nonumber  \\  \H^{(1,2,1)}_{\bullet\bullet}:\quad&   \frac{1}{8} {\bf I}{\bf P} - \frac{1}{2}{\bf P} [[p_1^2]]_{2,0,0} - \frac{1}{2} [[p_1^2 k_1^2]]_{2,2,0} - \frac{1}{4} ({\bf P} - {\bf I}) [[p_1^2 p^2]]_{3,0,0}  \nonumber \\  \H^{(1,1,1,1,0)}_{\bullet\bullet}:\quad& \frac{1}{2} {\bf P} [[p_1^2]]_{2,0,0} + [[p_1^2 k_1^2]]_{2,2,0} + \frac{1}{2} {\bf I} [[p_1^4]]_{3,0,0} + \frac{1}{4} {\bf P} [[p_1^2 p^2]]_{3,0,0} \nonumber \\ 
    & - [[p_1^3 k_1 k\cdot p]]_{3,1,1} + \frac{1}{2}[[p_1^2 k_1^2   p^2]]_{3,1,1}    -   [[(p_1^2 k_1^2  + p_1^3 k_1)  k^2]]_{2,2,1}    \nonumber\\ 
    \H^{(2)}_{\bullet\bullet} \Tr (\H^{(1,1)}):\quad& -\frac{1}{32}{\bf P} [[p^2]]_{2,0,0} + \frac{1}{8} [[(p+k)_1 k_1 (p\cdot k + k^2)  ]]_{2,1,1} \nonumber \\  
 \qquad  \H^{(1,1,0)}_{\bullet\bullet} \Tr(\H^{(1,1)}):\quad& \frac{1}{16}{\bf P}  [[p_1^2 p^2]]_{3,0,0} - \frac{1}{4} [[ (p_1 +k_1) p_1^2 k_1 (p\cdot k + k^2) ]]_{3,1,1} \, . 
\end{empheq}
This boxed result provides a determination of the contributions to the generalised metric two-loop counter term for which any subsequent regularisation scheme can be employed.   Occurrences of ${\bf P} = \frac{i}{2 \pi \epsilon}$ arise in these expressions from diagrams with one-loop counter-term insertions.    The explicit evaluation of the remaining integrals is a delicate matter and will be discussed at length in Section \ref{sec:evaluation}.
 
 In addition, if IR divergences are to be regulated by including an explicit mass term as described before, the background field expansion of the mass terms in the Lagrangian contained in eq. \eqref{eq:massLag} must be included as should one-loop diagrams with the mass counter-term of  eq. \eqref{eq:massCT} insertion. These contractions are detailed in Appendix \ref{app:Fibre}.
 
 \subsection{Base Contributions}
  
Contributions to the base manifold can be organised in terms of the type of external legs, $(\partial_0y)^2$, $(\partial_1 y)^2$ or $\partial_0 y \partial_1 y$, they come with. Given the chiral nature of the action \eqref{eq:TAction} these are better treated separately; eventually, we shall compare our findings for the various cases and comment on the (broken) Lorentz invariance of the final result. 

Restricting ourselves to  $(\partial_0y)^2$, we can discard all terms in which the classical backgrounds $\Omega_1$ and $\partial_1 \X$ are involved.  This leaves only a few contributors 
\bea
\mathcal{L}_{\mathrm{All}} &\supset&  -\frac{1}{2} \Omega_{0 A I} \partial_1 \xi^{A} \xi^I  \nonumber \\
&& -\frac{1}{2}\left(\H^{(1)} \zeta  +\frac{1}{2}\H^{(2)}\zeta^2  +X \right)_{AB}\partial_1 \xi^{A} \partial_1 \xi^{B}   \nonumber \\
&& + \frac{1}{2} Y  \partial_\mu \zeta  \partial^\mu \zeta +  Y^{(1)} \zeta \partial_\mu \zeta \partial^\mu y +\frac{1}{4}  Y^{(2)} \zeta^2 \partial_\mu y  \partial^\mu y \,  .
\eea

Similarly to the fibre, we group terms in the number of classical background fields they come with. Recalling that $\Omega_0$ counts as a $\partial_0y$ insertion, we have
 \bea
\mathcal{L}_{\mathrm{All}} &\supset& {\cal B}^{[0]} +\underbrace{B^{[1]}  \partial_0 y  }_{{\cal B}^{[1]}}    +  \underbrace{B^{[2]}(\partial_0 y)^2 }_{{\cal B}^{[2]}}  \, . 
\eea
 We denote by  ${\cal B}^{[n]}_i$  the term in ${\cal B}^{[n]}$ that contains $i$ derivatives of $\H$ or $\V$ such that 
\bea
{\cal B}^{[0]}  = {\cal B}^{[0]}_1 + {\cal B}^{[0]}_2 \, , \qquad 
{\cal B}^{[1]}   = {\cal B}^{[1]}_1 + {\cal B}^{[1]}_3\, , \qquad
{\cal B}^{[2]}   = {\cal B}^{[2]}_4 \, .
\eea  
First we consider only contractions that lead to exactly two occurrences of $\partial_0 y$ and four derivatives of four derivatives in $\H$  or $\V$.   These are given by
\bea
b_1 &=& \langle  {\cal B}^{[2]}_4   \rangle   \, ,  \quad  \non 
 b_2  =  i \langle   {\cal B}^{[1]}_1   {\cal B}^{[1]}_3   \rangle  \\
 b_3 &=& -\frac{1}{2} \langle {\cal B}^{[0]}_2   {\cal B}^{[1]}_1 {\cal B}^{[1]}_1    \rangle      \, ,  \quad 
b_4  =  -\frac{i}{4} \langle {\cal B}^{[0]}_1   {\cal B}^{[0]}_1  {\cal B}^{[1]}_1 {\cal B}^{[1]}_1   \rangle  \, . 
\eea
We detail the explicit evaluation of each of these in Appendix \ref{app:Base00}, but note here that $b_2$ vanishes outright since it contains no connected diagram. 

With the classical background fields viewed as {\em external} legs to the diagrams in $b_1-b_4$, the divergences are extracted at zero {\em external} momenta (i.e the momenta associated to the Fourier transform of the background field on external legs).    In addition to these contributions,  we must   take into account some diagrams for which the vertices contribute fewer than four derivatives of background fields, but for which the loop integrals carry divergences that are linear or quadratic in the  external momenta.   Fourier transforming these external momenta then produces a worldsheet derivative which acts on the  background fields $\H, \V,\Omega$.  The inclusion of such contributions is vital to the cancellation of terms involving $\Omega$ which could not otherwise be rewritten in terms of the generalised metric $\H$.  If vertices contain no occurrences of $\partial_0 y$ and two derivatives of $\H$ or $\V$,  the relevant contribution is
\bea
 b_5  =    \frac{i}{2} \langle   {\cal B}^{[0]}_1 {\cal B}^{[0]}_1    \rangle      \,    . 
\eea
If vertices  contain  $\partial_0y$ once  and   three derivatives of $\H$ or $\V$, the relevant contributions are
\bea
 b_6  =  i \langle   {\cal B}^{[1]}_1   {\cal B}^{[0]}_2   \rangle \, ,  \quad 
 b_7  =  -\frac{1}{2} \langle {\cal B}^{[1]}_1   {\cal B}^{[0]}_1 {\cal B}^{[0]}_1    \rangle      \,    . 
\eea
Note that $b_{5,6,7}$ do also source $(\partial_1 y)^2$ and  $\partial_1 y \partial_0 y$ terms, which we will carry forward for inclusion in  the relevant calculation later.  The evaluation of these contractions is detailed in the appendix.

Terms with $\partial_0 y \partial_1 y$ legs are somewhat simple to study. Adopting our by now familiar approach, we single out in $\mathcal{L}_{\mathrm{All}}$ the relevant terms and name them $\mathcal{C}_i^{[n; \sigma^\mu]}$ where $[n; \sigma^\mu]$ denotes the number of occurrences of $\partial_\mu y$ and $i$ the  number of derivatives of background fields.The relevant combinations are
  \begin{align}
     c_1 &= - \frac{i}{2} \langle  \mathcal{C}_2^{[2; \tau, \sigma]} \mathcal{C}_1^{[0]} \mathcal{C}_1^{[0]} \rangle \, , \quad
     c_2 = - \langle \mathcal{C}_2^{[2; \tau, \sigma]} \mathcal{C}_2^{[0]} \rangle \, , \quad 
     c_3 = \frac{1}{2} \langle \mathcal{C}_1^{[1; \tau]} \mathcal{C}_1^{[1; \sigma ]}  \mathcal{C}_1^{[0]} \mathcal{C}_1^{[0]}\rangle \, , \\
     c_4 &= - i \langle \mathcal{C}_1^{[1;\tau]} \mathcal{C}_1^{[1;\sigma]}  \mathcal{C}_2^{[0]} \rangle \, , \quad 
     c_5 = - i \langle \mathcal{C}_2^{[1;\sigma]} \mathcal{C}_1^{[1;\tau]} \mathcal{C}_1^{[0]}\rangle \, , \quad
     c_6 = - \langle \mathcal{C}_3^{[1;\tau]} \mathcal{C}_1^{[1;\sigma]}\rangle \,  .
 \end{align}
 
These $c$'s must be supplemented by some other contributions to get the full picture; in fact, as anticipated, integrals with external momentum insertion, such as the ones appearing in $b_{5,6,7}$, can give rise to terms with $\partial_0 y\partial_1y$ legs. We shall deal with them explicitly in Appendix \ref{app:Base01}.
 
Terms with $(\partial_1y)^2$ legs are high in number and complexity when compared to those we just analysed. To deal with them (and the others, too) more efficiently we have created an appropriate \texttt{Mathematica} notebook. We shall collect the results for all Wick contractions, including those on the fibre, in Appendix \ref{sec:preev}. Albeit involved, they are obtained with minimal assumptions\footnote{We have assumed in particular: momentum routing as per the figures in Appendix \ref{app:Integrals}, Taylor expansion of integrals involving external momentum insertion.} and could be employed as the starting point for testing new methods for the  evaluation of non-invariant integrals.

 \section{Evaluation of Integrals}\label{sec:evaluation}

 We  turn now to the evaluation of the momentum integrals for which we can follow two slightly different methods.  Adopting dimensional regularisation, a critical decision is {\em when} in the calculation one assumes $d=2+\epsilon$ or $d=2$, and what cancellations are made before the  evaluation of integrals:
\begin{itemize}
    \item {\bf Method 1:} We move immediately to $d=2+\epsilon$ and do not make any assumption on the relation between $p_0^2 - p^2_1$ and $p^2$ to combine integrals.  Instead, we use arguments of Lorentz invariance to evaluate the myriad of tensor integrals  $[[T(p_0,p_1,k_0,k_1)]]_{i, j, k}$ that are encountered.
    \item  {\bf Method 2:} We remain in $d=2$ for as long as possible, and simplify combinations of integrals by replacing   $p_0^2= p^2+ p^2_1$, $k_0^2= k^2+ k^2_1$ and $p_0 k_0 = p \cdot k + p_1 k_1$. The invariant combination are left to cancel against the denominators. Only once all such cancellations are made we continue to $d=2+\epsilon$.  This method dramatically simplifies the situation as the calculation can be reduced to the evaluation of just four loop integrals. 
\end{itemize}

\subsection{Method 1} 
 
The first strategy we consider is to use  $O(d)$ symmetry to relate the various integrals with non-scalar numerators in terms of a basis   of scalar integrals, i.e. those of form $[[f(p^2, k^2, p\cdot k  ]]_{i,j,k}$, which can be easily evaluated in terms of the basic dimensional regularised integrals (in Minkowski space)
\begin{equation}
    I_n = \int \frac{\mathrm{d}^d p}{(2 \pi)^d}  \frac{1}{(p^2 - m^2)^n} \,  .
\end{equation}

In particular, the integrals we shall encouter are
\be
I_1= {\bf I} = \frac{i}{2 \pi  \epsilon }+\frac{i \bar{\gamma }}{4 \pi } \, ,\qquad m^2 I_2 = \frac{i}{4 \pi  } \, ,\qquad
m^4 I_3 = -\frac{i}{8 \pi   } \, , 
\ee 
where the expressions have been truncated to the relevant order. To achieve this we operate as follows: given a non-invariant integral of the form $[[p_0^{n_1} p_1^{n_2}  k_0^{n_3} k_1^{n_4}]]_{i,j,k}$ for some $n_i \in \mathbb{N}$, we consider the associated integral where momenta are given $d$-dimensional Lorentz indices, $[[p_{\mu_1} \dots p_{\mu_{n_1 + n_2}}k_{\nu_1} \dots k_{\nu_{n_3 + n_4}}]]_{i,j,k}$. We then use $O(d)$ invariance to argue that the latter should equal a combination of Minkowski metrics multiplied by both a $d$-dependent finite factor and a scalar integral. Finally, we set the Lorentz indices so as to match our initial expression and recover the desired result. All of the relevant integrals are listed in Appendix \ref{app:Integrals}.

 Albeit standard in QFT,  this technique here necessarily involve explicitly the $\eta_{11}$, $\eta_{00}$ and $\eta_{01}$ components of the now $d$-dimensional worldsheet Minkowski metric.  A prescription for these needs to be given and might in principle depend on $\epsilon$.   To keep track of this possibility we consider\footnote{One can  consider a more general choice where $\eta_{00} = 1+ \mathfrak{g} \epsilon $ and $\eta_{11} = -1- \mathfrak{f} \epsilon $, however setting $\mathfrak{f} \neq \mathfrak{g}$ does not produce  simplifications of the   final result.  } $\eta_{00} = - \eta_{11} = 1+ \mathfrak{g} \epsilon$ for some $\mathfrak{g} \in \mathbb{R}$ (with higher orders in $\epsilon$ irrelevant to the two-loop calculation) and $\eta_{01} = \eta_{10}= 0$.  A simple  example encountered is 
 \be
[[p_\mu p_\nu]]_{2,0,0} = \int   \frac{\mathrm{d}^d p}{(2\pi)^d} \frac{p_\mu p_\nu} {(p^2- m^2)^2 } = \frac{\eta_{\mu \nu}}{d} ( I_1 + m^2 I_2 ) \, . 
\ee  
Specialising the Lorentz indices this prescription gives
\be
[[p_0 p_1 ]]_{2,0,0} = 0 \,  , \quad  [[p_0 p_0]]_{2,0,0} = -[[p_1 p_1]]_{2,0,0}  = \frac{1+  \mathfrak{g}\epsilon}{2 +\epsilon}  (I_1 + m^2 I_2) \,  .
\ee 
One might wonder if there is some preferred value of $\mathfrak{g}$ required by consistency.   A natural demand might be to set
\be 
 [[p_0 p_0]]_{2,0,0} -[[p_1 p_1]]_{2,0,0} \equiv [[p_\mu p^\mu ]]_{2,0,0}= (I_1 + m^2 I_2) \, , 
\ee 
 which is achieved for $\mathfrak{g}  = \frac{1}{2}$.   However, consider  now the ``triangle''   integral 
\begin{equation}
    I_\triangle = \int \frac{\mathrm{d}^d p}{(2 \pi)^d} \frac{p_1^2(p_0^2 - p_1^2)}{(p^2 -m^2)^3} =- 4 \frac{(  1+ \mathfrak{g} \epsilon)^2 }{d (d+2)} (I_1+2 m^2 I_2 + m^4 I_3)    \,  .
\end{equation}
In deriving this, we have prevented the numerator from cancelling against the denominator, thereby computing, in schematic form, $[[p_1^2 p_0^2]]_{3,0,0} - [[p_1^4]]_{3,0,0}$.   On the other hand, if we now replace $p_0^2 - p_1^2 = p^2$ {\em prior} to integrating we get 
\begin{equation}
    I_\triangle = [[p_1^2   ]]_{2,0,0} + m^2 [[p_1^2]]_{3,0,0}  = -\frac{1 + \mathfrak{g} \epsilon}{d } (I_1+2 m^2 I_2 + m^4 I_3)   \,  .
\end{equation}
These two results agree in their leading $\frac{1}{\epsilon}$ singularity but differ in the finite parts by $\frac{i(1-4 \mathfrak{g})}{16 \pi}$.   This shows that there is no universal unambiguous choice for $\mathfrak{g}$.   At one-loop this has no material impact on the $\beta$-functions, but at two-loops this ambiguity is dangerous because the $I_\triangle$ appears multiplied by a further $\frac{1}{\epsilon}$ (coming either from a counter-term insertion or from a factorised loop in a diagram).  The prescription we follow in this Method 1  is to {\em not} combine explicit factors of $p_0^2 - p_1^2$ into $p^2$ prior to performing the integral\footnote{With respect to the un-evaluated loop integrals reported in Appendix \ref{sec:preev}, we should re-expand $p^2 = p_0^2 - p_1^2$ etc. and then apply the rules for Method 1.}.   

 When evaluating the counter-term contributions on the base, one additional technical difficulty is posed by integrals with non-invariant denominators, containing explicit components, e.g. $p_1$, in place of invariant combinations. These we tackle by means of a Schwinger parametrisation 
     \begin{equation}
         \frac{1}{p_1} = \int_0^\infty \mathrm{d}u \, e^{- u p_1} \, .
     \end{equation} 
As described in detail in  Appendix \ref{app:Schwinger}, we then proceed formally by series expansion of this exponential to produce a sum of loop integrals with non-invariant numerators; each term  can be recast, using the same $O(d)$ symmetry technique, as some invariant integral multiplied by a combinatorial (and $d$-dependent) factor.  In the integrals encountered, we found that we could resum the series obtaining a hypergeometric function of the invariant combinations of momenta.
The Schwinger parameter can then be integrated using  standard identities for hypergeometric functions\footnote{For instance 
     \begin{equation}
    \int_0^\infty \mathrm{d}u \, u^{\alpha -1} \, {}_0\tilde{F}_1 (b; -u) = \frac{ \Gamma(\alpha)}{\Gamma(b-\alpha)} \,  .
\end{equation}}
to produce an expression for which the final loop momenta can then be integrated. 

At two-loop order we expect divergences of the form $\frac{1}{\epsilon^2}$ and $\frac{1}{\epsilon}$.  The latter contribute to the $\beta$-function whilst the former are constrained from the one-loop $\frac{1}{\epsilon}$ contribution by pole equations.    Terms of $\frac{1}{\epsilon^2}$ can be sourced in one of two ways; either as a two-loop diagram giving a contribution $I_1^2$, or as a one-loop diagram giving a contribution $I_1$ with a counter-term insertion carrying a further $\frac{1}{\epsilon}$.  Terms of $\frac{1}{\epsilon}$, instead, can arise in several ways:
\begin{enumerate}
    \item[(i)] First, expanding  $I_1^2$ and $\frac{1}{\epsilon} I_1$ produces sub-leading $\frac{1}{\epsilon}$ poles proportional to $\frac{\bar{\gamma}}{\epsilon} $.  These we anticipate should cancel out, and indeed the correct counter-term Lagrangian should make this the case.
    \item[(ii)] Second, we can find in a two-loop diagram a contribution proportional to either $ m^2I_1 I_2$ or $m^4I_1  I_3$.  The explicit mass  that enters here as a result of the IR regulator cancels the same in the finite integrals $I_2$ and $I_3$.   We anticipate that these divergences should also cancel as happens in the standard string $\beta$-function. 
    \item[(iii)] Finally we can have a pre-factor $f(d) I_1^2$ whose expansion results in a $\frac{1}{\epsilon}$ pole.  These are the terms responsible for the $\beta$-function.  
\end{enumerate}

We will now collect and present the results for our calculation using Method 1. In doing so, let us recall that terms with a double $\epsilon$-pole shall \emph{not} depend on our prescription for computing integrals and are thus unambiguous. 

Let us start from the fibre. The $\epsilon^{-2}$ counter-term turns out to be
\bea \label{eq:fct}
    T_2^{(2)}&&=\frac{1}{32 \pi^2 \lambda^2}\left(\H^{(4)}+4H^{(3,1,0)}-2\H^{(2,0,2)}-6\H^{(2,1,1)}-4\H^{(1,2,1)}+3\H^{(1,1,1,1,0)}\right)\non\\
&&+\frac{1}{128 \pi^2 \lambda^2}\Tr (\H^{(1,1)} )\left(\H^{(2)}+\H^{(1,1,0)}\right)\, ,
\eea
in exact agreement with  the pole equation   
\be 
  0 =  2  T^{(2)}_2  -  T^{(1)}_1  \circ  \frac{\delta   ~  }{\delta \H }T^{(1)}_1  +   \widetilde{T}^{(1)}_1  T^{(1)}_1 \, . 
 \ee
 
Regarding the single $\epsilon$-pole, $\bar{\gamma}$'s cancel out, as they should, among different $a$'s (equivalently: topologies). Contributions coming from IR regularisation - those involving $I_{2,3}$ - do not vanish on their own but can be removed with the addition of the appropriate mass term \eqref{eq:massLag} to the Lagrangian. 
Crucially this term has a non-trivial expansion so that interaction vertices with mass insertions are produced at all orders. These are relevant, as they can be used to precisely cancel off against $I_{2,3}$-dependent terms produced in the calculation. More concretely, as we explain in Appendix \ref{app:Fibre}, four different topologies are involved: triangle envelope, square envelope, decorated loop and decorated triangle. Their contributions respectively evaluate to 
 \begin{equation} \begin{gathered}
     m_1= \frac{m^2 I_1 I_2}{8 \lambda^2} \Tr(\H^{(1,1)})\H^{(2)}_{\bullet \bullet} \, , \qquad m_2=\frac{I_1 (m^2 I_2+ m^4 I_3)}{8 \lambda^2} \Tr(\H^{(1,1)})\H^{(1,1,0)}_{\bullet \bullet} \, , \\
     m_3 =  -\frac{m^2 I_1 I_2}{16 \lambda^2} \Tr(\H^{(1,1)})\H^{(2)}_{\bullet \bullet}  \, , \qquad m_4 =- \frac{I_1 (m^2 I_2+ m^4 I_3)}{16 \lambda^2} \Tr(\H^{(1,1)})\H^{(1,1,0)}_{\bullet \bullet}\,  .
 \end{gathered}\end{equation}
In summation these yield
 \begin{equation}
     \sum_{i=1}^4 m_i = -\frac{1}{256 \pi^2 \epsilon  \lambda^2}\Tr(\H^{(1,1)}) \left(\H^{(1,1,0)}_{\bullet \bullet} + 2 \H^{(2)}_{\bullet \bullet} \right) \,  .
 \end{equation}
 
Including these, which precisely cancel all $m^2 I_1 I_2$ and $m^4 I_1 I_3$ contributions,  we find the $\epsilon^{-1}$ counter-term 
\begin{align}
    T_1^{(2)} &= \frac{2 \mathfrak{g}-1}{32 \pi^2 \lambda^2 } \H^{(2,0,2)} + \frac{4 \mathfrak{g}-1}{16\pi^2 \lambda^2 } \H^{(2,1,1)} + \frac{3(8 \mathfrak{g}-1)}{128\pi^2 \lambda^2 } \H^{(1,1,1,1,0)} \nonumber  \\ 
    &+ \frac{1- 2 \mathfrak{g}}{128\pi^2 \lambda^2 } \Tr(\H^{(1,1)})\H^{(2)} + \frac{1 - 4 \mathfrak{g}}{512 \pi^2 \lambda^2} \Tr(\H^{(1,1)})\H^{(1,1,0)} \,  .
\end{align}

Let us consider the constraint that $T_1^{(2)} $ be compatible with the $O(n,n)$ structure.  We find that the consistency condition of eq.~\eqref{eq:consistencycond} is not obeyed and instead:
\begin{align} 
  T_1^{(2)}\eta^{-1} \H + \H \eta^{-1} T_1^{(2)} =&    \frac{4 \mathfrak{g}-3}{256 \pi^2 \lambda^2 }\Tr(\H^{(1,1)}) \H^{(1,1)} +  \frac{13 - 40 \mathfrak{g} }{64 \pi^2 \lambda^2 } \H^{(1,1,1,1)} \nonumber \\ & + \frac{2 \mathfrak{g}-1}{32 \pi^2 \lambda^2 }  \left(\H^{(2,0,2,0)} + \H^{(0,2,0,2)}  \right)\, , 
\end{align}  
which does not vanish for any choice of $\mathfrak{g}$. 

On the base we find  the   $\epsilon^{-2}$ pole 
\begin{equation} \label{eq:bct}
    \tilde{T}_2^{(2)} = - \frac{1}{64 \pi^2 \lambda} \left( 2 \Tr(\H^{(3,1)}) + \Tr(\H^{(2,2)}) + \Tr(\H^{(1,1,1,1)})\right) \, ,
\end{equation}
in which it is notable that we could combine all terms containing the connection $\Omega$ to give a final answer in terms of the generalised metric alone.   Moreover, we find that the $\epsilon^{-2}$ counter-term for $\partial_0 y \partial_0 y $ matches that of $\partial_1 y \partial_1 y$ (despite arising from a totally different set of diagrams and contractions) and that no $\epsilon^{-2}\partial_0 y \partial_1 y $ counter-term  is produced.   

Turning to the single $\epsilon$-pole on the base, $\bar{\gamma}$'s cancel out. Contributions coming from IR regularisation cancel entirely from diagrams with no momentum insertion (and we assume that this is also the case for diagrams that do have a momentum insertion).  For the remaining  $\epsilon^{-1}$  counter-terms,   indicated with $\tilde{T}_{1}^{(2)}|_{\mu \nu} \partial^\mu y \partial^\nu y $, we obtain 
\begin{align}
      \tilde{T}_{1}^{(2)}|_{01} &= \frac{1 - 6 \mathfrak{g}}{64 \pi^2 \lambda } \Tr(\H^{(1,1,0)}\Omega \Omega) - \frac{ \frac{16}{3} + 4 \mathfrak{g} }{64 \pi^2 \lambda } \Tr(\H^{(2)}\Omega \Omega) \, , \\
    \tilde{T}_{1}^{(2)}|_{00} &= \frac{1-  8 \mathfrak{g}}{128 \pi^2  \lambda} \Tr(\H^{(1,1,1,1)}) - \frac{1}{384\pi^2  \lambda}(13 - 10 \mathfrak{g}) \Tr(\H^{(2,2)}) \nonumber \\
    &+ \frac{35 - 8 \mathfrak{g}}{384\pi^2  \lambda} (\Tr(\H^{(1,1)}\Omega \Omega)  - \Tr(\H^{(1)}\Omega \H^{(1)} \Omega)) + \frac{1}{384\pi^2  \lambda} (56 + 8 \mathfrak{g}) \Tr(\H^{(2,1)}\Omega) \, , \\
    \tilde{T}_{1}^{(2)}|_{11} &= \frac{-3+4 \mathfrak{g}}{128 \pi^2  \lambda} \Tr(\H^{(1,1,1,1)}) - \frac{9 - 10 \mathfrak{g}}{384\pi^2 \lambda} \Tr(\H^{(2,2)}) \nonumber \\
    &+ \frac{13+16 \mathfrak{g}}{384\pi^2  \lambda} \Tr(\H^{(1,1)}\Omega \Omega)  - \frac{11-8 \mathfrak{g}}{384\pi^2  \lambda} \Tr(\H^{(1)}\Omega \H^{(1)} \Omega) + \frac{5(1- \mathfrak{g})}{48\pi^2  \lambda}  \Tr(\H^{(2,1)}\Omega) \,  .
\end{align}
It is clear that not only does the result depend on the connection $\Omega$ rather than $\H$ alone,  there is no value for $\mathfrak{g}$ for which  $\tilde{T}_{1}^{(2)}|_{01} =0$ and $   \tilde{T}_{1}^{(2)}|_{00}=   \tilde{T}_{1}^{(2)}|_{11}$.

\subsection{Method 2} 

In this method we shall perform the maximal simplifications that we can before actually evaluating any integral.  We make four key assumptions:  
\begin{enumerate}
    \item Integrals are first dealt with in $d =2$. In particular, we replace factors of   $p_0^2$ in the numerator of momentum integrals with $p^2  + p_1^2$ and cancel off against factors of $p^2$ between numerator and the denominator.\footnote{This step is potentially ambiguous as there can be multiple ways to implement such a simplification, e.g. in $k_0^3 p_0^3$ we could extract either $(k\cdot p)^3$ or $k^2 p^2 k\cdot p$. However, at least at the two-loop order we are working to, no such possible ambiguity occurs.} 
    \item After these cancellations have happened we continue the integral to $d = 2 + \epsilon$ dimensions and, in particular, we assume shift-symmetry in the momenta $k$ and $p$. 
    \item IR regulating should be done at the end of such simplifications, and based on the experience of Method 1 when done successfully will not be important for the $\frac{1}{\epsilon}$ pole. 
    \item Having done these simplifications, we will assume that the integration method is such that any integrand whose numerator contains an odd number of temporal components of momenta vanish.\footnote{One could do away this restriction, however this will not effect the broad conclusions we reach as 
    integrands with odd and even numbers of temporal   components of momenta are associated to different tensorial combinations of the background fields. This is because a factor of $p_1 p_0$ can only arise  in conjunction with an $\eta$ and a factor of $p_1 p_1$ comes with an $\H$. 
    } 
\end{enumerate} 
This methodology greatly assists in dealing with integrals that contain non-Lorentz invariant denominators. For instance, consider the following expression which is encountered in the computation
\begin{equation}
     {\cal J}  = \int \frac{\mathrm{d}^d k}{(2 \pi)^d} \frac{\mathrm{d}^d p}{(2 \pi)^d} \frac{p_1}{k^2 p^2 (k_1 + p_1)} \,  .
\end{equation}
Suppose in the numerator we sum and subtract $k_1$:
\begin{equation} \begin{gathered}
    {\cal J}  = \int \frac{\mathrm{d}^d k}{(2 \pi)^d} \frac{\mathrm{d}^d p}{(2 \pi)^d} \frac{k_1 + p_1}{k^2 p^2 (k_1 + p_1)} - \int \frac{\mathrm{d}^d k}{(2 \pi)^d} \frac{\mathrm{d}^d p}{(2 \pi)^d} \frac{k_1}{k^2 p^2 (k_1 + p_1)} \\
    = {\bf I}^2 - \int \frac{\mathrm{d}^d k}{(2 \pi)^d} \frac{\mathrm{d}^d p}{(2 \pi)^d} \frac{k_1}{k^2 p^2 (k_1 + p_1)} = {\bf I}^2  -    {\cal J}  \, ,
\end{gathered}\end{equation}
where in the last step we used the fact that the denominator is invariant under the swapping of $k$ and $p$. Hence, we see that the integral is easily solved as $ {\cal J}  = \frac{1}{2} {\bf I}^2 $. 

Further simplifications follow from the momenta shift-symmetry.  Consider $[[(k_1 + p_1)^2]]_{1,1,1}$;  expanding the square and using the  $k \leftrightarrow p$ symmetry of the integrand gives $[[(k_1 + p_1)^2]]_{1,1,1} = 2 [[k_1^2]]_{1,1,1} + 2[[k_1 p_1]]_{1,1,1}$. On the other hand,  shifting $k \rightarrow k - p$ followed by $p \rightarrow - p$ yields   $[[(k_1+p_1)^2]]_{1,1,1} = [[k_1^2]]_{1,1,1}$. Hence  $[[k_1 p_1 ]]_{1,1,1} = - \frac{1}{2} [[k_1^2]]_{1,1,1}$.

With these rules implemented the entire two-loop contributions can be  remarkably expressed in terms of only five (non-invariant)  integrals: 
\begin{align}
   {\bf I}  &= \int \frac{\mathrm{d}^d k }{(2 \pi)^d} \frac{1}{k^2} \, ,  \quad  {\bf L} = \int \frac{\mathrm{d}^d k }{(2 \pi)^d} \frac{k_1^2}{(k^2)^2} \, , \qquad    {\bf T}  = \int \frac{\mathrm{d}^d k }{(2 \pi)^d} \frac{k_1^4}{(k^2)^3} \, , \nonumber \\
     {\bf S}  &= \int \frac{\mathrm{d}^d k }{(2 \pi)^d} \frac{\mathrm{d}^d p }{(2 \pi)^d} \frac{k_1^2}{ p^2 k^2 (k+p)^2} \, , \qquad     {\bf TE} = \int \frac{\mathrm{d}^d k }{(2 \pi)^d} \frac{\mathrm{d}^d p }{(2 \pi)^d} \frac{k_1 p_1^2(k_1 + p_1)}{(p^2)^2k^2  (k+p)^2} \, ,
\end{align}
corresponding respectively to the fundamental integral {\bf I} and then integrals in the {\bf L}oop, {\bf T}riangle, {\bf S}unset and {\bf T}riangle {\bf E}nvelope topologies.   For each of these we denote by ${\bf X}_{(i)}$ the $\frac{1}{\epsilon^i}$ contribution to the integral ${\bf X}$.   
The leading divergence and $\bar{\gamma}$-dependence of the remaining integrals is unambiguous, such that we may express  
\begin{align} \label{eq:LTdef}
  {\bf I}&=    {\bf P}  + \frac{  i \bar{\gamma} }{4\pi} \, , \qquad \,   
   {\bf L} = - \frac{1}{2}  {\bf P}  - \frac{i \bar{\gamma} }{8 \pi }+  i {\bf L}_{(0)} \,  , \qquad    {\bf T}= \frac{3}{8} {\bf P}+ \frac{3 i \bar{\gamma} }{32 \pi }   + i  {\bf T}_{(0)}  \, ,  \nonumber \\   {\bf S}&= - \frac{1}{2}  {\bf P}^2 + \frac{ \bar{\gamma} }{8 \pi^2 \epsilon  } + \frac{  {\bf S}_{(1)}}{\pi \epsilon}
   \, , \qquad     {\bf TE}= \frac{1}{8}  {\bf P}^2 - \frac{ \bar{\gamma} }{32 \pi^2 \epsilon  }  + \frac{ {\bf TE}_{(1)}}{\pi \epsilon }   \,  ,
\end{align} 
in which we recall   ${\bf P}\equiv \frac{i}{2 \pi \epsilon}$ and have introduced    ${\bf L}_{(0)}, {\bf T}_{(0)},  {\bf S}_{(1)} $ and  $ {\bf TE}_{(1)}$ to signify the undetermined  contributions from these integrals.

In general, two-loop diagrams which can be factorised into the product of one-loop diagrams do not lead to simple $\frac{1}{\epsilon}$ poles once the appropriate one-loop diagrams with counter-term insertions are subtracted off \cite{Jack:1989vp}.   Here we see this through contributions of the form $({\bf I} - {\bf P} ) {\bf L}$  where ${\bf P}$ comes from a counter-term insertion in the MS scheme;  the simple pole part of $({\bf I} - {\bf P})$ drops such that only a term  proportional to $\frac{\bar{\gamma}}{\epsilon}$ is produced (such terms should cancel with an appropriate treatment of counter-terms). Using \eqref{eq:LTdef} we see similarly that the combination  $( {\bf P} + {\bf L} ) {\bf L} $ has no $\bar{\gamma}$-independent $\frac{1}{\epsilon}$ contribution. 

To expose the simplifications of this method one needs to organise the calculation by grouping all terms with the same tensorial structure.  In general these occur from different Wick contractions and across different topologies.  Let us highlight this method by examining the term $\H^{(1,1,1,1,0)}$  in more detail. The terms in $a_6$ require the most work and proceed as  follows.  The strategy is first remove all temporal components of momenta by replacing   e.g. $k_0 p_0 = k\cdot p + k_1 p_1$, and then cancel numerators and denominators.  For example
\bea
a_{6_b} 
 &=&   \frac{1}{2 \lambda^2} \H^{(1,1,1,1,0)}_{\bullet \bullet} \left(   - 2[[   k_1  p_1^3 k\cdot p ]]_{3,1,1} + [[k_1^2 p_1^2 p^2]_{3,1,1}  \right)  \nonumber \\
  &=&   \frac{1}{2 \lambda^2} \H^{(1,1,1,1,0)}_{\bullet \bullet} \left(   - [[  k_1  p_1^3   ]]_{3,1,0}+[[  k_1  p_1^3   ]]_{3,0,1}   +[[  k_1  p_1^3   ]]_{2,1,1} +  [[k_1^2 p_1^2  ]_{2,1,1}  \right)  \nonumber \\
    &=&   \frac{1}{2 \lambda^2} \H^{(1,1,1,1,0)}_{\bullet \bullet} \left(   - [[  k_1  p_1^3   ]]_{3,1,0}+[[  (k_1-p_1)  p_1^3   ]]_{3,1,0}   +[[  k_1  p_1^3   ]]_{2,1,1} +  [[k_1^2 p_1^2  ]_{2,1,1}  \right)  \nonumber \\
&=& \frac{1}{2 \lambda^2} \H^{(1,1,1,1,0)}_{\bullet \bullet} \left(  {\bf TE}   -  {\bf I} \times {\bf T}      \right)\, . 
\eea  

The contributions arising from the remaining Wick contractions relevant to  $\H^{(1,1,1,1,0)}$ are summarised in table 1. 
\vskip 0.25cm
\begin{equation*}
    \begin{array}{c|c }
 \text{} &   \lambda^{-2} \mathcal{H}^{(1,1,1,1,0)}_{\bullet \bullet } \\
\hline
 a_{3_c} &  \frac{1}{2} {\bf P}\times   {\bf L}  \\
 a_{5_{1a}} & {\bf L}^2  \\
 a_{5_{1b}} & \frac{1}{2} {\bf I} \times {\bf T}   \\
  a_{5_{1e}}  &  \frac{1}{4} {\bf P}\times {\bf L}   \\
a_{6_{b}}&   \frac{1}{2} \left( {\bf TE} - {\bf I} \times {\bf T}  \right)     \\
a_{6_{c}} & - {\bf TE}  \\
 \hline \hline
 \text{Tot.} &   {\bf L}^2   -\frac{1}{2}  {\bf TE}    + \frac{3}{4} {\bf P} \times {\bf L}  \\
\end{array}
\end{equation*} \begingroup\captionof{table}{
 Method 2 two-loop contributions to $\mathcal{H}^{(1,1,1,1,0)}_{\bullet \bullet }$.\\} \endgroup
 \vskip 0.25cm
 
The other tensorial structures can be treated in a similar fashion, both on the fibre and base manifold. We will refrain from detailing the discussion any further here and rather report our findings in tabular forms. The interested reader is referred to the appendices where the explicit steps for carrying out the calculation are shown. 
 \vskip 0.25cm
\begin{equation*}
    \begin{array}{c|c|c|c|c }
  \text{Tensor} &  \text{Result } & \frac{1}{64 \pi \epsilon^2 } & \frac{\bar{\gamma}}{ 4 \pi \epsilon }  &  \frac{1}{4 \pi \epsilon  }   \\ 
  \hline \hline  \H^{(4)}_{\bullet\bullet}  & \frac{1}{16} {\bf I}^2 - \frac{1}{8} {\bf I }{\bf P }& 1 & 0& 0 \\
   \H^{(3,1,0)}_{\bullet\bullet}  &  -\frac{1}{2} {\bf I} {\bf L} - \frac{1}{4}  {\bf I }{\bf P } + \frac{1}{2 } {\bf P}{\bf L}& -4 & 0& 0 \\
   \H^{(2,0,2)}_{\bullet\bullet}   &  \frac{1}{4} {\bf I }{\bf P}+ \frac{1}{4}{\bf S} & -2 & 0&  {\bf S}_1  \\
  \H^{(2,1,1)}_{\bullet\bullet} & \frac{1}{2} {\bf I }{\bf P} -\frac{1}{2} {\bf P} {\bf L} + \frac{1}{2 } {\bf I} {\bf L} + \frac{1}{4 }{\bf S}& -6 & 0&  {\bf S}_1 \\
  \H^{(1,2,1)}_{\bullet\bullet}  & \frac{1}{8} {\bf I}{\bf P} -\frac{3}{4}{\bf P}{\bf L} -\frac{1}{2} {\bf L }^2 +\frac{1}{4} {\bf I}{\bf L}& -4 & 0& 0 \\ 
  \H^{(1,1,1,1,0)}_{\bullet\bullet}  &  -\frac{1}{2}{\bf TE} + {\bf L}^2 +\frac{3}{4}{\bf P}{\bf L}& 3 & 0& -\frac{1}{2}(- {\bf L_0} + 4 {\bf TE}_1 )\\ 
 \H^{(2)}_{\bullet\bullet} \Tr (\H^{(1,1)})  &  -\frac{1}{32}{\bf I}{\bf P} - \frac{1}{32}{\bf S} & \frac{1}{4}& 0& - \frac{1}{8}   {\bf S}_1  \\ 
 \H^{(1,1,0)}_{\bullet\bullet} \Tr(\H^{(1,1)})  & \frac{1}{16} {\bf P}{\bf L} + \frac{1}{8} {\bf TE}& \frac{1}{4}& 0&  \frac{1}{8}( - {\bf L}_0 + 4 {\bf TE}_1 ) \\ 
\end{array}
\end{equation*}\begingroup\captionof{table}{Two-loop contribution for each tensorial structure on the fibre.} \endgroup
 \vskip 0.25cm
\begin{equation*}
    \begin{array}{c|c|c|c|c }
  \text{Tensor} &  \text{Result } & \frac{1}{64 \pi \epsilon^2 } & \frac{\bar{\gamma}}{ 4 \pi \epsilon }  &  \frac{1}{2 \pi \epsilon  }   \\ 
  \hline \hline  \Tr(\H^{(1,1,0)}\Omega \Omega)  & - \frac{1}{2}{\bf LP} - \frac{1}{2}{\bf L I} - \frac{1}{4}{\bf P I} -2 {\bf TE}& 0 & 0& {\bf L}_0 - 4 {\bf TE}_1 \\
   \Tr(\H^{(2)}\Omega \Omega)  &- \frac{1}{2} {\bf LP} + \frac{1}{2} {\bf LI} - \frac{1}{4}{\bf P I} + 4 {\bf T I} + \frac{1}{4}{\bf I}^2 + 2 {\bf S} -4 {\bf TE} &0 & 0& - \frac{1}{4} {\bf S}_1 - 2 {\bf TE}_1 
\end{array}
\end{equation*}\begingroup\captionof{table}{Two-loop contribution for each tensorial structure on the base with external legs $\partial_0 y \partial_1 y$.\\}\endgroup

\begin{equation*}
    \begin{array}{c|c|c|c|c }
  \text{Tensor} &  \text{Result } & \frac{1}{64 \pi \epsilon^2 } & \frac{\bar{\gamma}}{ 4 \pi \epsilon }  &  \frac{1}{4 \pi \epsilon  }   \\ 
  \hline \hline  \Tr(\H^{(3,1)})  & -\frac{3}{8} {\bf L P} + \frac{3}{8}{\bf L I} - \frac{1}{16}{\bf P I} + \frac{1}{8}{\bf I}^2& -1 & 0& 0 \\
   \Tr(\H^{(2,2)})  & - \frac{1}{16} {\bf S} - \frac{1}{2} {\bf TE} + \frac{1}{16} \bf{P I}& -\frac{1}{2} & 0& - \frac{1}{4} {\bf S}_{(1)} - 2 {\bf TE}_{(1)} \\
   \Tr(\H^{(1,1,1,1)}) &  -\frac{1}{8} {\bf L}^2 + \frac{1}{8} {\bf LP} + \frac{1}{8}{\bf S} + \frac{1}{2} {\bf TE} + \frac{1}{8} {\bf P I} & -\frac{1}{2} & 0&  - \frac{1}{2} {\bf L}_{(0)} + \frac{1}{2}{\bf S}_{(1)} +2 {\bf TE}_{(1)}  \\
  \Tr(\H^{(1,1)}\Omega \Omega) & 2 {\bf LI}+2{\bf T I} + \frac{1}{4} {\bf I}^2& 0 & 0&  -4 {\bf L}_{(0)} -4 {\bf T}_{(0)}  \\
  \Tr(\H^{(1)}\Omega \H^{(1)} \Omega)  &  \frac{1}{2}{\bf L}^2 - \frac{3}{2} {\bf L}{\bf I} - 2 {\bf T I} - \frac{1}{8} {\bf I}^2 & 0 & 0&4 {\bf L}_{(0)} +  4 {\bf T}_{(0)}  \\ 
  \Tr(\H^{(2,1)}\Omega)  & \frac{5}{2} {\bf L I} +2 {\bf T I}- {\bf L P} - \frac{1}{2} {\bf P I}+ \frac{1}{2} {\bf I}^2& 0 & 0& -3 {\bf L}_{(0)} - 4{\bf T}_{(0)}
\end{array}
\end{equation*}\begingroup\captionof{table}{Two-loop contribution for each tensorial structure on the base with external legs $\partial_1 y \partial_1 y$.\\}\endgroup

\begin{equation*}
    \begin{array}{c|c|c|c|c }
  \text{Tensor} &  \text{Result } & \frac{1}{64 \pi \epsilon^2 } & \frac{\bar{\gamma}}{ 4 \pi \epsilon }  &  \frac{1}{4 \pi \epsilon  }   \\ 
  \hline \hline  \Tr(\H^{(3,1)})  & \frac{1}{8} {\bf L P} - \frac{1}{8}{\bf L I} - \frac{1}{16}{\bf P I}& 1 & 0& 0 \\
   \Tr(\H^{(2,2)})  & - \frac{3}{16} {\bf S} - \frac{1}{2} {\bf TE} - \frac{1}{16} \bf{P I}& \frac{1}{2} & 0& - \frac{3}{4} {\bf S}_{(1)} - 2 {\bf TE}_{(1)} \\
   \Tr(\H^{(1,1,1,1)}) &  \frac{1}{8} {\bf L}^2 + \frac{1}{8} {\bf LP} & \frac{1}{2} & 0&  0  \\
  \Tr(\H^{(1,1)}\Omega \Omega) & \frac{1}{2} {\bf S} + {\bf L I} +2 {\bf T I}& 0 & 0&  -2 {\bf L}_{(0)} -4 {\bf T}_{(0)} + 2 {\bf S}_{(1)} \\
  \Tr(\H^{(1)}\Omega \H^{(1)} \Omega)  & - \frac{1}{2}{\bf L}^2 - \frac{1}{2} {\bf L}{\bf I} - 2 {\bf T I}- \frac{3}{4} {\bf S} +2 {\bf TE} & 0 & 0& 4 {\bf T}_{(0)} - 3 {\bf S}_{(1)} + 8 {\bf TE}_{(1)} \\ 
  \Tr(\H^{(2,1)}\Omega)  & \frac{1}{2} {\bf L I} +2 {\bf T I} + {\bf S}& 0 & 0& - {\bf L}_{(0)} + 4 {\bf S}_{(1)} -4 {\bf T}_{(0)}
\end{array}
\end{equation*}\begingroup \captionof{table}{Two-loop contribution for each tensorial structure on the base with external legs $\partial_0 y \partial_0 y$.\\}\endgroup
 
 \newpage 
In summary we find the results for the simple $\epsilon$-poles of counter-terms to be given by   
\begin{empheq}[box=\ovalbox]{align}  
  & \nonumber \\
    \pi \lambda^2 T^{(2)}_1 =& -\frac{1}{2} {\bf S}_{(1)} \left(\H^{(2,0,2)} + \H^{(2,1,1)} - \frac{1}{8} \Tr(\H^{(1,1)}) \H^{(2)}\right) \nonumber \\
    &+  \left(  {\bf TE}_{(1)} - \frac{1}{4} {\bf L}_{(0)}   \right) \left(  \H^{(1,1,1,1,0)}  -  \frac{1}{4}\Tr(\H^{(1,1)}) \H^{(1,1,0)}  \right) \,  , \nonumber\\ 
    & \nonumber\\ 
\pi  \lambda\widetilde{T}_{1}^{(2)}|_{01} =&  -2 \left(  {\bf TE}_{(1)}- \frac{1}{4} {\bf L}_{(0)} \right) \Tr(\H^{(1,1,0)}\Omega \Omega) + 2\left({\bf S}_{(1)} - 2{\bf TE}_{(1)}- {\bf T}_{(0)}\right) \Tr(\H^{(2)}\Omega \Omega)  \, , \nonumber \\
    & \nonumber\\ 
\pi  \lambda \widetilde{T}_{1}^{(2)}|_{11} =& - \left( \frac{1}{8}{\bf S}_{(1)} + {\bf TE}_{(1)}  \right) \Tr(\H^{(2,2)}) + \left( \frac{1}{4} {\bf S}_{(1)} + {\bf TE}_{(1)} - \frac{1}{4} {\bf L}_{(0)} \right) \Tr(\H^{(1,1,1,1)}) \nonumber \\
&- 2 \left({\bf L}_{(0)} + {\bf T}_{(0)} \right) \left(\Tr(\H^{(1,1)}\Omega \Omega) - \Tr(\H^{(1)}\Omega \H^{(1)}\Omega)\right) \nonumber \\
&- \left(\frac{3}{2} {\bf L}_{(0)} +2 {\bf T}_{(0)}\right)  \Tr(\H^{(2,1)}\Omega)\,  ,
\nonumber \\
    & \nonumber\\ 
\pi\lambda \widetilde{T}_1^{(2)}|_{00} =& \left(\frac{3}{8} {\bf S}_{(1)}+ {\bf TE}_{(1)} \right) \Tr(\H^{(2,2)}) + \left(- {\bf S}_{(1)} + {\bf L}_{(0)} + 2 {\bf T}_{(0)}\right) \Tr(\H^{(1,1)}\Omega \Omega)
\nonumber \\ &+ \left(\frac{3}{2} {\bf S}_{(1)}  - 4 {\bf TE}_{(1)}- 2 {\bf T}_{(0)}\right)\Tr(\H^{(1)}\Omega \H^{(1)}\Omega) \nonumber \\
&+ \left( - 2 {\bf S}_{(1)} + \frac{1}{2}{\bf L}_{(0)}+2 {\bf T}_{(0)}\right) \Tr(\H^{(2,1)}\Omega) \, .
\nonumber\\
    &  \nonumber 
\end{empheq}

\subsubsection{$O(n,n)$ Consistency Requirement}

We can now return to question of compatibility of the fibre counter-term with the $O(n,n)$ structure. At one-loop, it follows immediately 
that $T_1^{(1)}$ satisfies the required condition $T_1^{(1)} \cdot \eta^{-1} \cdot \H + \H \cdot \eta^{-1} \cdot T_1^{(1)} = 0$. 
At two-loop order $T_2^{(2)}$, we recall, ought to obey $0 =T_2^{(2) }\eta^{-1}\H + \H \eta^{-1} T_2^{(2) } + T_1^{(1)} \eta^{-1} T_1^{(1)}$. There are   four relevant independent tensors $X_a(\H,\eta)$  with four derivatives and total homogeneity one  that obey $\H \cdot \eta^{-1} \cdot X_a + X_a \cdot \eta^{-1} \cdot \H =0 $ given by 
\be
\begin{aligned} 
X_1  =&  \H^{(4) } + 2(\H^{(3,1,0)}+\H^{(0,1,3)})- 3 \H^{(2,0,2)}+ 6 \H^{(1,1,0,1,1)} \, , \\ 
X_2 =&   \H^{(2,1,1)}+\H^{(1,1,2)}+ 2\H^{(1,1,0,1,1)} \, , \\ 
X_3 =&   \H^{(1,2,1)}- \H^{(1,1,0,1,1)} \, , \\ 
 X_4 =&   \left(\H^{(2)} - \H^{(1,0,1)}  \right) \Tr(\H^{(1,1)} ) \, . 
\end{aligned} 
\ee 

It is useful to introduce these combinations in $T_2^{(2)}$ as they just drop when checking the compatibility condition:
\begin{equation}
    T_2^{(2)} = \frac{1}{128 \pi^2 \lambda^2}(4 X_1 - 12 X_2 - 16 X_3 + X_4) + \frac{1}{32 \pi^2 \lambda^2} (\H^{(2,0,2)} - \H^{(1,1,0,1,1)}) \,  .
\end{equation}

The rest of the proof is easy and only involves simple identities to recast 
\begin{equation}
    \H \cdot \eta^{-1} \cdot \H^{(2,0,2)} + \H^{(2,0,2)} \cdot \eta^{-1} \cdot \H = - 2(\H^{(2,2)} + \H^{(1,1,0,2)} + \H^{(2,0,1,1)}) \,  .
\end{equation}

As for $T_1^{(2)}$, the tensor $H^{(2,0,2)} $ enters into the result but there are no contributions of  $ \H^{(4) }$  and $H^{(3,1,0)}$ that allow for its completion into $X_1$.  As a result, to ensure $T_1^{(2)}$ is $O(n,n)$ compatible we are required to enforce ${\bf S}_{(1)} = 0$.  This on its own is not an unexpected conclusion, in fact is the case if we use the Lorentz invariant regularisation scheme with $\mathfrak{g} = \frac{1}{2}$.  However once ${\bf S}_{(1)} = 0$ is set, we are left with   
\begin{align}
    \pi \lambda^2 T^{(2)}_1 &\approx \left(  {\bf TE}_{(1)} - \frac{1}{4} {\bf L}_{(0)}   \right) \left(  \H^{(1,1,1,1,0)}  -  \frac{1}{4}\Tr(\H^{(1,1)}) \H^{(1,1,0)}  \right) \,  ,
\end{align}
 and again this is not $O(n,n)$ compatible unless $  {\bf L}_{(0)}  = 4{\bf TE}_{(1)} $.   Unlike ${\bf S}_{(1)}$ it is impossible to tune $\mathfrak{g}$ within the Lorentz invariant regularisation scheme to make this combination vanish since $  {\bf L}_{(0)}  - 4{\bf TE}_{(1)} = \frac{1}{32  \pi} $ is independent of $\mathfrak{g}$. 
 
The conclusion of this analysis is that the only way the counter-term $T^{(2)}_1$ is compatible with the $O(n,n)$-structure is that the prescription for evaluating the integrals be such that $ T^{(2)}_1  \approx 0$ and hence  $\beta^\H$  receives no contribution at two-loops.

\subsubsection{Lorentz Consistency Requirement}

 Turning now to the base, we examine if restoration of Lorentz invariance is possible.  To eliminate the mixed $\widetilde{T}_{1}^{(2)}|_{01}$ contribution we require again that  $  {\bf L}_{(0)}  = 4{\bf TE}_{(1)} $ and additionally ${\bf T}_{(0)} = {\bf S}_{(1)} - 2 {\bf TE}_{(1)}$. Notice also that $\Tr(\H^{(1,1,1,1)}) $ enters in $\widetilde{T}_{1}^{(2)}|_{11}$ and not in $ \widetilde{T}_1^{(2)}|_{00}$, so to eliminate this mismatch requires once again that ${\bf S}_{(1)}= 0$. Eliminating ${\bf S}_{(1)}$, ${\bf TE}_{(1)} $ and ${\bf T}_{(0)} $ in this way yields
\begin{align}
\pi  \widetilde{T}_{1}^{(2)}|_{11} &= -  \frac{1}{4} {\bf L}_{(0)}  \Tr(\H^{(2,2)})   -   {\bf L}_{(0)}   \left(\Tr(\H^{(1,1)}\Omega \Omega) - \Tr(\H^{(1)}\Omega \H^{(1)}\Omega)\right) \nonumber \\
&- \frac{1}{2} {\bf L}_{(0)}   \Tr(\H^{(2,1)}\Omega)\,  ,
\\
\pi  \widetilde{T}_1^{(2)}|_{00} &= +\frac{1}{4} {\bf L}_{(0)}  \Tr(\H^{(2,2)})    -   \frac{1}{2}{\bf L}_{(0)} \Tr(\H^{(2,1)}\Omega) \, .
\end{align}
Lorentz symmetry is restored only when also  ${\bf L}_{(0)}=0$ and the entire counter-term  $\widetilde{T}^{(2)}_1 =0$.

\subsubsection{Evaluation of Remaining Integrals}
We can invoke the $O(d)$ Lorentz invariant integration prescription employed throughout Method 1 to now evaluate  the remaining integrals to be: 
\be
{\bf TE}_{(1)} =- \frac{1}{6}{\bf T}_{(0)}  =  \frac{3 - 8  \frak{g} }{128 \pi }  \, , \quad {\bf S}_{(1)}= - \frac{1}{2} {\bf L}_{(0)} =  \frac{2 \frak{g}- 1 }{16 \pi } \,  .
\ee
Using these values we can eliminate   ${\bf TE}_{(1)}  $ and ${\bf S}_{(1)}$ to give:

\begin{align}
   \pi \lambda^2 T^{(2)}_1 &=   \frac{1}{4} {\bf L}_{(0)} \left(\H^{(2,0,2)} + \H^{(2,1,1)} - \frac{1}{8} \Tr(\H^{(1,1)}) \H^{(2)}\right) \nonumber \\
    &- \left(  \frac{1}{6}{\bf T}_{(0)}  + \frac{1}{4} {\bf L}_{(0)}   \right) \left(  \H^{(1,1,1,1,0)}  -  \frac{1}{4}\Tr(\H^{(1,1)}) \H^{(1,1,0)}  \right) \,  .
\end{align}

On the base we find 
\begin{align}
\pi  \widetilde{T}_{1}^{(2)}|_{01} &=    \left(    \frac{1}{3}{\bf T}_{(0)}+ \frac{1}{2} {\bf L}_{(0)} \right) \Tr(\H^{(1,1,0)}\Omega \Omega) - \left(   {\bf L}_{(0)}  + \frac{4}{3}{\bf T}_{(0)} \right) \Tr(\H^{(2)}\Omega \Omega)  \, , \\
\pi  \widetilde{T}_{1}^{(2)}|_{11} &=   \left(  \frac{1}{16} {\bf L}_{(0)} +   \frac{1}{6}{\bf T}_{(0)}  \right) \Tr(\H^{(2,2)}) -  \left(   \frac{3}{8} {\bf L}_{(0)}  + \frac{1}{6}{\bf T}_{(0)}  \right) \Tr(\H^{(1,1,1,1)}) \nonumber \\
&- 2 \left({\bf L}_{(0)} + {\bf T}_{(0)} \right) \left(\Tr(\H^{(1,1)}\Omega \Omega) - \Tr(\H^{(1)}\Omega \H^{(1)}\Omega)\right) \nonumber \\
&- \left(\frac{3}{2} {\bf L}_{(0)} +2 {\bf T}_{(0)}\right)  \Tr(\H^{(2,1)}\Omega)\,  ,
\\
\pi  \widetilde{T}_1^{(2)}|_{00} &=- \left(\frac{3}{16} {\bf L}_{(0)}+   \frac{1}{6}{\bf T}_{(0)}  \right) \Tr(\H^{(2,2)}) + \left(   \frac{3}{2}{\bf L}_{(0)} + 2 {\bf T}_{(0)}\right) \Tr(\H^{(1,1)}\Omega \Omega)
\nonumber \\ &- \left(\frac{3}{4}  {\bf L}_{(0)} +  \frac{4}{3}{\bf T}_{(0)} \right)\Tr(\H^{(1)}\Omega \H^{(1)}\Omega) \nonumber \\
&+ \left(   \frac{3}{2}{\bf L}_{(0)}+2 {\bf T}_{(0)}\right) \Tr(\H^{(2,1)}\Omega) \, .
\end{align}
The $\mathfrak{g} = \frac{1}{2}$ prescription could now be adopted to further simplify the results by setting ${\bf L}_{(0)} = 0$.

As a final simplification let us assume further that the background is such that the one-loop counter-terms all vanish, in which case 
\begin{align}
    T^{(2)}_1  \rightarrow   
    &  - \frac{1}{6}{\bf T}_{(0)}        \H^{(1,1,1,1,0)}  \,  , \\  
\pi  \widetilde{T}_{1}^{(2)}|_{01}\rightarrow   &      \frac{5}{3}{\bf T}_{(0)}    \Tr(\H^{(1,1,0)}\Omega \Omega)   \, , \\ 
\pi  ( \widetilde{T}_{1}^{(2)}|_{11} -\widetilde{T}_1^{(2)}|_{00})   \rightarrow &{\bf T}_{(0)} \left(\frac{1}{6} \Tr(\H^{(1,1,1,1)}) -4 \Tr(\H^{(1,1)}\Omega \Omega) + \frac{10}{3} \Tr(\H^{(1)}\Omega \H^{(1)}\Omega)\right) \,  .
\end{align}
These final expressions demonstrate that even with such additional assumptions, the fibre counter-term remains incompatible with $O(n,n)$ structure and Lorentz invariance of the base counter-terms can not be enforced without placing constraints. 

\subsection{Scheme Dependence and Field Re-definitions} 
  Suppose a theory depends on some set of couplings $\varphi^i$, $i= 1, \dots, N$ such that the $\beta$-function can be viewed as a vector $\beta =\beta^i \frac{\delta ~}{\delta \varphi^i}$ on the space of couplings.  Under a scheme transformation $\varphi^i \rightarrow \varphi^i + \delta \varphi^i$ the change in the $\beta$-function $\beta^i$ for the $i$-th coupling is given by the Lie derivative \cite{Metsaev:1987zx}
\begin{equation}
    \delta \beta^i = \delta \varphi^j \frac{\partial}{\partial \varphi^j} \beta^i - \beta^j \frac{\partial}{\partial \varphi^j} \delta \varphi^i \,  .
\end{equation}
 In our case the coupling space is parametrised by $\varphi^i=(\H, \lambda)$ such that the  effect of a scheme change on the two-loop $\beta^\H$ is given by,  
 \begin{align} \label{eq:schemchangerule}
     \delta \beta^\H_{(2)} &= \delta \H \circ \frac{\delta}{\delta \H} \beta^\H_{(1)} + \delta \lambda \frac{\partial}{\partial \lambda} \beta^\H_{(1)} - \beta^\H_{(1)} \circ \frac{\partial}{\delta \H} \delta \H - \beta^\lambda_{(1)} \frac{\partial}{\partial \lambda} \delta \H \,  ,
 \end{align}
 where $\beta^{\H}_{(1)}$ and $\beta^\lambda_{(1)}$ are the one-loop results of eq.~\eqref{eq:betaf1l}. 

The choices for $\delta \H$ and $\delta \lambda$ that i) are second order in derivatives and ii) do not produce tensor structures other than the ones we have encountered thus far are very limited. We can consider the following scheme change $\delta \H = \frac{a}{\lambda} \H^{(2)}+\frac{b}{\lambda} \H^{(1,1,0)}$, $\delta \lambda = \frac{c}{4} \Tr(\H^{(1,1)})$ for some real parameters $a,b,c$. We then obtain
 \begin{align}
     \delta \beta^{\H}_{(2)} &= \frac{a-b}{2 \pi \lambda^2} (\H^{(2,0,2)}+ \H^{(1,2,1)} - \H^{(2,1,1)}) + \frac{1}{16 \pi \lambda^2} \Tr(\H^{(1,1)})\left( (a+c) \H^{(2)} +(b+c) \H^{(1,1,0)}  \right) \,  .
 \end{align}

 As previously pointed out, adopting either Method 1 or Method 2, $T_1^{(2)}$ is not compatible with $O(n,n)$ symmetry. Since up to some constant factors $T_1^{(2)}$ determines $\beta^\H$, we might hope in a scheme change to restore the symmetry. Let us adopt Method 2 for definiteness and extract the $\beta$-function 
 \begin{align}
    \beta^{\mathcal{H}}_{(2)} =&
     \frac{1}{\pi\lambda^2} {\bf S}_{(1)} \left(\H^{(2,0,2)} + \H^{(2,1,1)} - \frac{1}{8} \Tr(\H^{(1,1)}) \H^{(2)}\right) \nonumber \\
    &- \frac{2}{\pi \lambda^2}  \left(  {\bf TE}_{(1)} - \frac{1}{4} {\bf L}_{(0)}   \right) \left(  \H^{(1,1,1,1,0)}  -  \frac{1}{4}\Tr(\H^{(1,1)}) \H^{(1,1,0)}  \right) \,  .
\end{align}

As mentioned in previous discussions, demanding compatibility with the $O(n,n)$ of $\beta^{\H \prime}_{(2)}\equiv   \beta^{\mathcal{H}}_{(2)} + \delta    \beta^{\mathcal{H}}_{(2)}  $  requires that the tensor $\H^{(2,0,2)}$  be absent and $\Tr(\H^{(1,1)}) \H^{(2)}$ and $\Tr(\H^{(1,1)}) \H^{(1,1,0)}$   have identical coefficients. These are necessary rather than sufficient conditions, for a precise fine-tuning of the remaining tensor structures is required too. 

If we accept $a \neq b$ (i.e. an $O(n,n)$-violating redefinition of $\H$) we could set the coefficient of $\H^{(2,0,2)}$ to zero; however, this makes it impossible to choose $b$ and $c$ so that $\Tr(\H^{(1,1)}) \H^{(2)}$ and $\Tr(\H^{(1,1)}) \H^{(1,1,0)}$ come with identical numerical pre-factor. Conversely, we could tune e.g. $a$ so that $\Tr(\H^{(1,1)}) \H^{(2)}$ and $\Tr(\H^{(1,1)}) \H^{(1,1,0)}$ share the same coefficient; but still, the expression for the remaining tensors turns out independent of $b,c$ and, moreover, not compatible with $O(n,n)$ symmetry. 
 
\section{Summary and Conclusions} \label{sec:summary}
In this manuscript we have computed the two-loop effective action for the T-duality symmetric bosonic string.  

As a first step we provided a complete calculation  of all contributions to the effective action arising from Wick contraction keeping loop integrals unevaluated. We then employed two methods to simplify and evaluate these (non-Lorentz covariant) loop integrals. The two methods agree with each other for the   $\epsilon^{-2}$ divergences, but are subtly different when it comes to the sub-leading $\epsilon^{-1}$ divergences that contribute to the $\beta$-functions at two-loop order. 

Of the two approaches, Method 2, in which the maximal number of simplifications is performed in $d=2$,  results in compelling simplifications such that the results can be phrased in terms of a basis of just five independent integrals. We are then able to analyse the results in a way that keeps the choice of regularisation method implicit giving  general conclusions that would hold with any choice of regularisation (dimensional or otherwise). For concreteness, here we completed Method 2 by employing   continuation to $d=2+\epsilon$ after all simplifications have been made to evaluate the remaining integrals. 

Both methods pass a number of  important consistency checks: 
 \begin{itemize}
    \item The $\epsilon^{-2}$ contributions  are in exact accordance with the expectations from the pole equation on the doubled fibre in which the T-duality acts.  
    \item The $\epsilon^{-2}$ contributions on the base are consistent with Lorentz invariance. This is to say, no $\partial_0 y \partial_1y$ legs are  produced (even though they do appear in intermediate steps) and the counter-terms for $\partial_0 y \partial_0 y$ and $\partial_1 y \partial_1 y$ coincide (even though they come from totally different sets of diagrams).
    \item For $\epsilon^{-2}$ poles, all occurrences of the Weitzenb\"ock connection $\Omega$ combine in a fashion to be expressible   in terms of   $\H$ alone.  
\item  After regularising IR divergence as described, mixed IR/UV divergences of the form $\frac{\log(m) }{\epsilon}$ are removed in the cancellation of $\bar\gamma$ terms.
\item On the doubled fibre, possible contributions  to the $\frac{1}{\epsilon}$ pole due to the IR mass regulator  giving divergences of the form  $m^2 I_1 I_2$ are cancelled with the introduction of an appropriate mass term and its background field expansion. 
\end{itemize}

Notwithstanding the dramatic simplifications afforded by Method 2 compared to Method 1, the results for the $\epsilon^{-1}$ pole that contribute to the two-loop $\beta$-function present some puzzles:  
\begin{itemize}  
    \item On the fibre, the counter-term does not  have the right structure to allow the $O(n,n)$ constraint $\H \eta \H = \eta$ to be preserved by RG flow. 
    \item The $\epsilon^{-1}$ pole on the base manifold is \emph{not}   Lorentz invariant. A new interaction vertex proportional to $\partial_0 y \partial_1 y$ is created, and the counter-terms for the legs $\partial_0y \partial_0y$ and $\partial_1 y \partial_1 y$ have   differences that can not be resolved by a scheme choice. 
    \item The $\epsilon^{-1}$ pole on the base manifold involves the connection $\Omega$ in a way that can not be combined into something expressible in terms of the generalised metric $\H$ alone.   
\end{itemize}

Method 2 does afford one possible avenue to resolve these puzzles.  Namely   the possibility that an integration prescription can be invoked such that the entire two-loop $\epsilon^{-1}$ counter- terms vanish.   This is the case if the undetermined subleading part of the integrals ${\bf L}, {\bf T},{\bf S},{\bf TE}$ of eq.  \eqref{eq:LTdef} vanish. Such a result would be equally surprising as it would be in contradiction to that of the conventional non-linear $\sigma$-model for the bosonic string.  

The calculation involved in arriving at these results is of considerable complexity (especially with regards to the counter-terms on the base) and so we can't rule out that these issues pointed out here may be resolvable.   
As methods other than ours might in principle be considered, we have collected the relevant loop integrals, prior to any evaluation, in a way which is suitable to further investigation. Even though it is possible that  different prescriptions might result in a non-vanishing $\beta$-function compatible with both $O(n,n)$- and Lorentz-symmetry, it seems likely that the resolution would be highly non-trivial\footnote{One might contend that an anomaly in double Lorentz transformations for the duality-symmetric string,   \cite{Bonezzi:2020ryb}, could play a role here. The  Green-Schwarz mechanism required to cancel this would doubtless be important in the most general setting, however in the present ``cosmological'' setting, there is no such anomaly to contend with as the base manifold has only one dimension.}  and would need to give a compelling non-ambiguous proposal for regulating the loop integrals involved.

As it stands, however, the results obtained cast some doubt as to the full validity of the doubled action in the form of eq. \eqref{eq:TAction} at the quantum level.   At the very least one can say that the power of invoking manifest T-duality on the worldsheet is far outweighed by the added complexities that the chiral nature of this formalism entails  at the quantum level.

\section*{Acknowledgments}
  Two of us, DCT and NBC, would like to thank the Isaac Newton Institute for hospitality during the  2012  programme The Mathematics and Applications of Branes in String and M-theory.  We thank also the organisers of this programme for providing a stimulating environment where a substantial amount of the results presented here were first obtained. DCT is supported by The Royal Society through a University Research Fellowship URF/150185 and   by the FWO-Vlaanderen through the project G006119N
and by the Vrije Universiteit Brussel through the Strategic Research Program “High-Energy Physics”.   GP is supported by The Royal Society Enhancement Award RGF/EA/180176 . 

We would like to thank David Berman, Chris Blair, Falk Hassler, Emanuel Malek, Alex Sevrin and Arkady Tseytlin for helpful discussion and communication on this work.

\begin{appendix}

 \section{Fibre Wick Contractions}\label{app:Fibre}
We report here Wick contractions which are relevant for the two-loop computation on the fibre. We use conventions as explained in the main text, namely we set
\bea
&& a_1= \langle  {\cal A}^{[2]}   \rangle \,  ,\quad  a_2 = i \langle  {\cal A}^{[0]}   {\cal A}^{[2]}    \rangle  \,  ,\quad 
a_3  = \frac{i}{2} \langle  {\cal A}^{[1]}   {\cal A}^{[1]}   \rangle  \, , \non \\
&& a_4= -\frac{1}{2} \langle  {\cal A}^{[0]}   {\cal A}^{[0]} {\cal A}^{[2]}  \rangle\, , \quad 
a_5  =  -\frac{1}{2} \langle {\cal A}^{[0]}   {\cal A}^{[1]} {\cal A}^{[1]}    \rangle \, , \quad  
a_6 = -\frac{i}{4} \langle {\cal A}^{[0]}   {\cal A}^{[0]}  {\cal A}^{[1]} {\cal A}^{[1]}   \rangle  \,  ,
\eea
with
\begin{align}
    \mathcal{A}^{[0]}_2 &=- \frac{1}{32 \pi \epsilon} \Tr(\H^{(1,1)}) \partial_\mu \zeta \partial^\mu \zeta - \frac{1}{8 \pi \epsilon \lambda} (\H^{(2)}_{AB} - \H^{(1,0,1)}_{AB}) \partial_1 \xi^A \partial_1 \xi^B - \frac{1}{4} \H^{(2)}_{AB} \zeta^2 \partial_1 \xi^A \partial_1 \xi^B \, \\
     \mathcal{A}^{[0]}_1 &= - \frac{1}{2} \H^{(1)}_{AB} \zeta \partial_1 \xi^A \partial_1 \xi^B  \, , \\
    \mathcal{A}^{[1]}_1&= - \H^{(1)}_{A \bullet} \zeta \partial_1 \xi^A \, , \\
    \mathcal{A}^{[1]}_2&= - \frac{1}{2}\H^{(2)}_{A \bullet} \zeta^2 \partial_1 \xi^A \, , \\
    \mathcal{A}^{[1]}_3&= \frac{1}{4 \pi \epsilon \lambda} \left(-\H^{(3)}  +\H^{(1,1,1)}  +  \H^{(1,0,2)} + \H^{(2,0,1)}\right)_{A \bullet} \zeta \partial_1 \xi^A - \frac{1}{6} \H^{(3)}_{A \bullet} \zeta^3 \partial_1 \xi^A\, , \\
    \mathcal{A}^{[2]}_2 &=- \frac{1}{4} \mathcal{H}^{(2)}_{\bullet \bullet} \zeta^2 \, ,  \\
    \mathcal{A}^{[2]}_{3}&= - \frac{1}{12} \H^{(3)}_{\bullet \bullet} \zeta^3 \, , \\
    \mathcal{A}^{[2]}_{4}&= \frac{1}{16 \pi \epsilon \lambda} \left(-\H^{(4)} + \H^{(1,2,1)} + 4 \H^{(1,1,2)} + 2 \H^{(2,0,2)} + 2 \H^{(1,0,3)}\right)_{\bullet \bullet}\zeta^2 - \frac{1}{48} \H^{(4)}_{\bullet \bullet} \zeta^4 \,  .
\end{align}
  Owing to the fact that $\mathcal{H}$ is idempotent, and thus $\dot{\mathcal{H}} \cdot \mathcal{H} = -\mathcal{H} \cdot  \dot{\mathcal{H}}$,   at two-loop order a basis for the relevant independent tensors without traces is 
\begin{equation}
    \H^{(4)}_{\bullet\bullet}\, , \quad \H^{(3,1,0)}_{\bullet\bullet}\, , \quad \H^{(2,0,2)}_{\bullet\bullet} \, , \quad \H^{(2,1,1)}_{\bullet\bullet}\, , \quad \H^{(1,2,1)}_{\bullet\bullet}\,  , \quad \H^{(1,1,1,1,0)}_{\bullet\bullet}\, ,
\end{equation}
which can be extended by the ones with trace
\begin{equation}
  \H^{(2)}_{\bullet\bullet} \Tr (\H^{(1,1)}) \, , \quad  \H^{(1,1,0)}_{\bullet\bullet} \Tr(\H^{(1,1)}) \, .
\end{equation}
We first extract the coefficients of this basis in terms of the unevaluated tensorial integrals $[[f(p_0, p_1, q_0, q_1 )]]_{i,j,k} $ which can be evaluated using the Method 1 rules. We then use Method 2 rules to present a final answer in the  ${\bf I},{\bf L},{\bf T},{\bf S},{\bf TE}$ basis of integrals. When dealing with counter-term insertions we adopt the shorthands
\begin{equation}
    X = \frac{1}{8 \pi \epsilon \lambda} (\H^{(2)} + \H^{(1,1,0)}) \, , \qquad Y =  \frac{1}{32 \pi \epsilon} \Tr(\H^{(1,1)}) \,  .
\end{equation}

\subsection*{${a_1}$} 
The    contributing  diagrams are either bubbles or decorated bubbles and  evaluate  to
\bea
a_1 &=& \langle  {\cal A}^{[2]}_4   \rangle  = -\frac{1}{48} \H^{(4)}_{\bullet \bullet } \langle \zeta^4  \rangle   -\frac{1}{2} X^{(2)}_{\bullet \bullet} \langle \zeta^2  \rangle \nonumber  \\
&=&   \frac{1}{16\lambda^2} \H^{(4)}_{\bullet \bullet }  [[1]]_{1,1,0}- \frac{i}{2\lambda } X^{(2)}_{\bullet \bullet} [[1]]_{1,0,0}
= \frac{1}{16 \lambda^2} \H^{(4)}_{\bullet \bullet} {\bf I}^2   -\frac{i}{2\lambda} X^{(2)}_{\bullet \bullet } {\bf I} \, .
\eea
Expanding the derivatives of the counter-term insertion yields
\bea
a_1 
= \frac{1}{16 \lambda^2}{\bf I}^2   \H^{(4)}_{\bullet \bullet}   -\frac{1}{8\lambda^2 } {\bf I}  {\bf P }  \left(\H^{(4)} -4 \H^{(2,1,1)} -2 \H^{(2,0,2)} + 2 \H^{(3,1,0)} -\H^{(1,2,1)} \right)_{\bullet \bullet } \, . 
\eea

\subsection*{$a_2$} 
After discarding non-1PI graphs we obtain
\bea
a_2 &=& i \langle   {\cal A}^{[0]}_1 {\cal A}^{[2]}_3 +  {\cal A}^{[0]}_2 {\cal A}^{[2]}_2 \rangle  
=  \frac{i}{16} \H^{(2)}_{A B} \H^{(2)}_{\bullet \bullet} \langle \zeta_{\sigma_1}^2 \partial_1 \xi_{\sigma_1}^{A} \partial_1 \xi_{\sigma_1}^{B} \zeta^2_{\sigma_2} \rangle 
 + \frac{i}{4} Y \H^{(2)}_{\bullet \bullet} \langle   \partial_\mu \zeta_{\sigma_1} \partial^\mu \zeta_{\sigma_1} \zeta^2_{\sigma_2} \rangle \nonumber    \\
 &=& -\frac{1}{8 \lambda^2 }\Tr(\H^{(1,1)}) \H^{(2)}_{\bullet \bullet} [[ k_1^2]]_{2,1,0}  - \frac{i}{2 \lambda^2 } Y [[p^2]]_{2,0,0} \,  .
\eea

The potentially linearly divergent $[[k_1^2]]_{2,1,0}  $ term cancels with the same from $a_4$ and can be set to zero.  
In Method 2 we cancel the $p^2$ in numerator and denominator of $[[p^2]]_{2,0,0}$ to yield   
\be 
a_2  =     - \frac{  {\bf I} {\bf P}} {32 \lambda^2 }  \Tr(\H^{(1,1)}) \H^{(2)}_{\bullet \bullet}  \,  .
\ee

\subsection*{$a_3$} 

There are two contributions in $a_3$ without counter-term insertion, of which the first comes from  
\bea 
a_{3a}&=& i \langle  {\cal A}_1^{[1]} {\cal A}_3^{[1] } \rangle  = \frac{i}{6} \H_{\bullet A }^{(1) }\H_{\bullet B}^{(3) }  \langle \zeta_{\sigma_1} \zeta_{\sigma_2}^3\partial_1 \xi^{A}_{\sigma_1} \partial_1 \xi_{\sigma_2}^{B } \rangle   
\nonumber \\
&=&- \frac{1}{2    \lambda^2} \H^{(3,1,0)}_{\bullet \bullet}    [[p_1^2]]_{2,1,0}  + \frac{1}{2    \lambda^2} \H^{(3,1 )}_{\bullet \bullet}    [[p_1 p_0]]_{2,1,0}  \nonumber \\
&=& - \frac{ {\bf I}{\bf L}}{2    \lambda^2} \H^{(3,1,0)}_{\bullet \bullet}   \, .  
\eea
The second contribution arises from
\bea
a_{3b}&=& \frac{i}{2} \langle  {\cal A}_2^{[1]}  {\cal A}_2^{[1]} \rangle  = \frac{i}{8} \H^{(2)}_{\bullet A }\H^{(2)}_{\bullet B }  \langle \zeta_{\sigma_1}^2 \zeta_{\sigma_2}^2 \partial_1 \xi^{A}_{\sigma_1} \partial_1 \xi^{B}_{\sigma_2}  \rangle 
  \nonumber \\
 &=&  \frac{1}{4 \lambda^2} \H^{(2,0,2)}_{\bullet \bullet} [[p_1^2]]_{1,1,1}+ \frac{1}{4 \lambda^2} \H^{(2, 2)}_{\bullet \bullet} [[p_1 p_0]]_{1,1,1} \nonumber \\ 
 & =&     \frac{{\bf S}}{4 \lambda^2} \H^{(2,0,2)}_{\bullet \bullet}   \, .  
\eea 
We also have a one-loop diagram with a counter-term insertion 
\bea  
a_{3c} &=& i \H^{(1)}_{\bullet A} X^{(1)}_{B \bullet }    \langle \zeta_{\sigma_1}  \zeta_{\sigma_2}  \partial_1 \xi^{A}_{\sigma_1} \partial_1 \xi^{B}_{\sigma_2}  \rangle \nonumber   
  =  - i   \H^{(1)}_{\bullet A} \H^{ A  B }  X^{(1)}_{B \bullet }   [[p_1^2]]_{2,0,0} - i   \H^{(1)}_{\bullet A} \eta^{ A  B }  X^{(1)}_{B \bullet }   [[p_1 p_0]]_{2,0,0}    \\
  &=& \frac{ {\bf P} {\bf L} }{2 \lambda^2} \left( \H^{(3,1,0)} + H^{(1,1,1,1,0)}  -\H^{(1,2,1)} - \H^{(1,1,2)} \right)  \, . 
\eea 

In the final steps we have invoked that $[[p_1 p_0]]_{2,0,0} = [[p_1 p_0]]_{2,1,0} = [[p_1 p_0]]_{1,1,1}  = 0$. 

\subsection*{$a_4$}
This   triangle envelope topology diagram  evaluates to  
\bea
a_4 &=&  -\frac{1}{2} \langle {\cal A}^{[2]}_2 {\cal A}^{[0]}_1 {\cal A}^{[0]}_1  \rangle  
= \frac{1}{32} \H^{(2)}_{\bullet \bullet} \H^{(1)}_{A B }\H^{(1)}_{C D } \langle \zeta^2_{\sigma_1} \zeta_{\sigma_2}\zeta_{\sigma_3} \partial_1 \xi^{A}_{\sigma_2} \partial_1 \xi^{B}_{\sigma_2}  \partial_1 \xi^{C}_{\sigma_3} \partial_1 \xi^{D}_{\sigma_3}   \rangle \nonumber \\ 
 &=& \frac{1}{8 \lambda^2}  \H^{(2)}_{\bullet \bullet} \Tr (\H^{(1,1)})  \left( [[(p_1+k_1) (p_0+k_0) k_1 k_0   ]]_{2,1,1}-  [[(p_1+k_1)^2 k_1^2  ]]_{2,1,1}  \right) \,  .
\eea 
Under Method 2 we proceed by replacing e.g. $p_0^2 = p^2 + p_1^2$ to give  
 \bea
 a_4  
 &=& \frac{1}{8 \lambda^2}  \H^{(2)}_{\bullet \bullet} \textrm{tr}\H^{(1,1)}  \left(-\frac{1}{4}[[p_1^2]]_{1,1,1} + [[k_1^2]]_{2,1,0} \right) \nonumber \\
 &=& -\frac{ {\bf S} }{32 \lambda^2}  \H^{(2)}_{\bullet \bullet} \textrm{tr}\H^{(1,1)} \, . 
\eea 
In the last line we dispensed with the potentially linearly divergent contribution $[[k_1^2]]_{2,1,0}$ which in a case cancels against the same from $a_2$. 
 
\subsection*{$a_5$} 
First we consider the part of 
\begin{equation}
a_{5_1}  =-\frac{1}{2} \langle {\cal A}^{[1]}_{1}{\cal A}^{[1]}_{1}  {\cal A}^{[0]}_{2}    \rangle
\end{equation}
that does not involve the insertion of one-loop counter-term operators.  There are three topologies involved here giving contributions $a_{5_{1a}}$, $a_{5_{1b}}$, $a_{5_{1c}}$:
\bea 
a_{5_{1a}}  &=& \frac{1}{2} \H^{(1)}_{\bullet A} \H^{(2)}_{C D} \H^{(1)}_{B \bullet } \langle \zeta_{\sigma_1} \zeta_{\sigma_3}   \rangle \langle \zeta_{\sigma_2} \zeta_{\sigma_3}   \rangle  \langle  \partial_1 \xi^{A}_{\sigma_1}   \partial_1 \xi^{C}_{\sigma_3}  \rangle     \langle  \partial_1 \xi^{B}_{\sigma_2}   \partial_1 \xi^{D}_{\sigma_3}  \rangle \nonumber  \\ 
&=& \frac{1}{2\lambda^2} \left(  - \H^{(1,2,1)} + 2\H^{(1,1,1,1,0)} \right)_{\bullet \bullet}  [[p_1^2 k_1^2]]_{2,2,0}  +  \frac{1}{2\lambda^2} \H^{(1,2,1)}_{\bullet \bullet} [[p_1p_0 k_1 k_0]]_{2,2,0}    \nonumber \\
&=& \frac{{\bf L}^2}{2\lambda^2} \left(  - \H^{(1,2,1)} + 2\H^{(1,1,1,1,0)} \right)_{\bullet \bullet}  \, . \eea

Here there is a small subtlety; in Method 2 one could have made a replacement such as $[[p_1p_0 k_1 k_0]]_{2,2,0} \rightarrow[[p_1 k_1 p\cdot k ]]_{2,2,0} + [[p_1p_1 k_1 k_1]]_{2,2,0} $ and   produced a  $\frac{1}{\epsilon }$ pole; however, as this is factorised diagram, general arguments \cite{Jack:1989vp}  imply that the counter-term contribution must cancel such a pole.   Hence the correct procedure is to   replace  $ [[p_1p_0 k_1 k_0]]_{2,2,0} \rightarrow ([[p_1 p_0]]_{2,0,0})^2 \rightarrow 0$. The second and third parts are
\bea
a_{5_{1b}}  &=&  \frac{1}{4} \H^{(1)}_{\bullet A} \H^{(2)}_{CD} \H^{(1)}_{B \bullet } \langle \zeta_{\sigma_3}   \zeta_{\sigma_3}   \rangle   \langle \zeta_{\sigma_1} \zeta_{\sigma_2}   \rangle    \langle  \partial_1 \xi^{A}_{\sigma_1}   \partial_1 \xi^{C}_{\sigma_3}  \rangle     \langle  \partial_1 \xi^{B}_{\sigma_2}   \partial_1 \xi^{D}_{\sigma_3}  \rangle \nonumber  \\
 &=& \frac{1}{4\lambda^2}   \left( 2 \H^{(1,1,1,1,0)}_{\bullet \bullet} [[p_1^4]]_{3,1,0} + \H^{(1,2,1)}_{\bullet \bullet} ( [[p_1^2 p_0^2]]_{3,1,0} -[[p_1^4]]_{3,1,0}) \right) \nonumber \\  
 &=&  \frac{1}{4\lambda^2}   \left( 2 {\bf I }{\bf T} \H^{(1,1,1,1,0)}_{\bullet \bullet}   + {\bf I}{\bf L} \H^{(1,2,1)}_{\bullet \bullet}  \right)  \, , \\ \nonumber \\
  a_{5_{1c}}  &=&  \frac{1}{4} \H^{(1)}_{\bullet A } \H^{(2)}_{C D} \H^{(1)}_{B \bullet }   \langle  \partial_1 \xi^{C}_{\sigma_3}   \partial_1 \xi^{D}_{\sigma_3}  \rangle    \langle \zeta_{\sigma_1} \zeta_{\sigma_3}   \rangle \langle \zeta_{\sigma_2} \zeta_{\sigma_3}   \rangle  \langle  \partial_1 \xi^{A}_{\sigma_1}  \partial_1 \xi^{B}_{\sigma_2}   \rangle \nonumber  \\ 
  &=& \frac{1}{2\lambda^2} \Tr(\H^{(1,1)} ) \H^{(1,1,0) }_{\bullet \bullet} [[p_1^2 k_1^2]]_{3,1 ,0}   \, . 
 \eea

Then there are two contributions,  $a_{5_{1d}}$ and $a_{5_{1e}}$, from  one-loop triangles with counter-term insertions:
\bea
a_{5_{1d}}  &=&  \frac{1}{2} \H^{(1)}_{\bullet A}  \H^{(1)}_{B \bullet }  Y  \langle   \zeta_{\s_1} \partial_1 \xi_{\s_1}^{A } \zeta_{\s_2} \partial_{1} \xi^{B}_{\s_2} \partial_\mu\zeta_{\s_3} \partial^\mu\zeta_{\s_3}    \rangle \nonumber \\
&=& - \frac{i }{2\lambda^2} \H^{(1,1,0) }_{\bullet \bullet}  Y [[p_1^2 p^2 ]]_{3,0,0} +   \frac{i }{\lambda^2} \H^{(1,1 ) }_{\bullet \bullet}  Y [[p_1 p_0 p^2 ]]_{3,0,0} \nonumber \\
&=&   \frac{ {\bf P} {\bf L} }{16\lambda^2} \Tr\H^{(1,1)}  \H^{(1,1,0) }_{\bullet \bullet}   \, , \\ \nonumber \\
a_{5_{1e}}  &=&  \frac{1}{2} \H^{(1)}_{\bullet A}  \H^{(1)}_{B \bullet }  X_{CD}   \langle   \zeta_{\s_1} \partial_1 \xi_{\s_1}^{A} \zeta_{\s_2} \partial_{1} \xi^{B}_{\s_2} \partial_1 \xi^{C}_{\s_3}\partial_1 \xi^{D}_{\s_3}    \rangle \nonumber  \\
&=&  - \frac{i}{\lambda } \left(  (\H^{(1,0)}   X  \H^{(0, 1)})_{\bullet \bullet} [[p_1^4]]_{3,0,0}  +   (\H^{(1)}   X    \H^{(1)})_{\bullet \bullet} [[p_1^2 p_0^2 ]]_{3,0,0}  \right) \nonumber  \\ 
  & =&  -\frac{ {\bf P }{\bf L}}{4 \lambda^2 } \left(   \H^{(1,2,1)} - \H^{(1,1,1,1,0)}\right)_{\bullet \bullet} \, . 
\eea

Finally we have a second contraction with three vertices given by
\bea
a_{5_2}  &=& - \langle {\cal A}^{[1]}_{1}{\cal A}^{[1]}_{2}  {\cal A}^{[0]}_{1}  \rangle    =  \frac{1}{4}\H^{(1)}_{\bullet A} \H^{(1)}_{CD }\H^{(2 )}_{B \bullet} \langle \zeta_{\sigma_1} \zeta_{\sigma_2}^2 \zeta_{\sigma_3}  \partial_1 \xi^{A}_{\s_1} \partial_1 \xi^{B}_{\s_2}   \partial_1 \xi^{C}_{\s_3} \partial_1 \xi^{D}_{\s_3}  \rangle \nonumber \\
&=&\frac{1}{\lambda^2} \H^{(1,1,2)}_{\bullet \bullet} \left(-[[p_1^2 (p+k)_1^2]]_{2,1,1} + [[p_1 p_0 (p+k)_1 (p+k)_0]]_{2,1,1} \right) \nonumber  \\    &=& \frac{{\bf S} + 2 {\bf I}{\bf L }}{4\lambda^2} \H^{(1,1,2)}_{\bullet \bullet} - \frac{1}{4\lambda^2} \textrm{tr}(\H^{(1,1)} ) \H^{(1,1,0) }_{\bullet \bullet} [[p_1^2 k_1^2]]_{3,1 ,0}   \, . 
\eea 

The final  $[[p_1^2 k_1^2]]_{3,1 ,0} $ contribution  (which we expect not  to contain  divergent terms in $\frac{1}{\epsilon}$ or $\frac{1}{\epsilon^2}$) cancel between $a_{5_{2}}$ and $a_{5_{1c}}$, and $a_{6a}$.

\subsection*{$a_6$} 
For the last contraction, 
\be 
a_6 = -\frac{i}{4} \langle {\cal A}^{[0]}_{1}{\cal A}^{[0]}_{1}  {\cal A}^{[1]}_{1} {\cal A}^{[1]}_{1}     \rangle 
\, , \ee
there are three different topologies of diagrams to consider: 
\bea
a_{6_a} &=& -\frac{i}{4}\H^{(1)}_{AB} \H^{(1)}_{CD} \H^{(1)}_{\bullet E} H^{(1)}_{\bullet F } \langle\zeta_{\s_1} \zeta_{\s_3}    \rangle \langle\zeta_{\s_2} \zeta_{\s_4}    \rangle  \langle \partial_1\xi^{A}_{\s_1} \partial_1 \xi^{C}_{\s_2}    \rangle   \langle \partial_1\xi^{B}_{\s_1} \partial_1 \xi^{D}_{\s_2}    \rangle   \langle \partial_1\xi^{E}_{\s_3} \partial_1 \xi^{F}_{\s_4}    \rangle \\   &=&  -\frac{1}{4 \lambda^2} \textrm{tr}(\H^{(1,1)})\H^{(1,1,0)}_{\bullet \bullet} \times \left( -[[(p+k)_1^2 k_1^2 p_1^2 ]]_{3,1,1}+ [[(p+k)_1 (p+k)_0 k_1 k_0  p_1^2 ]]_{3,1,1}  \right) \, ,   \nonumber \\ \nonumber \\
a_{6_b} &=& - \frac{i}{4}\H^{(1)}_{AB} \H^{(1)}_{CD} \H^{(1)}_{\bullet E} H^{(1)}_{\bullet F } \langle\zeta_{\s_1} \zeta_{\s_2}    \rangle \langle\zeta_{\s_3} \zeta_{\s_4}    \rangle  \langle \partial_1\xi^{A}_{\s_1}  \partial_1\xi^{C}_{\s_2}    \rangle   \langle \partial_1\xi^{B}_{\s_1}  \partial_1\xi^{E}_{\s_3}    \rangle   \langle \partial_1\xi^{D}_{\s_2}  \partial_1\xi^{F}_{\s_4}    \rangle  \\ 
&=&  \frac{1}{2 \lambda^2} \H^{(1,1,1,1,0)}_{\bullet \bullet} \left( [[(p+k)_1^2 p_1^4]]_{3,1,1} - 2[[(p+k)_1 (p+k)_0  p_1^3 p_0]]_{3,1,1} + [[(p+k)_1^2 p_1^2 p_0^2]]_{3,1,1}  \right) \, ,    \nonumber \\ \nonumber \\
a_{6_c} &=& - \frac{i}{4}\H^{(1)}_{AB} \H^{(1)}_{CD} \H^{(1)}_{\bullet E} H^{(1)}_{\bullet F } \langle\zeta_{\s_1} \zeta_{\s_3}    \rangle \langle\zeta_{\s_2} \zeta_{\s_4}    \rangle  \langle \partial_1\xi^{A}_{\s_1}  \partial_1\xi^{C}_{\s_2}    \rangle   \langle \partial_1\xi^{B}_{\s_1} \partial_1 \xi^{F}_{\s_4}    \rangle   \langle \partial_1\xi^{D}_{\s_2}  \partial_1\xi^{E}_{\s_3}    \rangle   \\ 
&=& \frac{ \textbf{}1}{2 \lambda^2} \H^{(1,1,1,1,0)}_{\bullet \bullet} \biggl( [[(p+k)_1^2 p_1^2 k_1^2]]_{2,2,1} - 2[[(p+k)_1 (p+k)_0  p_1 p_0 k_1^2]]_{2,2,1} \nonumber \\
&{}&+ [[(p+k)_1^2 p_1  p_0 k_1 k_0 ]]_{2,2,1}  \biggr) \nonumber \, .    
\eea 

As detailed in the main body for the case of $a_{6_b}$, under Method 2 each of these can be simplified to yield
\bea 
a_{6_a} &=& -\frac{1}{8 \lambda^2 } \left(2 [[p_1^2 q_1^2 ]]_{3,1,0} - {\bf T E} \right) \Tr(\H^{(1,1)} )   \H^{(1,1,0)}_{\bullet \bullet } \, ,  \\
a_{6_b} &=&  \frac{1}{2 \lambda^2 } \left(  {\bf T E}  - {\bf I} {\bf T}\right) \H^{(1,1,1,1,0)}_{\bullet \bullet } \, ,
\\
a_{6_c} &=&  - \frac{1}{\lambda^2} {\bf T E} \,   \H^{(1,1,1,1,0)}_{\bullet \bullet } \, . 
\eea 

\subsection{IR Regularisation in Method 1}
Let us explore here in some detail the way our IR regularisation prescription deals with the cancellation of loop integrals proportional to $I_{2,3}$ in the final result. It is fairly easy to tackle this problem explicitly once some observations are made. First, if we are interested in $I_1 I_2$ or $I_1 I_3$ contributions only, we can safely neglect the counter-term insertions, as by definition they would not give rise to any such term at this loop order. Obviously, we will also drop in $\mathcal{L}_{\mathrm I}$ any term proportional to $\Omega$.  Another important remark is that, when considering $\exp( i S_{\mathrm{I}})$, we can discard any term which is not proportional to $m^2$, as these do not originate from the expansion of the mass term and have thus been previously considered. 
Finally, to keep things simple, we restrict ourselves to combinations that  lead eventually to the desired tensor structures, namely $\Tr(\H^{(1,1)})\H^{(2)}$ and $\Tr(\H^{(1,1)})\H^{(1,1,0)}$. While other tensors might arise in the full calculation, they eventually cancel in the final result. 
The first contribution belongs to what we call triangle envelope topology. It is given by
\begin{align}
    m_1 &= \frac{m^2}{32} \H^{(2)}_{\bullet \bullet} \H^{(1)}_{AB} \H^{(1)}_{CD} \langle \zeta_{\sigma_1}^2 \zeta_{\sigma_2} \zeta_{\sigma_3} \rangle \langle \xi^A_{\sigma_2} \xi^B_{\sigma_2} \partial_1 \xi^C_{\sigma_3} \partial_1 \xi^D_{\sigma_3}\rangle \nonumber \\
    &= \frac{m^2}{8 \lambda^2} [[k^2 + k \cdot p]]_{2,1,1} \Tr(\H^{(1,1)}) \H^{(2)}_{\bullet \bullet} \,  . 
\end{align}

The second contribution comes from the square envelope topology and evaluates to
\begin{align}
    m_2 &= - \frac{i m^2}{16 } \H^{(1)}_{AB} \H^{(1)}_{C \bullet} \H^{(1)}_{D \bullet} \H^{(1)}_{EF} \langle \zeta_{\sigma_1}\zeta_{\sigma_2} \zeta_{\sigma_3}\zeta_{\sigma_4}\rangle \langle \xi^A_{\sigma_4} \xi^B_{\sigma_4} \partial_1 \xi^C_{\sigma_1} \partial_1 \xi^D_{\sigma_2} \partial_1 \xi^E_{\sigma_3} \partial_1 \xi^F_{\sigma_3}\rangle \nonumber \\
    &=\frac{m^2}{4 \lambda^2} [[p_1^2 (k^2 + k \cdot p)]]_{3,1,1} \Tr(\H^{(1,1)})\H^{(1,1,0)}_{\bullet \bullet}\,  .
\end{align}
The third possibility is a decorated loop diagram, coming from
\begin{equation}
    m_3 =  \frac{i}{32} m^2 \H^{(2)}_{AB} \H^{(2)}_{\bullet \bullet} \langle \zeta_{\sigma_1}^2 \zeta_{\sigma_2}^2\rangle  \langle \xi_{\sigma_2}^A \xi_{\sigma_2}^B\rangle  =- \frac{m^2}{16 \lambda^2} {\bf I} [[1]]_{2,0,0} \Tr(\H^{(1,1)})\H^{(2)}_{\bullet \bullet} \,  .
\end{equation}

Finally, we have a decorated triangle (we neglect the double loop part, as it is not relevant for the tensor structure we are interested in)
\begin{equation}
    m_4=  \frac{m^2}{16} \H^{(2)}_{AB} \H^{(1)}_{C \bullet} \H^{(1)}_{D \bullet} \langle \zeta_{\sigma_1} \zeta_{\sigma_2} \zeta_{\sigma_3}^2\rangle \langle \xi^A_{\sigma_3} \xi^B_{\sigma_3} \partial_1 \xi^C_1 \partial_1 \xi^D_2\rangle = \frac{m^2}{8 \lambda^2} I_1 [[p_1^2]]_{3,0,0} \Tr(\H^{(1,1)}) \H^{(1,1,0)}_{\bullet \bullet} \,  .
\end{equation}

\section{Base $(\partial_0 y)^2$ Wick Contractions} \label{app:Base00}
We are now to evaluate in detail the Wick contractions associated to the base $(\partial_0 y)^2$  term. Completing combinations which are already fourth-order in derivatives, namely
\bea
b_1 &=& \langle  {\cal B}^{[2]}_4   \rangle  \, , \quad 
 b_2 =  i \langle   {\cal B}^{[1]}_1   {\cal B}^{[1]}_3   \rangle  \, , \quad 
 b_3 = -\frac{1}{2} \langle {\cal B}^{[0]}_2   {\cal B}^{[1]}_1 {\cal B}^{[1]}_1    \rangle \, , \quad 
b_4 = -\frac{i}{4} \langle {\cal B}^{[0]}_1   {\cal B}^{[0]}_1  {\cal B}^{[1]}_1 {\cal B}^{[1]}_1   \rangle   \, , 
\eea
 we have three which are second- or third-order
 \begin{equation}
     b_5 = \frac{i}{2} \langle \mathcal{B}_1^{[0]}\mathcal{B}_1^{[0]} \rangle \, , \quad b_6 = i \langle \mathcal{B}_1^{[1]} \mathcal{B}_2^{[0]}\rangle  \, , \quad b_7 = - \frac{1}{2} \langle \mathcal{B}_1^{[1]} \mathcal{B}_1^{[0]} \mathcal{B}_1^{[0]} \rangle \, .
 \end{equation}
 We need the following identifications 
 \begin{align}
     \mathcal{B}^{[0]}_1 &= - \frac{1}{2} \H^{(1)}_{AB} \zeta \partial_1 \xi^A \partial_1 \xi^B \, , \\
     \mathcal{B}_1^{[1]} &= \frac{1}{2} \Omega_{0 AB} \xi^A \partial_1 \xi^B \, , \\
     \mathcal{B}^{[1]}_3 &= - \frac{1}{8 \pi \epsilon}  \Tr(\H^{(2,1)}) \zeta \partial^\mu \zeta \partial_\mu y \, \\
     \mathcal{B}^{[2]}_0 &= - \frac{1}{32\pi \epsilon} \Tr(\H^{(1,1)}) \partial_\mu \zeta \partial^\mu \zeta - \frac{1}{8 \pi \epsilon \lambda} \left( \H^{(2)} - \H^{(1,0,1)}\right)_{AB} \partial_1 \xi^A \partial_1 \xi^B - \frac{1}{4} \H^{(2)}_{AB} \zeta^2 \partial_1 \xi^A \partial_1 \xi^B \,  , \\
     \mathcal{B}^{[2]}_4 &= -\frac{1}{32 \pi \epsilon \lambda} \left(- \Tr(\H^{(2,2)})   +\Tr(\H^{(3,1)}) \right) \zeta^2 \partial_\mu y \partial^\mu y \,  .
 \end{align}
The combinations $b_{5,6,7}$ require us to evaluate integrals with insertion of external momenta. When looking at terms on the base manifold with legs $\partial_0 y \partial_0 y$ or $\partial_1 y \partial_1 y$, the relevant basis of tensors turns out to be 
\be
\quad \Tr(\H^{(3,1)})\, , \quad \Tr(\H^{(2,2)}) \, , \quad  \Tr(\H^{(1,1,1,1)})   \, , \quad \Tr(\H^{(2,1)}\Omega  ) \, , \quad  \Tr(\H^{(1,1)}\Omega \Omega ) \, , \quad \Tr(\H^{(1)} \Omega \H^{(1)}  \Omega  )\, . 
\ee

\subsection*{$b_1$}
The first case is immediately solved as
\bea
b_1= \langle \mathcal{B}_4^{[2]}\rangle=- Y^{(2)} (\partial_0 y)^2 \langle \zeta^2 \rangle   =  - \frac{{\bf I} {\bf P} }{16\lambda} \left( \Tr(\H^{(2,2)}) + \Tr(\H^{(3,1)})   \right) (\partial_0 y)^2  \, .
\eea    

\subsection*{$b_2$}
This contribution consists of diagrams in which one vertex contains only $\zeta$ and the other only $\xi$ fields and hence upon Wick contraction no relevant 1PI graphs are produced. 

\subsection*{$b_3$}
 $b_3$ has a simple structure
\begin{align}
    b_3 &=  -\frac{1}{2} \langle {\cal B}^{[0]}_2   {\cal B}^{[1]}_1 {\cal B}^{[1]}_1    \rangle= \frac{1}{16} \left(4 X +\frac{i}{2 \lambda} {\bf I} \mathcal{H}^{(2)}\right)_{EF} \Omega_{0AB} \Omega_{0CD} \langle\partial_1 \xi^E_{\sigma_1} \partial_1 \xi^F_{\sigma_1} \xi^A_{\sigma_2} \partial_1 \xi^B_{\sigma_2} \xi^C_{\sigma_3} \partial_1 \xi^{D}_{\sigma_3} \rangle \nonumber \\
  &= \frac{1}{2} \left(4X +\frac{i}{2 \lambda} {\bf I} \mathcal{H}^{(2)}\right)_{EF} \Omega_{0AB} \Omega_{0CD} \langle\partial_1 \xi^E_{\sigma-1} \partial_1 \xi^B_{\sigma_2} \rangle \langle \partial_1 \xi^F_{\sigma_1} \partial_1 \xi^{D}_{\sigma_3}  \rangle \langle \xi^A_{\sigma_2}  \xi^C_{\sigma_3} \rangle \,  ,
\end{align}
where we have already contracted $\langle \zeta_\sigma^2\rangle = i \lambda^{-1} {\bf I}$ for simplicity and used the symmetries of the tensorial part to simplify the Wick contraction. We immediately recognise a triangle diagram, possibly decorated in the case of $\H^{(2)}$. The contractions are easily calculated as
\begin{equation} \begin{gathered}
    i \int \mathrm{d} \sigma_2 \mathrm{d} \sigma_3\langle\partial_1 \xi_{\sigma_1} \partial_1 \xi_{\sigma_2} \rangle \otimes \langle \partial_1 \xi_{\sigma_1} \partial_1 \xi_{\sigma_3}  \rangle \otimes\langle \xi_{\sigma_2}  \xi_{\sigma_3} \rangle \\ =  [[(p^1)^4]]_{3,0,0} \mathcal{H} \otimes \mathcal{H} \otimes \mathcal{H}  + [[(p^0)^2 (p^1)^2]]_{3,0,0} (\eta \otimes \eta \otimes \H + \eta \otimes \H \otimes \eta + \H \otimes \eta \otimes \eta) \,  . 
\end{gathered} \end{equation}
in which we have once again omitted the vanishing $[[p^0 (p^1)^3]]_{3,0,0} $.   
Using Method 2 we replace $ [[(p^0)^2 (p^1)^2]]_{3,0,0} \rightarrow {\bf T} +{\bf L}$ and $ [[(p^1)^4]]_{3,0,0} \rightarrow {\bf T}$.  This produces after simplification of the tensors
\be
\lambda b_3 = \left(\frac{3}{4}  {\bf I}{\bf L} +   {\bf I}{\bf T} \right)\Tr(\H^{(1,1)} \Omega \Omega) +  \frac{1}{4}( {\bf L}   {\bf I} - {\bf L}  {\bf P})   \Tr(\H^{(2,1)}\Omega) + \frac{1}{8} \left(  2 {\bf I }{\bf T}+  {\bf L}{\bf P} \right) \Tr(\H^{(1,1,1,1)})  \, .
\ee
Two remarks are in order.   Assuming that integration by parts holds, we can write 
\be 
{\bf L} = \int \frac{\mathrm{d}^2k}{(2 \pi)^2}\frac{ k_1^2}{(k^2)^2} = - \int \frac{\mathrm{d}^2k}{(2 \pi)^2} k_1 \partial_{k_1}\frac{ k_1^2}{(k^2)^2}  = -2 {\bf L} -4 {\bf T }
\ee
 which implies that the $ \Tr(\H^{(1,1)} \Omega \Omega)$ term vanishes.  However we will not enforce this directly but allow $3{\bf  L}_0 + 4{\bf T}_0 \neq 0 $ to keep track of any ambiguity.   The $\Tr(\H^{(2,1)}\Omega)$ coefficient gives rise only to a $\frac{\bar{\gamma}}{\epsilon} $ that will cancel against a counter-term insertion. 
    
\subsection*{$b_4$}
Within
\begin{align}
    b_4 &= -\frac{i}{4} \langle {\cal B}^{[0]}_1   {\cal B}^{[0]}_1  {\cal B}^{[1]}_1 {\cal B}^{[1]}_1   \rangle \nonumber \\
    &= - \frac{i}{64} \H^{(1)}_{EF} \H^{(1)}_{GH} \Omega_{0 AB} \Omega_{0CD} \langle \zeta_{\sigma_1} \zeta_{\sigma_2}\rangle \langle \xi^A_{\sigma_3} \xi^{C}_{\sigma_4}  \partial_1 \xi^E_{\sigma_1} \partial_1 \xi^F_{\sigma_1} \partial_1 \xi^G_{\sigma_2} \partial_1 \xi^H_{\sigma_2} \partial_1 \xi^B_{\sigma_3} \partial_1 \xi^D_{\sigma_4}\rangle \,  ,
\end{align}
there are two different topologies of   Wick contractions to consider. First is a {\em diamond sunset} arising from  
 \begin{align}
   \mathrm{DS} = 
  \int \mathrm{d}\sigma_{2}\mathrm{d}\sigma_{3}\mathrm{d}\sigma_4  \langle \zeta_{\sigma_1} \zeta_{\sigma_2}\rangle 
    \langle \partial_1 \xi_{\sigma_1} \xi_{\sigma_3}\rangle  \otimes
    \langle \partial_1 \xi_{\sigma_3}  \partial_1 \xi_{\sigma_2} \rangle \otimes
     \langle \partial_1 \xi_{\sigma_2} \partial_1 \xi_{\sigma_4}\rangle \otimes
         \langle \partial_1 \xi_{\sigma_1}   \xi_{\sigma_4}  \rangle \,  .
 \end{align}

 Setting $[[\dots ]]_{2,2,1}$   integrals  
  with odd number of timelike or spacelike components of momenta to zero, we get the contribution to $b_4$ from diamond sunset diagrams
 \begin{align}
     b_4|_{\mathrm{DS}}  =&  \frac{1}{4 \lambda} \Tr(\Omega \H^{(1)} \Omega \H^{(1)}) [[k^1 p^1 (k^0 p^0 - k^1 p^1)^2]]_{2,2,1} \nonumber \\
     &  - \frac{1}{2 \lambda} \Tr(\Omega \H^{(1,0)} \Omega \H^{(1,0)}) [[(k^1)^2 p^0 p^1 (k^1 p^0 - k^0 p^1)]]_{2,2,1} 
    \, . 
 \end{align}
 
 Proceeding now to Method 2 we obtain
 \be
 \lambda    b_4|_{\mathrm{DS}} = \left(\frac{1}{8}{\bf L}^2 - \frac{1}{4} {\bf TE}  \right) \Tr(\H^{(1,1,1,1)} )  + \left(-\frac{1}{2}{\bf L}^2+{\bf TE} - \frac{1}{8}{\bf S} \right) \Tr( \H^{(1)}\Omega  \H^{(1)}\Omega  ) \,  .
 \ee 

In addition the {\em square envelope} topology is given by  
\begin{equation}
    \mathrm{SE} = \int \mathrm{d} \sigma_2\mathrm{d} \sigma_3\mathrm{d} \sigma_4
    \langle \zeta_{\sigma_1} \zeta_{\sigma_2}\rangle 
    \langle \partial_1\xi_{\sigma_1}\partial_1\xi_{\sigma_2}\rangle \otimes \langle\partial_1\xi_{\sigma_2}\xi_{\sigma_3}\rangle \otimes  \langle \partial_1 \xi_{\sigma_3} \xi_{\sigma_4} \rangle
    \otimes \langle  \partial_1 \xi_{\sigma_4} \partial_1 \xi_{\sigma_1}\rangle \,  .
\end{equation}
Once the contractions are carried out, and basic identities applied, two tensors only appear. The final result for the square envelopes evaluates to 
\begin{align}
    \lambda b_4|_{\mathrm{SE}}  =&\frac{1}{2} \Tr(\H^{(1,1)}\Omega \Omega) [[k_1 p_1(3 k_1p_1 p_0^2 - 3 k_0 p_1^2 p_0  + k_1   p_1^3 -k_0 p_0^3)]]_{3,1,1} \nonumber \\
    &+ \frac{1}{4} \Tr(\H^{(1,1,1,1)}) [[k_1 p_1^2 (k_1 p_0^2 + k_1 p_1^2 - 2 k_0 p_0 p_1)]]_{3,1,1} \,  .
\end{align} 

Proceeding with Method 2 we have 
 \be
 \lambda    b_4|_{\mathrm{SE}} = \left(  \frac{1}{4} {\bf TE} - \frac{1}{4} {\bf I}{\bf T} \right) \Tr(\H^{(1,1,1,1)} ) -\left(- {\bf TE} + \frac{1}{4}{\bf I }{\bf L} +\frac{1}{8} {\bf S} + {\bf I}{\bf T} \right) \Tr( \H^{(1,1)}\Omega  \Omega  ) \,  .
 \ee

In summation we obtain
\bea
\lambda b_4 &=&  \left(  \frac{1}{8} {\bf L}^2 - \frac{1}{4} {\bf I}{\bf T} \right) \Tr(\H^{(1,1,1,1)} )  - \left(- {\bf TE} + \frac{1}{4}{\bf I }{\bf L} +\frac{1}{8} {\bf S} + {\bf I}{\bf T} \right) \Tr( \H^{(1,1)}\Omega  \Omega  ) \nonumber  \\&& 
+ \left(-\frac{1}{2}{\bf L}^2+{\bf TE} - \frac{1}{8}{\bf S} \right) \Tr( \H^{(1)}\Omega  \H^{(1)}\Omega  ) \, . 
\eea

\subsection*{$b_5$}
As it stands, $b_5$ contains tensors which, in total, are second-order in derivatives. We are thus prompted to extract terms quadratic in the external momentum $q$ from the loop integral.  We have
\begin{align}
     b_5  &=  \frac{i}{2} \langle   {\cal B}^{[0]}_1 {\cal B}^{[0]}_1    \rangle   
   = \frac{i}{8}  \int \mathrm{d}^2 \sigma_2  \H^{(1)}_{AB}(\sigma_1) \H^{(1)}_{CD}(\sigma_2) \langle \zeta_{\sigma_1} \zeta_{\sigma_2} \rangle \langle \partial_1 \xi^A_{\sigma_1} \partial_1 \xi^B_{\sigma_1} \partial_1 \xi^C_{\sigma_2} \partial_1 \xi^D_{\sigma_2}    \rangle  \nonumber\\
    &=\frac{1}{4} \H_{AB}^{(1)}(\sigma_1) \int \frac{\mathrm{d}^2 q}{(2 \pi)^2} \H_{CD}^{(1)}(q) e^{-i q \sigma_1} \int \frac{\mathrm{d}^2 k}{(2 \pi)^2}\frac{\mathrm{d}^2 p}{(2 \pi)^2} \frac{ p_1^2 k_1^2\H^{AC}\H^{BD} + p_1 p_0k_1 k_0 \eta^{AC}\eta^{BD} }{ p^2 k^2[q-(k+p)]^2} \, .
\end{align}
 Now, to second order, the denominator with external momentum insertion is expanded as
\begin{equation}
    \frac{1}{(q-(k+p))^2} = \dots + 4 \frac{[(k+p) \cdot q]^2}{[(k+p)^2]^3} - \frac{q^2}{[(k+p)^2]^2} + \dots \, .
\end{equation}

In the previous expression, we restrict ourselves to terms involving $q_0^2$, as we want to concentrate on external legs $(\partial_0y)^2$. In passing from momentum to position space, we turn $q_0^2$ into $-\partial_0^2$ acting upon $\H_{CD}^{(1)}$. However, as a second derivative would necessarily produce a derivative of $\Omega$,  integration by parts is in order. Specifically we let $\H_{AB}^{(1)} \partial_0^2 (\H_{CD}^{(1)}) = - \partial_0 (\H^{(1)}_{AB}) \partial_0 (\H^{(1)}_{CD})$. We eventually arrive at
\begin{align}
    b_5  =& -\frac{1}{2}\left(\Tr(\H^{(1,1)}\Omega \Omega ) - \Tr(\H^{(1)}\Omega \H^{(1)} \Omega) +2 \Tr(\H^{(2,1)}\Omega) - \frac{1}{2} \Tr(\H^{(2,2)})\right) \nonumber \\
    & \times \left( -[[p_1 k_1(p_0k_0 - p_1 k_1)]]_{1,1,2}  + 4 [[(k_0+ p_0)^2 p_1 k_1 (p_0 k_0-p_1k_1)]]_{1,1,3}\right) (\partial_0 y)^2 \,  .
\end{align}

With Method 2, the above expression boils down to 
\begin{equation}
    b_5 = \left( \frac{3}{4} {\bf S} + 2 {\bf TE}\right) \left(\Tr(\H^{(1,1)}\Omega \Omega ) - \Tr(\H^{(1)}\Omega \H^{(1)} \Omega) +2 \Tr(\H^{(2,1)}\Omega) - \frac{1}{2} \Tr(\H^{(2,2)})\right)(\partial_0 y)^2 \,  .
\end{equation}

\subsection*{$b_6$}

Although ${\cal B}_2^{[0]} = - Y \partial_\mu \zeta \partial^\mu \zeta - \frac{1}{4}\H^{(2)}_{AB} \zeta^2 \partial_1 \xi^A \partial_1 \xi^B -  X_{AB}\partial_1 \xi^A \partial_1 \xi^B  $ we can simplify the evaluation of $b_6 = i  \langle {\cal B}^{[1]}_1 {\cal B}_2^{[0]}  \rangle $ by noting the term with $Y$ will not contribute (it is a  disconnected diagram) and the $\zeta$ loop on the term with $\H^{(2)} $ is evaluated to $i {\bf I}$; hence effectively we use
\begin{equation}
  {\cal B}_2^{[0]} = A_{AB}\partial_1 \xi^A \partial_1 \xi^B   \, , \quad A_{AB} = -\frac{i}{4}{\bf I} \H^{(2)}_{AB} - X_{AB} \, .   
\end{equation}
 The contraction gives 
\begin{align} b_6  =&    -\frac{i}{2}\Omega_{0 AB} (\sigma_1) A_{CD}(\sigma_2) \langle \partial_1 \xi^A_{\sigma_1} \xi^B_{\sigma_1} \partial_1\xi^C_{\sigma_2} \partial_1 \xi^D_{\sigma_2} \rangle \nonumber \\
 = &    - \Omega_{0 AB} (\sigma_1) \int \frac{\mathrm{d}^2q }{(2 \pi)^2} A_{CD}(q) e^{-i q \sigma_1 }  \int \frac{\mathrm{d}^2 p  }{(2 \pi)^2}   \frac{  1 }{p^2 (p-q)^2}  \non\\
   &\times   \biggl( p_1 p_1 (q-p)_1  \H^{AC} \H^{BD} + p_0 p_1 (q-p)_0 \eta^{AC}\eta^{BD} \non \\
   &+  p_1 p_1 (q-p)_0\H^{AC}\eta^{BD}    +  p_0 p_1 (q-p)_1 \eta^{AC} \H^{BD}   \biggr) \,  .
\end{align}

To proceed one simply Taylor expands to extract the linear dependence on $q_0$ (and $q_1$, even though we will omit that part here) from the integrands.  In this case it is not even really necessary to use the specific rules for Method 2, for no $p_0^2$ appears once the dust settles. Still, in the language of Method 2 we can rephrase the result as 
\be
b_6 =\frac{1}{4} {\bf L}( {\bf I} - {\bf P})   \Tr(\H^{(2,1 )}\Omega ) \partial_0 y \partial_0y    - \frac{1}{8} {\bf L}( {\bf I} -{\bf P})  \Tr(\H^{(3,1)} ) \partial_0 y \partial_0y  \,  .
\ee  
Note that the final contribution from this diagram to $\partial_0 y \partial_1 y $ cancels out.

\subsection*{$b_7$}
This is by far the most complicated case and we shall provide the reader with additional details. The Wick contractions are easily simplified exploiting the symmetries of the tensorial structure 
 \begin{align}
     b_7 &= - \frac{1}{2} \langle  \mathcal{B}_1^{[0]} \mathcal{B}_1^{[0]}\mathcal{B}_1^{[1]}\rangle \nonumber \\
     &=\frac{1}{16} \mathcal{H}^{(1)}_{EF}(\sigma_1)
     \int \mathrm{d}\sigma_2 \mathrm{d} \sigma_3 \, \mathcal{H}^{(1)}_{CD}(\sigma_2) \Omega_{0AB}(\sigma_3) \langle \partial_1 \xi^A_{\sigma_3} \xi^{B}_{\sigma_3} \partial_1 \xi^{C}_{\sigma_2} \partial_1 \xi^D_{\sigma_2}\partial_1 \xi^{E}_{\sigma_1} \partial_1 \xi^F_{\sigma_1}\rangle \langle \zeta_{\sigma_1} \zeta_{\sigma_2}\rangle \nonumber \\
     &= \frac{1}{2} \mathcal{H}^{(1)}_{EF}(\sigma_1) \int \mathrm{d}\sigma_2 \mathrm{d} \sigma_3 \, \mathcal{H}^{(1)}_{CD}(\sigma_2) \Omega_{0AB}(\sigma_3)  
    \Delta_{12} \partial_{12}\Delta_{12}^{CE} \partial_{23}
\Delta_{23}^{AD} \partial_1 \Delta_{13}^{BF} \, , 
 \end{align} 
 where we have shortened the expression using $\Delta_{ij} \equiv \Delta(\sigma_i - \sigma_j)$ and $\Delta_{ij}^{AB} \equiv \H^{AB}\Delta(\sigma_i - \sigma_j) + \eta^{AB} \theta(\sigma_i - \sigma_j)$. The momentum routing is slightly subtle. First: two external momenta, $q$ and $l$ (Fourier partners of $\sigma_2$ and $\sigma_3$), have to be introduced. Second: we are entitled to choose the routing that will best suite our purpose. 

Killing every instance of $q_1$ and $l_1$ (they would eventually produce some $\partial_1 y$ leg) we arrive at
\begin{align}
    b_7  =& -\frac{i}{2} \mathcal{H}^{(1)}_{EF}(\sigma_1) \int \frac{\mathrm{d}^2 q}{(2 \pi)^2} \Omega_{0AB}(q)e^{-iq \sigma_1} \int \frac{\mathrm{d}^2 l}{(2 \pi)^2}\mathcal{H}^{(1)}_{CD}(l) e^{-il \sigma_1}  \nonumber \\
    &\times \int \frac{\mathrm{d}^2 p}{(2 \pi)^2}\frac{\mathrm{d}^2 k}{(2 \pi)^2} \frac{p_1(p_1+k_1)}{k^2 (p+k)^2 (p+l)^2 (p + l + q)^2}  \nonumber \\
    &\times\left((k_1+p_1) \H^{CE} + (p_0 + k_0) \eta^{CE}\right) \left( p_1\H^{AD} + (p_0 + l_0) \eta^{AD}\right) \left( p_1 \H^{BF}+ (p+l+q)_0 \eta^{BF}\right) \,  .
\end{align}

Judging from the tensorial structures already at our disposal, we need to extract from the loop integral linear terms in either $q_0$ or $l_0$

\begin{align}
b_7 =& -\frac{i}{2} \mathcal{H}^{(1)}_{EF}(\sigma_1)  \int \frac{\mathrm{d}^2 q}{(2 \pi)^2} \Omega_{0AB}(q)e^{-iq \sigma_1} \int \frac{\mathrm{d}^2 l}{(2 \pi)^2}\mathcal{H}^{(1)}_{CD}(l) e^{-il \sigma_1} \nonumber \\
&\times \int \frac{\mathrm{d}^2 p}{(2 \pi)^2}\frac{\mathrm{d}^2 k}{(2 \pi)^2}\frac{1}{(p^2)^3}\frac{1}{k^2}  \frac{p_1(p_1+k_1)}{(p+k)^2} \left((p_1+k_1) \H^{CE} + (p_0 + k_0) \eta^{CE}\right) \nonumber \\
    &\times \biggl( l_0 \left[ 4 p_0 p_1^2 \H^{AD} \H^{BF} + 2 p_0 (p_0^2+p_1^2) \eta^{AD} \eta^{BF} +  \left(2 p_0^2 p_1 + p_1 (p_0^2 + p_1^2) \right) \left(\H^{AD} \eta^{BF}+ \eta^{AD} \H^{BF} \right)\right] \nonumber \\
  &+ q_0 \left[ 2 p_0 p_1^2 \H^{AD} \H^{BF} + p_0(p_0^2+ p_1^2)\eta^{AD} \eta^{BF} + 2 p_0^2 p_1 \eta^{AD}\H^{BF} + p_1(p_0^2 + p_1^2) \H^{AD}\eta^{BF}\right] \biggr) \,  .
\end{align} 

We shall now pass to position space and turn $l_0,q_0$ into derivatives. While the former would hit a generalised metric, the latter would act upon $\Omega$. As we would like to avoid that situation we integrate by parts. Massaging the expression we obtain a little further, dropping terms linear in $p_0$ or $k_0$ 
\begin{align}
b_7 =& \frac{1}{2} \Omega_{0AB} \left(\H_{EF}^{(1)}  \partial_0 \H^{(1)}_{CD} - \partial_0\H^{(1)}_{EF}\H^{(1)}_{CD} \right) \int \frac{\mathrm{d}^2 p}{(2 \pi)^2}\frac{\mathrm{d}^2 k}{(2 \pi)^2} \frac{1}{(p^2)^3 k^2 (p+k)^2} \nonumber \\
&\times \biggl[p_1^2(p_1+k_1)^2 \left( 2 p_0^2 \eta^{AD} \H^{BF} + (p_0^2+p_1^2)\H^{AD} \eta^{BF} \right) \H^{CE} \nonumber \\
&+p_1 p_0 (p_1+k_1)(p_0+k_0) \left(2 p_1^2 \H^{AD} \H^{BF} + (p_0^2 + p_1^2) \eta^{AD} \eta^{BF} \right) \eta^{CE}\biggr] \, .
\end{align} \,  .

We can now perform the tensor contractions. The crucial observation here is that, when contracted, $\H_{AB}$  anti-commutes  with both $\H^{(1)}_{AB}$ and $\partial_0 \H^{(1)}_{AB}$. To see this, it is important to keep in mind that $\H^{(1)}_{AB} \equiv \V_A{}^I \H^{(1)}_{IJ} \V_B{}^J$.  We obtain
\begin{equation}
    b_7 =  I \left(\Tr(\H^{(1,1)}\Omega \Omega)  - \Tr(\H^{(1)}\Omega \H^{(1)}\Omega) + \Tr(\H^{(2,1)} \Omega)\right) \partial_0y \partial_0 y \, , 
\end{equation}
where $I$ stands for the integral we are left with, namely
\begin{equation}
    I = [[p_1 (k_1+p_1) \left(k_0 \left(p_0^3+3 p_0 p_1^2\right)-3 k_1 p_0^2 p_1-p_1^3 (k_1+p_1)+p_0^4\right)]]_{3,1,1} \, .
\end{equation}

Using Method 2 the latter becomes 
\begin{equation}
    I = \frac{1}{4} {\bf S} -2 {\bf TE} +\frac{1}{2} {\bf L I} + 2 {\bf T I} \,  .
\end{equation}
 
 \section{Base $\partial_0 y \partial_1 y$ Wick Contractions} \label{app:Base01}

In order to evaluate Wick contractions associated to the $\partial_0 y \partial_1 y$ legs we single out the following terms in the action
\begin{align}
    \mathcal{C}_1^{[0]} &= - \frac{1}{2} \H^{(1)}_{AB} \zeta \partial_1 \xi^A \partial_1 \xi^B \, , \\
    \mathcal{C}_2^{[0]} &= - \frac{1}{4} \H^{(2)}_{AB} \zeta^2 \partial_1 \xi^A \partial_1 \xi^B 
    - Y \partial_\mu \zeta \partial^\mu \zeta - X_{AB}\partial_1 \xi^A \partial_1 \xi^B 
    \, , \\
   \mathcal{C}_3^{[1;\tau]} &= -2 Y^{(1)} \zeta \partial_0\zeta \partial_0 y \, \\
    \mathcal{C}_3^{[1; \sigma]} &= - \frac{1}{2} \H^{(2)}_{BC} \Omega_{1 A}{}^C \zeta^2 \xi^A \partial_1 \xi^B  +2 Y^{(1)} \zeta \partial_1\zeta \partial_1 y - 2 X_{BC} \Omega_{1 A}{}^C \xi^A \partial_1 \xi^B \, , \\
    \mathcal{C}_{2}^{[1;\sigma]} &= - \H^{(1)}_{BC} \Omega_{1 A}{}^C \zeta \xi^A \partial_1 \xi^B\, , \\
    \mathcal{C}_{1}^{[1;\sigma]} &= -\H_{BC} \Omega_{1 A}{}^C \xi^A \partial_1 \xi^B + \frac{1}{2} \Omega_{1 AB} \xi^A \partial_0 \xi^B  \, , \\
    \mathcal{C}_1^{[1; \tau]} &=\frac{1}{2} \Omega_{0 AB} \xi^A \partial_1 \xi^B \, , \\
     \mathcal{C}_2^{[2;\tau,\sigma]} &= \frac{1}{2} \Omega_{1AC} \Omega_{0 B}{}^C \xi^A \xi^B \,  ,
\end{align}
where $\mathcal{C}_p^{[n;\sigma^\mu]}$ indicates a term with $p$ derivatives, $n$ external legs of type $\sigma^\mu$. They can be used to be create the following combinations
 \begin{align}
     c_1 &= - \frac{i}{2} \langle  \mathcal{C}_2^{[2; \tau, \sigma]} \mathcal{C}_1^{[0]} \mathcal{C}_1^{[0]} \rangle \, , \quad
     c_2 = - \langle \mathcal{C}_2^{[2; \tau, \sigma]} \mathcal{C}_2^{[0]} \rangle \, , \quad 
     c_3 = \frac{1}{2} \langle \mathcal{C}_1^{[1; \tau]} \mathcal{C}_1^{[1; \sigma ]}  \mathcal{C}_1^{[0]} \mathcal{C}_1^{[0]}\rangle \, , \\
     c_4 &= - i \langle \mathcal{C}_1^{[1;\tau]} \mathcal{C}_1^{[1;\sigma]}  \mathcal{C}_2^{[0]} \rangle \, , \quad 
     c_5 = - i \langle \mathcal{C}_2^{[1;\sigma]} \mathcal{C}_1^{[1;\tau]} \mathcal{C}_1^{[0]}\rangle \, , \quad
     c_6 = - \langle \mathcal{C}_3^{[1;\tau]} \mathcal{C}_1^{[1;\sigma]}\rangle \,  .
 \end{align}
 
These should be supplemented with additional terms coming from integral with external momenta insertion. The relevant basis of tensors is simply
\begin{equation}
    \Tr(\H^{(1,1,0)}\Omega \Omega) \, , \quad \Tr(\H^{(2)}\Omega \Omega) \,  .
\end{equation}
 
 \subsection*{$c_1$}
 In this case
 \begin{align}
     c_1 &= - \frac{1}{16} \H^{(1)}_{CD} \H^{(1)}_{EF} \Omega_{1AI} \Omega_{0B}{}^I\langle \zeta_{\sigma_2} \zeta_{\sigma_3}\rangle \langle \xi^A_{\sigma_1} \xi^B_{\sigma_1} \partial_1 \xi^C_{\sigma_2} \partial_1 \xi^D_{\sigma_2} \partial_1 \xi^E_{\sigma_3} \partial_1 \xi^F_{\sigma_3}\rangle \nonumber \\
     &= - \frac{1}{2 \lambda} [[k_1^2 (p_0^2+ p_1^2) - 2 k_1 k_0 p_1 p_0]]_{2,1,1} \Tr(\H^{(1,1,0)}\Omega \Omega) \partial_0 y \partial_1 y \nonumber  \\
     &= \frac{1}{2 \lambda} \left( {\bf L I} - \frac{1}{2} {\bf S}\right)\Tr(\H^{(1,1,0)}\Omega \Omega) \partial_0y \partial_1 y \,  .
 \end{align}
 
 \subsection*{$c_2$}
 
 In $c_2$ we discard a disconnected contraction and obtain
 \begin{align}
     c_2  &=  - \frac{i}{2} X_{CD} \Omega_{1AI} \Omega_{0B}{}^I \langle \xi^A_{\sigma_1} \xi_{\sigma_2}^B \partial_1 \xi^C_{\sigma_2} \partial_2 \xi^D_{\sigma_2}\rangle - \frac{i}{8} \H^{(2)}_{CD} \Omega_{1AI} \Omega_{0B}{}^I\langle \zeta_{\sigma_2} \zeta_{\sigma_2}\rangle \langle \xi^A_{\sigma_1} \xi^B_{\sigma_1} \partial_1 \xi^C_{\sigma_2} \partial_1 \xi^D_{\sigma_2}\rangle \nonumber\\
    & = - \frac{i}{2} \Omega_{1AI} \Omega_{0B}{}^I \left(X +\frac{i}{4} {\bf I} \H^{(2)}\right)_{CD} \langle \xi^A_1 \xi^B_1 \partial_1 \xi^C_2 \partial_1 \xi^D_2\rangle \nonumber \\
    &= \frac{1}{4\lambda} ({\bf I} - {\bf P})[[p_0^2 - p_1^2]]_{2,0,0}  \Tr(\H^{(2)}\Omega \Omega)\partial_0y \partial_1 y \nonumber \\
    &\phantom{=}  - \frac{1}{2 \lambda} \left({\bf I} [[p_1^2]]_{2,0,0}  + \frac{1}{2} {\bf P} [[p_0^2 - p_1^2]]_{2,0,0}\right) \Tr(\H^{(1,1,0)}\Omega \Omega) \partial_0 \partial_1 y\nonumber \\
    &= -\frac{1}{2 \lambda} {\bf I}\left( \bf{L } + \frac{1}{2} {\bf P } \right) \Tr(\H^{(1,1,0)}\Omega \Omega) \partial_0y \partial_1 y + \frac{1}{4 \lambda} {\bf I} \left(\bf{I} - {\bf P} \right) \Tr(\H^{(2)} \Omega \Omega) \partial_0y \partial_1 y \,  .
 \end{align}
 
 \subsection*{$c_3$}
 This combination comprises of square envelope and diamond sunset topologies, making it the hardest to compute. However, one can show that the diamond sunset topology does not ultimately contribute to the final result, as the corresponding integral always contains odd powers of 0-components. 
 
 \begin{align}
     c_3  &= - \frac{i}{16} \Omega_{0AB}\H^{(1)}_{CD} \H^{(1)}_{EF} \H_{ HI}\Omega_{1}{}^{I}{}_G  \langle \xi^G_{\sigma_1} \partial_1 \xi^{H}_{\sigma_1}  \xi^{A}_{\sigma_2} \partial_1 \xi^B_{\sigma_2} \partial_1 \xi^{C}_{\sigma_3} \partial_1 \xi^{D}_{\sigma_3} \partial_1 \xi^{E}_{\sigma_4} \partial_1 \xi^F_{\sigma_4}  \rangle \langle \zeta_{\sigma_3} \zeta_{\sigma_4} \rangle \nonumber \\
     &\phantom{=}- \frac{i}{32} \H^{(1)}_{CD} \H^{(1)}_{EF} \Omega_{1GH} \Omega_{0AB} \langle  \xi^G_{\sigma_1} \partial_0 \xi^H_{\sigma_1} \xi^A_{\sigma_2} \partial_1 \xi^B_{\sigma_2} \partial_1 \xi^C_{\sigma_3} \partial_1 \xi^D_{\sigma_3} \partial_1 \xi^E_{\sigma_4} \partial_1 \xi^{F}_{\sigma_4} \rangle \langle \zeta_{\sigma_3} \zeta_{\sigma_4}\rangle \nonumber \\
     & = \frac{1}{\lambda} [[(k_1^2(p_0^2 + 2 p_1^2) - k_0k_1p_0 p_1)(p_0^2 - p_1^2) + 4 k_1 p_0 p_1^2 (k_0 p_1 -k_1 p_0)]]_{3,1,1} \Tr(\H^{(1,1,0)}\Omega \Omega) \partial_0 y \partial_1 y \nonumber \\
     & =\frac{1}{\lambda} \left(\frac{3}{4} {\bf S} - 2 {\bf TE} - \frac{1}{2} {\bf L I} +2 {\bf T I}\right)\Tr(\H^{(1,1,0)}\Omega \Omega) \partial_0 y \partial_1 y \, .
 \end{align}
 
 \subsection*{$c_4$}
 
 In this case the topologies involved are triangle (with counter-term insertion) and decorated triangle. 
 \begin{align}
     c_4 &= \frac{1}{2} \H_{DI} \Omega_1{}^{I}{}_{C} X_{EF} \Omega_{0AB} \langle \xi^C_{\sigma_1} \xi^A_{\sigma_2}\partial_1 \xi^D_{\sigma_1} \partial_1 \xi^B_{\sigma_2} \partial_1 \xi^E_{\sigma_3} \partial_1 \xi^F_{\sigma_3}\rangle \nonumber \\
     &\phantom{=}+\frac{1}{8} \H^{(2)}_{EF} \H_{DI} \Omega_1{}^I{}_C \Omega_{0AB} \langle \zeta_{\sigma_3}^2\rangle \langle \xi_{\sigma_1}^C \xi_{\sigma_2}^A \partial_1 \xi^D_{\sigma_1} \partial_1 \xi^B_{\sigma_2} \partial_1 \xi^E_{\sigma_3} \partial_1 \xi^F_{\sigma_3}\rangle \nonumber \\
     &\phantom{=}+ \frac{1}{4} X_{EF} \Omega_{1CD} \Omega_{0AB} \langle \xi_{\sigma_1}^C \xi_{\sigma_2}^A \partial_0 \xi^D_{\sigma_1} \partial_1 \xi^B_{\sigma_2} \partial_1 \xi^E_{\sigma_3} \partial_1 \xi^F_{\sigma_3}\rangle \nonumber \\
     & \phantom{=} + \frac{1}{16} \H^{(2)}_{EF} \Omega_{1CD} \Omega_{0AB} \langle \zeta_{\sigma_3}^2\rangle \langle \xi_{\sigma_1}^C \xi_{\sigma_2}^A \partial_0 \xi^D_{\sigma_1} \partial_1 \xi^B_{\sigma_2} \partial_1 \xi^E_{\sigma_3} \partial_1 \xi^F_{\sigma_3}\rangle \nonumber \\
     &= \frac{1}{\lambda} \left(\frac{1}{2} {\bf P}[[(p_0^2 - p_1^2)^2]]_{3,0,0} -\frac{1}{2} {\bf I} [[p_1^2 (3 p_1^2 + p_0^2)]]_{3,0,0}  \right) \Tr(\H^{(1,1,0)}\Omega \Omega) \partial_0y \partial_1 y \nonumber \\
     &\phantom{=}+ \frac{1}{2\lambda} ({\bf P} - {\bf I}) [[(p_0^2 - p_1^2)^2]]_{3,0,0}\Tr(\H^{(2)}\Omega \Omega) \partial_0y \partial_1 y \nonumber \\
     &= \frac{{\bf I}}{2 \lambda} \left( {\bf P } - {\bf L }  - 4 {\bf T }\right) \Tr(\H^{(1,1,0)}\Omega \Omega) \partial_0y \partial_1 y + \frac{{\bf I}}{2 \lambda} ({\bf P } - {\bf I}) \Tr(\H^{(2)} \Omega \Omega)\partial_0y \partial_1 y \,  .
 \end{align}

\subsection*{$c_5$}
The only topology involved here are triangle envelopes. 
\begin{align}
    c_5 &=  \frac{1}{4} \H^{(1)}_{B L} \H^{(1)}_{EF} \Omega_{1}{}^{L}{}_A \Omega_{0CD} \langle \zeta_{\sigma_1} \zeta_{\sigma_3}\rangle \langle \xi^{A}_{\sigma_1} \xi^C_{\sigma_2} \partial_1 \xi^B_{\sigma_1} \partial_1 \xi^F_{\sigma_2} \partial_1 \xi^E_{\sigma_3} \partial_1 \xi^F_{\sigma_3}\rangle \nonumber \\
    & = -\frac{1}{\lambda} [[k_1^2 (p_0^2 + p_1^2) -2 k_0 k_1 p_0 p_1]]_{2,1,1} \Tr(\H^{(1,1,0)}\Omega \Omega) \partial_0 y \partial_1 y \nonumber \\
    &=  \frac{1}{\lambda} \left( {\bf LI} - \frac{1}{2} {\bf S} \right) \Tr(\H^{(1,1,0)}\Omega \Omega) \partial_0 y \partial_1 y \,  .
\end{align}

\subsection*{$c_6$}

This one contains only loop and decorated loop topologies. 

\begin{align}
    c_ 6 &= i X_{DE} \Omega_{0AB} \Omega_1^{E}{}_C \langle \xi^C_{\sigma_1} \xi^A_{\sigma_2} \partial_1 \xi^D_{\sigma_1} \partial_1 \xi^B_{\sigma_2}\rangle + \frac{i}{4} \H^{(2)}_{DI} \Omega_{1}{}^{I}{}_C \Omega_{0AB} \langle \zeta_{\sigma_1} \zeta_{\sigma_1}\rangle \langle \xi^C_{\sigma_1} \xi^A_{\sigma_2} \partial_1 \xi^D_{\sigma_1} \partial_1 \xi^B_{\sigma_2}\rangle  \nonumber \\
    & =- \frac{1}{2 \lambda} \left( {\bf P} [[p_0^2]]_{2,0,0} + {\bf I} [[p_1^2]]_{2,0,0}\right)  \Tr(\H^{(1,1,0)}\Omega \Omega) \partial_0y \partial_1 y \nonumber\\
    &\phantom{=} +\frac{1}{2 \lambda} [[p_0^2]]_{2,0,0} ({\bf I} - {\bf P}) \Tr(\H^{(2)}\Omega \Omega) \partial_0y \partial_1 y \nonumber \\
    &= - \frac{1}{2 \lambda} \left({\bf I P } + \bf{L I} + {\bf L P} \right)  \Tr(\H^{(1,1,0)}\Omega \Omega) \partial_0y \partial_1 y + \frac{1}{2 \lambda} ({\bf I} + {\bf L })({\bf I} - {\bf P}) \Tr(\H^{(2)}\Omega \Omega) \partial_0y \partial_1 y  \, .
\end{align}

\subsection*{Contributions From External Momentum Insertion}

There are a few contributors to $\partial_0 y \partial_1 y$ legs that $c_1,\dots, c_6$ have missed. These arise from loop integral with non-vanishing external momentum. A careful analysis reveals that in fact only the sunset topology with a linear insertion of external momentum is responsible for such contributions. Also, we shall only get a correction, call it $c_7$, to $\Tr(\H^{(2)}\Omega \Omega)$. We find
\begin{align}
    c_7 &= \frac{2}{\lambda} [[k_1 p_0 (k_1 p_0 - k_0 p_1)]]_{2,1,1} \Tr(\H^{(2)}\Omega \Omega) \partial_0 y \partial_1 y \nonumber \\
    & = \frac{1}{\lambda} \left( \frac{3}{2} \bf S - \bf{L I} \right)\Tr(\H^{(2)}\Omega \Omega) \partial_0 y \partial_1 y  \,  . 
\end{align}
 
\section{Results of Wick Contractions}\label{sec:preev}

 Before we evaluate the loop integrals,  we collect and present in this section the results of the standard Wick contractions.   To slightly reduce the length of expressions, each time a numerator of a loop integrand contains an even number of $\tau$-components of momenta, these are swapped for the invariant combination e.g. $p_0^2 \rightarrow p^2 + p_1^2$, though at this stage no assumption about how this holds in $d=2+\epsilon$ is made and, in particular, no cancellations of factors of $p^2$ between numerator and denominator are employed up to this point. 

For the counter-term attached to $\lambda^{-2}\partial_1 \X^I \partial_1 \X^J$ we obtain

  \begin{align}
    \H^{(4)}_{\bullet\bullet}:\quad&  \frac{1}{16} {\bf I}^2 - \frac{1}{8} {\bf I} {\bf P} \nonumber \\  \H^{(3,1,0)}_{\bullet\bullet}:\quad&  \frac{1}{2} ( {\bf P} - {\bf I}) [[p_1^2]]_{2,0,0}  - \frac{1}{4} {\bf I}{\bf P}\nonumber \\  \H^{(2,0,2)}_{\bullet\bullet}:\quad& \frac{1}{4} {\bf I }{\bf P} + \frac{1}{4} [[p_1^2]]_{1,1,1}\nonumber  \\  \H^{(2,1,1)}_{\bullet\bullet}:\quad& \frac{1}{2}{\bf I} {\bf P} - \frac{1}{2} {\bf P} [[p_1^2]]_{2,0,0} + [[p_1 k_1 p \cdot k]]_{2,1,1}\nonumber  \\  \H^{(1,2,1)}_{\bullet\bullet}:\quad&   \frac{1}{8} {\bf I}{\bf P} - \frac{1}{2}{\bf P} [[p_1^2]]_{2,0,0} - \frac{1}{2} [[p_1^2 k_1^2]]_{2,2,0} - \frac{1}{4} ({\bf P} - {\bf I}) [[p_1^2 p^2]]_{3,0,0}  \nonumber \\  \H^{(1,1,1,1,0)}_{\bullet\bullet}:\quad& \frac{1}{2} {\bf P} [[p_1^2]]_{2,0,0} + [[p_1^2 k_1^2]]_{2,2,0} + \frac{1}{2} {\bf I} [[p_1^4]]_{3,0,0} + \frac{1}{4} {\bf P} [[p_1^2 p^2]]_{3,0,0} \nonumber \\ 
    & - [[p_1^3 k_1 k\cdot p]]_{3,1,1} + \frac{1}{2}[[p_1^2 k_1^2   p^2]]_{3,1,1}    -   [[(p_1^2 k_1^2  + p_1^3 k_1)  k^2]]_{2,2,1}    \nonumber\\ 
     \H^{(2)}_{\bullet\bullet} \Tr (\H^{(1,1)}):\quad& -\frac{1}{32}{\bf P} [[p^2]]_{2,0,0} + \frac{1}{8} [[(p+k)_1 k_1 (p\cdot k + k^2)  ]]_{2,1,1} \nonumber \\  \H^{(1,1,0)}_{\bullet\bullet} \Tr(\H^{(1,1)}):\quad& \frac{1}{16}{\bf P}  [[p_1^2 p^2]]_{3,0,0} - \frac{1}{4} [[ (p_1 +k_1) p_1^2 k_1 (p\cdot k + k^2) ]]_{3,1,1} 
 \end{align} 
For counter-term attached to $\partial_0 y \partial_0 y$ external legs we find:
 \begin{align}
     \Tr(\H^{(2,2)}):\quad& [[k_1(k_1+p_1) (p^2 + p_1^2) (k^2 + k \cdot p)]]_{3,1,1} - \frac{1}{4} [[k_1(k_1+p_1)  (k^2 + k \cdot p)]]_{2,1,1} \nonumber \\
      \Tr(\H^{(3,1)}):\quad & \frac{1}{8}\left({\bf P } - {\bf I} \right)[[p_1^2]]_{2,0,0} \nonumber \\
      \Tr(\H^{(1,1,1,1)}):\quad&  \frac{1}{8}{\bf P } [[p_1^2 p^2]]_{3,0,0} + \frac{1}{4} {\bf I} [[p_1^4]]_{3,0,0} + \frac{1}{4} [[k_1p_1^2(k_1 p^2 - 2 p_1 k \cdot p)]]_{3,1,1} \nonumber \\
     &- \frac{1}{16} [[k_1^2 p^2 - 4 p_1 k_1 k \cdot p +3 p_1^2 k^2]]_{2,2,1} \nonumber \\
      \Tr(\H^{(1,1)}\Omega \Omega):\quad& 
     \frac{1}{4} {\bf I} [[p_1^2 (3 p^2+ 4 p_1^2)]]_{3,0,0} + \frac{1}{2} [[k_1 (k \cdot p (k_1 - p_1) + k^2 (k_1 + p_1))]]_{2,1,1} \nonumber \\
    &- \frac{1}{2} [[k_1 (2 p^2 k_1 p_1^2 + 4 k^2 (k_1 + p_1) (p^2 + p_1^2) + k\cdot p (4 k_1 p_1^2 + p^2(4 k_1 + p_1)))]]_{3,1,1}
     \nonumber \\
     \Tr(\H^{(1)}\Omega \H^{(1)}\Omega):\quad &- \frac{1}{2} [[k_1 (k \cdot p(k_1 - p_1) + k^2 (k_1 + p+1))]]_{2,1,1} \nonumber \\
     &2 [[k_1 (k_1(k \cdot p + k^2) p^2 +p_1 k^2 p^2 + k_1p_1^2(k \cdot p +k^2 + p^2) + p_1^3 (k^2 - k \cdot p)  )]]_{3,1,1} \nonumber \\
     \Tr(\H^{(2,1)}\Omega):\quad & 
     \frac{1}{4} ({\bf P }- {\bf I}) [[p_1^2]]_{2,0,0} - \frac{1}{4} ({\bf P} - {\bf I}) [[p_1^2 p^2]]_{3,0,0} + [[k_1 (k^2 (k_1+p_1)+ k_1 k \cdot p)]]_{2,1,1} \nonumber \\
     &-2 [[k_1 (k_1 p_1^2 p^2 +2 (k_1 + p_1) k^2 (p^2 + p_1^2) +(2 k_1p_1^2 + 2 k_1 p^2 + p_1p^2) k \cdot p)]]_{3,1,1}
 \end{align}
  
 For the counter-term attached to mixed derivatives of the background field $\partial_0 y \partial_1 y$ we obtain:
 \begin{align}
      \Tr(\H^{(1,1,0)} \Omega \Omega):\quad &  -\frac{1}{4}{\bf P}[[3 p^2 -2 p_1^2]]_{2,0,0} - {\bf I}[[p_1^2]]_{2,0,0} + \frac{1}{2}{\bf I} [[(p^2)^2]]_{3,0,0}\nonumber \\
     &- \frac{1}{2} {\bf I} [[p^2 p_1^2-4 p_1^4]]_{3,0,0} -\frac{3}{2} [[k_1^2 p^2 -2 k_1 p_1 k \cdot p]]_{2,1,1}\nonumber \\
     & + [[k_1(k_1(p^2)^2 +4 p_1^3 k \cdot p -p_1 p^2 (k \cdot p + 2 k_1 p_1))]]_{3,1,1} \nonumber \\
    \Tr(\H^{(2)}\Omega \Omega):\quad &  
     - \frac{1}{4} ({\bf P} - {\bf I})[[3p^2 + 2p_1^2]]_{2,0,0} + \frac{1}{2} ({\bf P}-{\bf I})[[(p^2)^2]]_{3,0,0} \nonumber \\
     &+2 [[k_1(k_1(p^2)^2 -2 k_1 p_1^2 p^2 + 4 p_1^3 k \cdot p)]]_{3,1,1}
 \end{align}
 For the counter-term attached to   $\partial_1 y \partial_1 y$ we obtain
 \begin{align}
      \Tr(\H^{(2,2)}):\quad & [[k_1(k_1+p_1) p_1^2 (k^2 + k \cdot p)]]_{3,1,1} + \frac{1}{4} [[k_1(k_1+p_1)  (k^2 + k \cdot p)]]_{2,1,1} \nonumber \\
      \Tr(\H^{(3,1)}):\quad & \frac{1}{8} {\bf I}^2- \frac{1}{8} {\bf P} [[p^2 -  p_1^2]]_{2,0,0} + \frac{3}{8}{\bf I} [[p_1^2]]_{2,0,0} - \frac{1}{2}{\bf P} [[p_1^2 p^2]]_{3,0,0} \nonumber \\
      \Tr(\H^{(1,1,1,1)}):\quad & \frac{1}{8}{\bf P} [[p^2+2 p_1^2]]_{2,0,0} + \frac{1}{4}{\bf I}[[p_1^2]]_{2,0,0} -
     \frac{1}{8}{\bf P } [[p_1^2 p^2+ (p^2)^2]]_{3,0,0} + \frac{1}{4} {\bf I} [[p_1^4]]_{3,0,0}  \nonumber \\
     &+ \frac{1}{4}[[k_1^2 + k_1 p_1]]_{1,1,1} + \frac{1}{4} [[k_1^2p^2 - 2p_1(p_1+2k_1) k \cdot p]]_{2,1,1} \nonumber \\
     &- \frac{1}{4} [[k_1(k_1 p^2 - 2 p_1 k \cdot p)(p^2 - p_1^2)]]_{3,1,1} 
     + \frac{1}{16} [[k_1^2 p^2 (k \cdot p + k_1 p_1)]]_{2,2,1} \nonumber \\
      \Tr(\H^{(1,1)}\Omega \Omega) :\quad & 
    \frac{1}{4} {\bf I} [[p^2 + 6p_1^2]]_{2,0,0}- \frac{1}{4} {\bf I} [[(p^2)^2 + p_1^2 p^2 - 4 p_1^4]]_{3,0,0}\nonumber \\
    &- \frac{1}{2} [[\frac{(k_1+p_1)(p^2 + k \cdot p)}{p_1}]]_{1,1,1}\nonumber \\
    &- \frac{1}{2} [[\frac{k_1}{p_1} (- 2 p^2 k \cdot p + k_1 p_1 (k \cdot p + k^2 -3 p^2) +5 p_1^2 (k^2 + k \cdot p))]]_{2,1,1} \nonumber \\
     &- \frac{1}{2} [[\frac{k_1}{p_1} ((p^2)^2 k \cdot p + 2 p_1 k_1 (p^2)^2 + p_1^2 p^2 k \cdot p  ]]_{3,1,1}
      \nonumber \\
      &- [[ \frac{k_1}{p_1}(k_1p_1^3 (p^2 + 2 k \cdot p + 2k^2) + 2 k^2 p_1^4)]]_{3,1,1} \nonumber \\
     \Tr(\H^{(1)}\Omega \H^{(1)}\Omega):\quad & - [[p^2 + k \cdot p]]_{1,1,1} \nonumber \\
     &+ \frac{1}{2} [[(4 p_1^2 - k_1 p_1 + k_1^2 - 2 p^2)k \cdot p + k_1 (k^2 k_1 + 5 p_1 k^2 - 2 p_1 p^2)]]_{2,1,1} \nonumber \\
     &+2 [[k_1 p_1 (p^2 k \cdot p) + k_1 p_1 (k \cdot p + k^2 +p^2) +(k^2 - k \cdot p) p_1^2]]_{3,1,1}   \nonumber \\
    &-\frac{1}{8} [[k_1 p_1 (-2 k^2 p^2 + k_1^2 p^2 + 7 p_1^2 k^2) ]]_{2,2,1}  \nonumber \\
    &- \frac{1}{8} [[k \cdot p (k_1^2 (3p^2 - 8 p_1^2) + k^2 (-2 p^2 + 5 p_1^2))]]_{2,2,1}\nonumber \\
      \Tr(\H^{(2,1)}\Omega):\quad &  
     - \frac{1}{4} {\bf P} [[2p^2 + p_1^2]]_{2,0,0} + \frac{1}{4} {\bf I}[[p^2 + 5 p_1^2]]_{2,0,0} \nonumber \\
     &- \frac{1}{4} ({\bf P} - {\bf I}) [[(p^2)^2]]_{3,0,0} - \frac{3}{4} {\bf P} [[p^2 p_1^2]]_{3,0,0} - \frac{1}{4} {\bf I}[[p_1^2 p^2]]_{3,0,0} 
     \nonumber \\ &+ [[k_1 (k^2 (k_1+p_1)+ k_1 k \cdot p)]]_{2,1,1} \nonumber \\
     &-2 [[k_1 (k_1 p_1^2 p^2 +2 (k_1 + p_1) k^2 (p^2 + p_1^2) +(2 k_1p_1^2 + 2 k_1 p^2 + p_1p^2) k \cdot p)]]_{3,1,1}
 \end{align}
 
 \section{Loop Integrals via $O(d)$-invariance (Method 1)} \label{app:Integrals}
 
 For a $L$-loop calculation, $2L$ copies of the worldsheet are needed. We shall label each copy by a number and indicate the $n$-th with $\mathbf{n}$. Propagators stretching from $\mathbf{n}$ to $\mathbf{m}$ will be reported schematically as $\mathbf{n} \rightarrow \mathbf{m}$, where $\mathbf{n}$ and $\mathbf{m}$ are allowed to coincide. Accordingly, a sequence with the same extrema shall indicate a closed loop.

 At one-loop order, two topologies only contribute to the $\beta$-function: having two copies of the worldsheet,  propagators can either stretch from $\mathbf{1}$ to $\mathbf{2}$ or close on the same copy. All possible loops are exhausted by $\mathbf{1} \rightarrow \mathbf{1}$ and $(\mathbf{1} \rightarrow \mathbf{2})^2$. We shall name them \textit{bubble} and \textit{loop}, respectively.

 At two-loop order,  divergences can originate from either  i) one-loop integrals multiplied by a $1/\epsilon$ pole due to one-loop counter-term insertions; ii) products of one-loop integrals; iii) genuinely new two-loop integrals. New topologies for one-loop diagrams arise in i) and ii).  
 
 For a two-loop calculation, triangle-shaped and square-shaped loops, corresponding schematically to $\mathbf{1} \rightarrow \mathbf{2} \rightarrow \mathbf{3} \rightarrow \mathbf{1}$ and $\mathbf{1} \rightarrow \mathbf{2} \rightarrow \mathbf{3} \rightarrow \mathbf{4} \rightarrow \mathbf{1}$, are in fact allowed. 
 For the sake of simplicity, we will refer to them  as \textit{triangle} and \textit{square} diagrams. On top of that, $\mathbf{1} \rightarrow \mathbf{1}$ and $(\mathbf{1} \rightarrow \mathbf{2})^2$ can still appear, even though their finite $O(\epsilon^0)$ parts are now to be kept\footnote{Actually, square diagrams are only required in calculating one-loop diagrams with external quantum fields. }. 
 
 Graphs of type ii)  can all be seen graphically as dressings of the previous diagrams (up to the squares) with an extra loop or bubble. When a bubble is added, we call the resulting diagram ``decorated'': for example, adding a bubble to a triangle results in a \textit{decorated triangle}. 
 
 Scenario iii) is the most intricate as it allows for many different topologies.   The simplest instance of genuine two-loop diagram is the \textit{sunset} diagram, corresponding to $(\mathbf{1} \rightarrow \mathbf{2})^3$. The only way to non-trivially extend the triangle diagram to two-loop is by adding an internal line, resulting in $(\mathbf{1} \rightarrow \mathbf{2})^2 \rightarrow \mathbf{3} \rightarrow \mathbf{1}$. We shall call it \textit{triangle envelope} diagram. Squares allow for two extensions: we can either add a line joining two adjacent vertices, $(\mathbf{1}\rightarrow \mathbf{2})^2 \rightarrow \mathbf{3} \rightarrow \mathbf{4}$ (\textit{square envelope}), or let is stretch to the opposite vertex, $(\mathbf{1} \rightarrow \mathbf{2} \rightarrow \mathbf{3} \rightarrow \mathbf{4} \rightarrow \mathbf{1})(\mathbf{1} \rightarrow \mathbf{3})$ (\textit{diamond sunset}).

  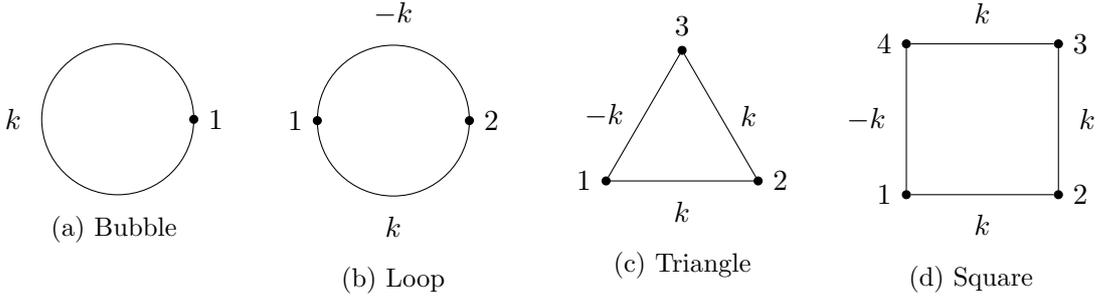
\begin{figure}[!h]
 \begin{subfigure}{.23\textwidth}
   \centering
    \begin{tikzpicture}
    \node[draw,circle,fill=black,minimum size=3pt,inner sep=0pt] (one) at (1,0) [label=right:$1$]{};
    \node (mome) at (-1,0) [label=left:$k$] {};
    \draw (0,0) circle [radius=1];
        \end{tikzpicture}
  \caption{Bubble}
  \label{fig:sfig1}
\end{subfigure}%
\begin{subfigure}{.23\textwidth}
  \centering
       \begin{tikzpicture}
    \node[draw,circle,fill=black,minimum size=3pt,inner sep=0pt] (one) at (0,0) [label=left:$1$]{};
    \node[draw,circle,fill=black,minimum size=3pt,inner sep=0pt] (two) at (2,0) [label=right:$2$]{};
    \node (mome1) at (1,-1) [label=below:$k$] {};
     \node (mome2) at (1,1) [label=above:$-k$] {};
    \draw (1,0) circle [radius=1];
        \end{tikzpicture}
  \caption{Loop}
  \label{fig:sfig2}
\end{subfigure}
\begin{subfigure}{.23\textwidth}
  \centering
   \begin{tikzpicture}
    \node[draw,circle,fill=black,minimum size=3pt,inner sep=0pt] (one) at (0,0) [label=left:$1$]{};
    \node[draw,circle,fill=black,minimum size=3pt,inner sep=0pt] (two) at (2,0) [label=right:$2$]{};
    \node[draw,circle,fill=black,minimum size=3pt,inner sep=0pt] (three) at (1,1.73205) [label=above:$3$]{};
    \draw (one) to node [label=below:$k$]{} (two);
    \draw (two) to node [label=right:$k$]{} (three);
    \draw (three) to node [label=left:$-k$]{} (one);
        \end{tikzpicture}
  \caption{Triangle}
  \label{fig:sfig3}
\end{subfigure}
\begin{subfigure}{.23\textwidth}
  \centering
   \begin{tikzpicture}
    \node[draw,circle,fill=black,minimum size=3pt,inner sep=0pt] (one) at (0,0) [label=left:$1$]{};
    \node[draw,circle,fill=black,minimum size=3pt,inner sep=0pt] (two) at (2,0) [label=right:$2$]{};
    \node[draw,circle,fill=black,minimum size=3pt,inner sep=0pt] (three) at (2,2) [label=right:$3$]{};
     \node[draw,circle,fill=black,minimum size=3pt,inner sep=0pt] (four) at (0,2) [label=left:$4$]{};
    \draw (one) to node [label=below:$k$]{} (two);
    \draw (two) to node [label=right:$k$]{} (three);
    \draw (three) to node [label=above:$k$]{} (four);
    \draw (four) to node [label=left:$-k$]{} (one);
        \end{tikzpicture}
  \caption{Square}
  \label{fig:sfig4}
\end{subfigure}
\caption{One-loop diagrams.}
\label{fig:fig}
\end{figure}

\subsection{Combinatorics} \label{sec:combinatorics}
  The basic integrals  in Minkowski space are given by
\begin{equation}
I_n= \int \frac{\mathrm{d}^d k}{(2\pi)^d}   \frac{1 }{(k^2 -m^2)^n} = \frac{(-1)^n i}{(4\pi)^\frac{d}{2}} \frac{\Gamma(n-\frac{d}{2})} {\Gamma(n)} (m^2)^{\frac{d}{2}- n} \, . 
\end{equation}

Our aim is to recast divergent 1-loop and 2-loop integrals as combinations of the basic integrals $I_n$, postponing the evaluation of their precise dependence on the dimensional-regulator $\epsilon$. At 2-loop order we  require only $I_1, I_2$ and $I_3$ which are given  by 
\be
I_1\approx\frac{i}{2 \pi  \epsilon }+\frac{i \bar{\gamma }}{4 \pi }+  \frac{i \epsilon  \left(6 \bar{\gamma }^2+\pi ^2\right)}{96 \pi } \, ,\quad m^2 I_2 \approx +\frac{i}{4 \pi  } + \frac{i \epsilon  \bar{\gamma }}{8 \pi  }\, ,\quad
m^4 I_3\approx -\frac{i}{8 \pi   } -\frac{i \epsilon  \left(\bar{\gamma }-1\right)}{16 \pi   }\, , 
\ee 
where 
\be
\bar{\gamma}= \gamma_E  + \log\left( \frac{m^2}{4 \pi} \right) \, . 
\ee
  Integrals with non-positive $n$ can also be dropped: $\Gamma(n)$ has poles for $n \in \mathbb{Z}^-$, thus forcing $I_{n \leq 0} \rightarrow 0$.

Prior to venturing into the explicit evaluation of loop integrals, let us pause and analyse their combinatorial structure. Fix some integer number $q \in \mathbb{N}$ and consider the one-loop integral
\begin{equation}
    I_{\lambda_1 \dots \lambda_{2q}}= \int \frac{\mathrm{d}^d p}{(2 \pi)^d} \frac{p_{\lambda_1} \dots p_{\lambda_{2q}}}{D}
\end{equation}
for some denominator $D=D(p^2)$. Integrals of this form are usually evaluated assuming $O(d)$ symmetry in the final result; that is, we postulate that the right-hand side can be recast in the form
\begin{equation} \label{eq:etaperm}
    \int \frac{\mathrm{d}^d p}{(2 \pi)^d} \frac{p_{\lambda_1} \dots p_{\lambda_{2q}}}{D} = A (\eta_{\lambda_1 \lambda_2} \dots \eta_{\lambda_{2q-1} \lambda_{2q}} + \mathrm{perms.})  \int \frac{\mathrm{d}^d p}{(2 \pi)^d} \frac{(p^2)^{q}}{D} \,  ,
\end{equation}
where the ``scalar'' integral is in general known, and the prefactor $A$ is to be fixed by taking appropriate contractions of both the left- and right-hand side with the $d$-dimensional Minkowski metric. 

Let us indicate by $P_{\lambda_1 \dots \lambda_{2q}}$ the rank $2q$ totally symmetric tensor made up of all possible permutations of tensor products of the $d$-dimensional Minkowski metric 
\begin{equation}
    P_{\lambda_1 \dots \lambda_{2q}} = \eta_{\lambda_1 \lambda_2} \dots \eta_{\lambda_{2q-1} \lambda_{2q}} + \mathrm{perms.} 
\end{equation}
As can be easily seen, a total number of $(2q-1)!!$ terms have to be accounted for in the permutations. Particularly important is the trace of this object, $P^{(2q)}$,  
\begin{equation} \label{eq:P2q}
   P^{(2q)} \equiv \eta^{\lambda_1 \lambda_2} \dots \eta^{\lambda_{2q-1} \lambda_{2q}} P_{\lambda_1 \dots \lambda_{2q}}  = \prod_{j=0}^{q-1} (d+2j) = 2^{q-1} d  \left(1+ \frac{d}{2}\right)_{q-1} \,  .
\end{equation}
 To be concrete,  the constant $A$ in \eqref{eq:etaperm} would be  $1/P^{(2q)}$. 

In the case of two-loop integrals, the situation is more involved as there are two distinct momenta to deal with. A prototypical example is for instance
\begin{equation} \label{eq:secondI} \begin{gathered}
       \int \frac{\mathrm{d}^d p}{(2 \pi)^d}\frac{\mathrm{d}^d k}{(2 \pi)^d} \frac{k_\mu k_\nu p_{\lambda_1} \dots p_{\lambda_{2q}}}{D} = A \eta_{\mu \nu}(\eta_{\lambda_1 \lambda_2} \dots \eta_{\lambda_{2q-1} \lambda_{2q}} + \mathrm{perms.}) \\
       + B (\eta_{\mu \lambda_1} \eta_{\nu \lambda_2} \eta_{\lambda_3 \lambda_4}\dots \eta_{\lambda_{2q-1} \lambda_{2q}} + \mathrm{perms.}) \,  ,
\end{gathered}\end{equation}
where $A,B$ will now be combinations of scalar loop integrals and some combinatorial factors.
Contracting both sides with the tensor $\eta^{\mu \nu} \eta^{\lambda_1 \lambda_2} \dots \eta^{\lambda_{2q-1} \lambda_{2q}}$ we obtain a first equation
\begin{equation} \label{eq:AB1}
     \int \frac{\mathrm{d}^d p}{(2 \pi)^d}\frac{\mathrm{d}^d k}{(2 \pi)^d} \frac{k^2 (p^2)^{q}}{D} =    P^{(2q)} ( d A+ 2 q B) \,  .
\end{equation}

   A second equation is deduced in a similar manner by contracting with $\eta^{\mu \lambda_1} \eta^{\nu \lambda_2} \dots \eta^{\lambda_{2q-1} \lambda_{2q}}$
\begin{equation} \label{eq:AB2}
     \int \frac{\mathrm{d}^d p}{(2 \pi)^d}\frac{\mathrm{d}^d k}{(2 \pi)^d} \frac{(k \cdot p)^2 (p^2)^{q-1}}{D} =   P^{(2q)} \left[ A+ (2q+d-1) B \right] \,  .
\end{equation}
Solving \eqref{eq:AB1} and \eqref{eq:AB2} for $A$ and $B$, one finds the correct rewriting of \eqref{eq:secondI} in terms of scalar loop integrals and combinatiorial factors. 

In a theory with non-manifest Lorentz invariance, $P_{\lambda_1 \dots \lambda_{2q}}$ is most often encountered with explicit values assigned to its indices (either 0 or 1). As we shall make large use of this formula, let us mention that if $P_{\lambda_1 \dots \lambda_{2q}}$ comes with $2q_0$ 0-indices and $2q_1$ 1-indices (so that $2q = 2q_0 + 2q_1$), the tensor specifically evaluates to 
\begin{equation}
    P_{0_1 \dots 0_{2q_0} 1_1 \dots 1_{2q_1}} = (2q_0-1)!! (2q_1-1)!! (\eta_{00})^{q_0} (\eta_{11})^{q_1}= (2q_0-1)!!(2q-2q_0-1)!! (\eta_{00})^{q_0} (\eta_{11})^{q_1} \,  .
\end{equation}

 \subsection{One-loop Integrals}

\subsubsection*{Bubble Diagrams}

Integrals for bubble diagrams are trivial, as they simply coincide with $ I_1$. 

\subsubsection*{Loop Diagrams}

Loop diagrams are only slightly more involved. In general we are interested in both the divergent and convergent (at least $O(\epsilon^0)$) parts, as the latter is important for counter-terms. The divergent diagrams we shall encounter are of the form
\begin{equation}
    \int \frac{\mathrm{d}^d k}{(2 \pi)^d} \frac{k_\mu k_\nu}{(k^2 - m^2)^2} = \frac{\eta_{\mu \nu}}{d} (I_1 + m^2 I_2)\,  . 
\end{equation}

Diagrams with an odd number of momenta in the numerator vanish by symmetry. Fully convergent integrals are necessary of the form
\begin{equation}
    \int \frac{\mathrm{d}^d k}{(2 \pi)^d} \frac{1}{(k^2 - m^2)^2} = I_2 \,  .
\end{equation}

\subsubsection*{Triangle Diagrams}

Divergent integrals stemming from triangle diagrams are of the form
\begin{equation}
    \int \frac{\mathrm{d}^dk}{(2 \pi)^d} \frac{k_\mu k_\nu k_\rho k_\sigma}{(k^2-m^2)^3} = \frac{1}{d(d+2)} (\eta_{\mu \nu} \eta_{ \rho \sigma} + \eta_{\mu \rho}\eta_{\nu \sigma} + \eta_{\mu \sigma} \eta_{\nu \rho}) (I_1 + 2m^2 I_2 + m^4 I_3) \,  ,
\end{equation}
where we have made use of $O(d)$ symmetry to perform the integral. By the remarks above, the right-hand side is mass independent at zero-th order in $\epsilon$. Other finite results are
\begin{equation}
     \int \frac{\mathrm{d}^dk}{(2 \pi)^d} \frac{1}{(k^2-m^2)^3} = I_3 \, , \qquad \int \frac{\mathrm{d}^dk}{(2 \pi)^d} \frac{k_\mu k_\nu}{(k^2-m^2)^3} = \frac{\eta_{\mu \nu}}{d} (I_2+m^2 I_3) \,  .
\end{equation}
\subsection{Two-loop Integrals}
 A two-loop integral with vanishing external momenta is of the form 
\begin{equation}
\left[\left[\dots \right]\right]_{i j k}  =   \int \frac{\mathrm{d}^{d}k}{(2\pi)^{d}} \frac{\mathrm{d}^{d}p}{(2\pi)^{d}}  \frac{(\dots )}{(p^{2} - m^{2})^{i}(k^{2} - m^{2})^{j}[(k+p)^{2} - m^{2}]^{k}  } \, , 
\end{equation}
where the dots will be specified on a case-by-case basis. In performing manipulation we discard finite terms, for instance:
\begin{equation}
    [[1]]_{i,j,k} = 0 \, , \quad \text{for $i,j,k \geq 1$} \, .
\end{equation}
Typical UV divergent integrals encountered are 
\begin{equation}
\left[\left[1 \right]\right]_{1,0,1}=\left[\left[1 \right]\right]_{1,1, 0}=\left[\left[1 \right]\right]_{0,1, 1}  =   I_{1}^{2} \ , \quad \left[\left[1 \right]\right]_{2,0,1}   =  I_{1} I_{2} \ .
\end{equation}

 \begin{figure}[!h]
 \begin{subfigure}{.3\textwidth}
   \centering
     \begin{tikzpicture}
    \node[draw,circle,fill=black,minimum size=3pt,inner sep=0pt] (four) at (0,0) [label=left:$4$]{};
    \node[draw,circle,fill=black,minimum size=3pt,inner sep=0pt] (three) at (3,0) [label=right:$3$]{};
    \node[draw,circle,fill=black,minimum size=3pt,inner sep=0pt] (two) at (3,3) [label=right:$2$]{};
     \node[draw,circle,fill=black,minimum size=3pt,inner sep=0pt] (one) at (0,3) [label=left:$1$]{};
    \draw (four) to node [label=below:$-p$]{} (three);
    \draw (three) to node [label=right:$-p$]{} (two);
    \draw (two) to node [label=above:$k$]{} (one);
    \draw (one) to node [label=left:$p$]{} (four);
    \draw(one) to [bend right=60] node [label=below:$-(k+p)$]{} (two);
        \end{tikzpicture}
  \caption{Square Envelope}
  \label{fig:SquareEnvelope}
\end{subfigure}%
\begin{subfigure}{.3\textwidth}
  \begin{center}
        \begin{tikzpicture}
    \node[draw,circle,fill=black,minimum size=3pt,inner sep=0pt] (one) at (0,0) [label=left:$1$]{};
    \node[draw,circle,fill=black,minimum size=3pt,inner sep=0pt] (two) at (3,0) [label=right:$2$]{};
    \node[draw,circle,fill=black,minimum size=3pt,inner sep=0pt] (three) at (1.5,-1.5) [label=below:$3$]{};
     \node[draw,circle,fill=black,minimum size=3pt,inner sep=0pt] (four) at (1.5,1.5) [label=above:$4$]{};
    \draw (one) to node [label=above:$k+p$]{} (two);
    \draw (two) to node [label=right:$p$]{} (three);
    \draw (two) to node [label=right:$k$]{} (four);
    \draw (one) to node [label=left:$-p$]{} (three);
    \draw (one) to node [label=left:$-k$]{} (four);
        \end{tikzpicture}
  \caption{Diamond Sunset}
  \label{fig:DiamondSunset}
  \end{center}
\end{subfigure}
\begin{subfigure}{.3\textwidth}
  \centering
   \begin{tikzpicture}
    \node[draw,circle,fill=black,minimum size=3pt,inner sep=0pt] (one) at (0,0) [label=left:$1$]{};
    \node[draw,circle,fill=black,minimum size=3pt,inner sep=0pt] (two) at (3,0) [label=right:$2$]{};
     \node (mome1) at (1.5,-1.5) [label=below:$k$] {};
     \node (mome2) at (1.5,1.5) [label=above:$p$] {};
    \draw (1.5,0) circle [radius=1.5];
    \draw (one) to node [label= above:$-(k+p)$]{} (two);
        \end{tikzpicture}
  \caption{Sunset}
  \label{fig:sfig5}
\end{subfigure}
\begin{subfigure}{.3\textwidth}
  \centering
   \begin{tikzpicture}
    \node[draw,circle,fill=black,minimum size=3pt,inner sep=0pt] (one) at (0,0) [label=left:$1$]{};
    \node[draw,circle,fill=black,minimum size=3pt,inner sep=0pt] (two) at (3,0) [label=right:$2$]{};
    \node[draw,circle,fill=black,minimum size=3pt,inner sep=0pt] (three) at (1.5,-2.59808) [label=below:$3$]{};
    \draw (one) to node [label=above:$-(k+p)$]{} (two);
    \draw (two) to node [label=right:$-p$]{} (three);
    \draw (three) to node [label=left:$p$]{} (one);
    \draw (one) to [bend right=60]node [label=below:$k$]{} (two);
        \end{tikzpicture}
  \caption{Triangle Envelope}
  \label{fig:sfig6}
  \end{subfigure}
  \begin{subfigure}{.3\textwidth}
  \centering
   \begin{tikzpicture}
    \node[draw,circle,fill=black,minimum size=3pt,inner sep=0pt] (one) at (1,0) [label=right:$1$]{};
    \node (mome) at (-1,0) [label=left:$k$] {};
    \node (mome) at (2.2,0) [label=right:$p$] {};
    \draw (0,0) circle [radius=1];
    \draw (1.6,0) circle [radius=0.6];
        \end{tikzpicture}
  \caption{Decorated bubble}
  \end{subfigure}
   \begin{subfigure}{.3\textwidth}
    \centering
       \begin{tikzpicture}
    \node[draw,circle,fill=black,minimum size=3pt,inner sep=0pt] (one) at (0,0) [label=left:$1$]{};
    \node[draw,circle,fill=black,minimum size=3pt,inner sep=0pt] (two) at (2,0) [label=right:$2$]{};
    \node (mome1) at (1,-1) [label=below:$k$] {};
     \node (mome2) at (1,1) [label=above:$-k$] {};
     \node (mome3) at (3.2,0) [label=right:$p$] {};
    \draw (1,0) circle [radius=1];
    \draw (2.6,0) circle [radius=0.6];
        \end{tikzpicture}
  \caption{Decorated loop}
  \label{fig:sfig9}
  \end{subfigure}
  \begin{subfigure}{.3\textwidth}
  \centering
   \begin{tikzpicture}
    \node[draw,circle,fill=black,minimum size=3pt,inner sep=0pt] (one) at (0,0) [label=left:$1$]{};
    \node[draw,circle,fill=black,minimum size=3pt,inner sep=0pt] (two) at (2,0) [label=right:$2$]{};
    \node[draw,circle,fill=black,minimum size=3pt,inner sep=0pt] (three) at (1,1.73205) [label=above:$3$]{};
     \node (mome1) at (-0.853553,-0.353553) [label=left:$p$] {};
    \draw (one) to node [label=below:$k$]{} (two);
    \draw (two) to node [label=right:$k$]{} (three);
    \draw (three) to node [label=left:$-k$]{} (one);
    \draw (-0.353553,-0.353553) circle [radius=0.5];
        \end{tikzpicture}
  \caption{Decorated triangle}
  \label{fig:sfig10}
\end{subfigure}
   \begin{subfigure}{.3\textwidth}
    \centering
       \begin{tikzpicture}
    \node[draw,circle,fill=black,minimum size=3pt,inner sep=0pt] (one) at (0,0) [label=left:$1$]{};
    \node[draw,circle,fill=black,minimum size=3pt,inner sep=0pt] (two) at (1.7,0) [label=right:$2$]{};
    \node[draw,circle,fill=black,minimum size=3pt,inner sep=0pt] (two) at (3.4,0) [label=right:$3$]{};
    \node (mome1) at (0.85,-0.85) [label=below:$k$] {};
     \node (mome2) at (0.85,0.85) [label=above:$-k$] {};
         \node (mome3) at (2.55,-0.85) [label=below:$p$] {};
     \node (mome4) at (2.55,0.85) [label=above:$-p$] {};
    \draw (0.85,0) circle [radius=0.85];
    \draw (2.55,0) circle [radius=0.85];
        \end{tikzpicture}
  \caption{Double loop}
  \label{fig:sfig13}
  \end{subfigure}
  
\caption{Two-loop diagrams. Momentum flows are aligned with numerical ordering of vertices. }
\label{fig:fig1}
\end{figure}
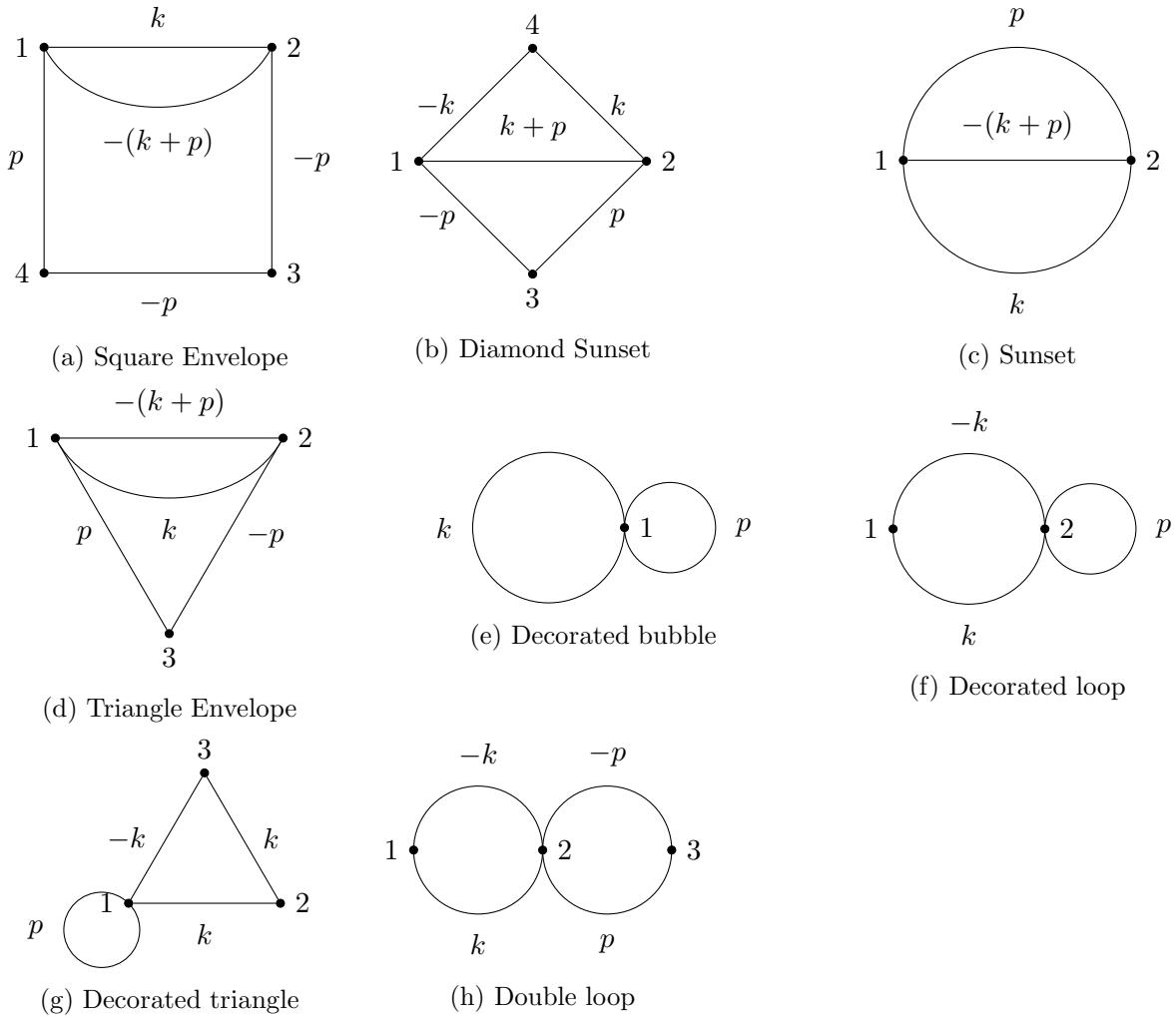

\subsubsection*{Sunset Diagrams}

Sunset diagrams are equivalent to $[[\dots]]_{1,1,1}$ integrals. Given the exchange symmetry $k \leftrightarrow p$ in the denominator, the only relevant integrals are
\begin{equation}
    [[p^2]]_{1,1,1} = I_1^2 \, , \qquad [[(p \cdot k)]]_{1,1,1} = - \frac{1}{2}I_1^2 \,  .
\end{equation}
It immediately follows that
\begin{equation}
    [[p_{\mu}p_\nu]]_{1,1,1} = \frac{1}{d}\eta_{\mu \nu} I_1^2 \, , \qquad   [[p_{\mu}k_\nu]]_{1,1,1} = -\frac{1}{2d}\eta_{\mu \nu} I_1^2 \, .
\end{equation}

 \subsubsection*{Diamond Sunset Diagrams}

Diamond sunset diagrams correspond to $[[\dots]]_{2,2,1}$.
Notice that this class is symmetric under $k \leftrightarrow p$; hence, we will omit  integrals which can be deduced from symmetry arguments. Again, we begin with the scalar integrals
\begin{subequations}
  \begin{align}
      [[(p^2)^3]]_{2,2,1} &=I_1^2 + 4 m^2 I_1 I_2 \, ,\\
      [[(p^2)^2 (p \cdot k)]]_{2,2,1}&=-I_1^2 - m^2 I_1 I_2 \, , \\
      [[(p^2)^2 k^2]]_{2,2,1} &=I_1^2 + m^2 I_1 I_2 \, , \\
      [[p^2 (p \cdot k) k^2]]_{2,2,1} &= -\frac{1}{2}I_1^2  \, , \\
      [[p^2 (p \cdot k)^2]]_{2,2,1}&=\frac{3}{4}I_1^2 + \frac{1}{2}m^2 I_1 I_2  \, , \\
      [[(p \cdot k)^3]]_{2,2,1} &= \frac{2-3d}{4d}I_1^2 + \frac{2-d}{2d} m^2 I_1 I_2 \,  .
  \end{align}
\end{subequations}

Tensorial integrals can now be deduced
\begin{subequations}
  \begin{align}
      [[p_\mu p_\nu p_\rho p_\sigma p_\kappa p_\lambda]]_{2,2,1} &= F_1 (\eta_{\mu \nu} \eta_{\rho \sigma} \eta_{\kappa \lambda} + 14 \, \mathrm{perms}) (I_1^2 +4 m^2 I_1 I_2) \,  , \\
      [[p_\mu p_\nu p_\rho p_\sigma p_\kappa k_\lambda]]_{2,2,1} &= F_1 (\eta_{\mu \nu} \eta_{\rho \sigma} \eta_{\kappa \lambda} + 14 \, \mathrm{perms}) \left(-I_1^2 - m^2 I_1 I_2  \right) \,  , \\
      [[p_\mu p_\nu p_\rho p_\sigma k_\kappa k_\lambda]]_{2,2,1} &=\frac{F_1}{d-1}(\eta_{\mu \nu} \eta_{\rho \sigma} + 2 \, \mathrm{perms}) \eta_{\kappa \lambda}(d I_1^2 + (d+1)m^2 I_1 I_2)  \nonumber \\
      &+ \frac{F_1}{4(d-1)} (\eta_{\mu \kappa} \eta_{\nu \lambda} \eta_{\rho \sigma} + 11 \, \mathrm{perms})
      ((3d-4)I_1^2 + 2 (d-2)m^2 I_1 I_2 )  \, , \\
       [[p_\mu p_\nu p_\rho k_\sigma k_\kappa k_\lambda]]_{2,2,1} &=\frac{F_1}{4 d(d-1)} (\eta_{\mu \nu} \eta_{\rho \sigma} \eta_{\kappa \lambda} + 8 \, \mathrm{perms}) \biggl[-2(d^2-2d+2)I_1^2+4(d-2)m^2 I_1 I_2  \biggr] \nonumber \\
       &+ \frac{F_1}{8d(d-1)} (\eta_{\mu \sigma} \eta_{\nu \kappa} \eta_{\rho \lambda}  +5 \, \mathrm{perms})\biggl[-2(3d^2-2d-4)I_1^2 -4(d^2-4) m^2 I_1 I_2   \biggr] \,  .
  \end{align}
\end{subequations}

 \subsubsection*{Triangle Envelope Diagrams}

This class is equivalent to $[[\dots]]_{2,1,1}$. Scalar integrals are
\begin{subequations}
  \begin{align}
      [[(p^2)^2]]_{2,1,1} &= I_1^2 \, , \\
      [[p^2 k^2]]_{2,1,1} &= I_1^2 + m^2 I_1 I_2 \, , \\
      [[(p \cdot k)^2]]_{2,1,1} &= \frac{3}{4}I_1^2 + \frac{1}{2}m^2 I_1 I_2 \, , \\
      [[p^2 (p \cdot k)]]_{2,1,1}&=-\frac{1}{2}I_1^2 \, , \\
      [[(p \cdot k) k^2]]_{2,1,1}&=- I_1^2 - m^2 I_1 I_2 \, , \\
       [[ k^4]]_{2,1,1}&=I_1^2 + 3 m^2 I_1 I_2 \, .
  \end{align}
\end{subequations}
Adopting the symbol $F_2 = \frac{1}{d (d+2)}$, tensorial integrals evaluate to
\begin{subequations}
  \begin{align}
      [[p_\mu p_\nu p_\rho p_\sigma]]_{2,1,1} &= F_2 (\eta_{\mu \nu} \eta_{\rho \sigma}+ 2 \, \mathrm{perms}) I_1^2 \, , \\
      [[p_\mu p_\nu p_\rho k_\sigma]]_{2,1,1} &= -\frac{1}{2}F_2 (\eta_{\mu \nu} \eta_{\rho \sigma}+ 2 \, \mathrm{perms}) I_1^2 \, , \\
      [[p_\mu p_\nu k_\rho k_\sigma]]_{2,1,1}&= \frac{F_2}{2(d-1)} \eta_{\mu \nu} \eta_{\rho \sigma} I_1 [(2d-1)I_1+ 2 d m^2 I_2] \nonumber \\
      &+ \frac{F_2}{4(d-1)}( \eta_{\mu \rho} \eta_{\nu \sigma} + \eta_{\mu \sigma} \eta_{\nu \rho}) I_1 [(3d-4)I_1+ 2 (d-2) m^2 I_2] \,  , \\ 
      [[p_{\mu}k_\nu k_\rho k_\sigma]]_{2,1,1}&=  -F_2(\eta_{\mu \nu} \eta_{\rho \sigma}+ 2 \, \mathrm{perms}) I_1(I_1 + m^2 I_2) \, , \\
        [[k_{\mu}k_\nu k_\rho k_\sigma]]_{2,1,1}&=F_2(\eta_{\mu \nu} \eta_{\rho \sigma}+ 2 \, \mathrm{perms}) (I_1^2 + 3 m^2 I_1 I_2) \,  .
  \end{align}
\end{subequations}

\subsubsection*{Square Envelope Diagrams}

Square envelope diagrams correspond to $[[\dots]]_{3,1,1}$, and the scalar integrals we shall need for this topology are
\begin{subequations}
	\begin{align}
	[[(p^2)^3]]_{3,1,1} &= I_1^2 \, , \\
	[[(p^2)^2 k^2]]_{3,1,1} &= I_1(I_1+ 2 m^2 I_2 + m^4 I_3) \, , \\
	[[p^2 (k^2)^2]]_{3,1,1}&=I_1(I_1 + 4 m^2 I_2 + 3 m^4 I_3)  \, ,\\
	[[(p^2)^2(p \cdot k)]]_{3,1,1} &=- \frac{1}{2}I_1^2 \, , \\
	[[p^2 (p \cdot k)^2]]_{3,1,1} &= \frac{3}{4}I_1^2 +m^2 I_1 I_2 + \frac{m^4}{2} I_1 I_3 \, ,\\
	[[p^2 (p \cdot k)k^2]]_{3,1,1} &= -I_1^2 - 2m^2 I_1 I_2-m^4 I_1 I_3 \, , \\
	[[(p \cdot k)^2 k^2]]_{3,1,1} &=I_1^2  +\frac{7d+2}{2d}m^2 I_1( I_2 + m^2 I_3)  \, , \\
	[[(p \cdot k)^3]]_{3,1,1} &= - \frac{7}{8}I_1^2- \frac{3}{2}m^2 I_1 I_2- \frac{3}{4}m^4 I_1 I_3 \,   .
	\end{align}
\end{subequations}

Integrals with ``open'' Lorentz indices can be built out of these. They all share a common pre-factor $F_1 = [d(d+2)(d+4)]^{-1}$ coming from the contraction of indices. The results are:
\newpage
\begin{subequations}
  \begin{align}
      [[p_\mu p_\nu p_\rho p_\sigma p_\kappa p_\lambda]]_{3,1,1} &= F_1 (\eta_{\mu \nu} \eta_{\rho \sigma} \eta_{\kappa \lambda} + 14 \, \mathrm{perms}) I_1^2 \, , \\
      [[p_\mu p_\nu p_\rho p_\sigma p_\kappa k_\lambda]]_{3,1,1} &= -\frac{1}{2}F_1 (\eta_{\mu \nu} \eta_{\rho \sigma} \eta_{\kappa \lambda} + 14 \, \mathrm{perms})I_1^2 \, , \\
      [[p_\mu p_\nu p_\rho p_\sigma k_\kappa k_\lambda]]_{3,1,1} &= \frac{F_1}{d-1}(\eta_{\mu \nu} \eta_{\rho \sigma} + 2 \, \mathrm{perms}) \eta_{\kappa \lambda} I_1\left[d I_1 + m^2(d+1)(2I_2 + m^2 I_3) \right] \nonumber \\
      &+ \frac{F_1}{4(d-1)} (\eta_{\mu \kappa} \eta_{\nu \lambda} \eta_{\rho \sigma} + 11 \, \mathrm{perms}) I_1 \left[\left(3d-4 \right)I_1+ 2\left(d-2\right)m^2(2 I_2 + m^2 I_3)\right] \,  , \\
      [[p_\mu p_\nu p_\rho k_\sigma k_\kappa k_\lambda]]_{3,1,1}&=\frac{F_1}{4(d-1)} (\eta_{\mu \nu}\eta_{\rho \sigma} \eta_{\kappa \lambda} + 8 \, \mathrm{perms}) I_1 \left[ (3-4d)I_1 + 2(1-2d)m^2(2I_2 + m^2 I_3)\right] \nonumber \\
      &+\frac{F_1}{8(d-1)}(\eta_{\mu \sigma} \eta_{\nu \kappa} \eta_{\rho \lambda} + 5 \, \mathrm{perms})I_1\left[(10-7d)I_1+ 6(2-d)m^2(2I_2+m^2 I_3)\right] \, ,\\
      [[p_\mu p_\nu k_\rho k_\sigma k_\kappa k_\lambda]]_{3,1,1}&=\frac{F_1}{d(d-1)}(\eta_{\mu \nu}\eta_{\rho \sigma} \eta_{\kappa \lambda} + 2 \, \mathrm{perms}) \biggl[d(d-1)I_1^2+(4d^2-2d-4)m^2 I_1I_2 \nonumber \\
      &+(3d^2-5d-4)m^4 I_1 I_3  \biggr] \nonumber \\
      &+ \frac{F_1}{2(d-1)}(\eta_{\mu \rho}\eta_{\nu \sigma} \eta_{\kappa \lambda}+ 11 \, \mathrm{perms}) \biggl[2(d-1)I_1^2+(7d-6)m^2 I_1 I_2 \nonumber \\
      &+(7d-4)m^4 I_1 I_3 \biggr] \,  .
  \end{align}
\end{subequations}

\subsection{Schwinger Parametrisation}\label{app:Schwinger}

When considering the renormalisation of the $(\partial_1y)^2$ component of the metric, the integrals above are not sufficient since we also encounter diagrams which give rise to integrands in Fourier space wit non-scalar denominators.

 The first integral we shall be concerned with arises from a particular instance of the sunset topology. Consider
 \begin{equation} \label{eq:J1def}
     J_1 =  \int \frac{\mathrm{d}^d k}{(2 \pi)^d} \frac{\mathrm{d}^d p}{(2 \pi)^d} \frac{k^1 (p^0)^2}{p^1} \frac{1}{(k^2-m^2) (p^2-m^2) [(k+p)^2-m^2]} \,  .
 \end{equation}

To address the   $p^1$ momentum in the denominator we employ a   Schwinger parametrisation 
\begin{equation}
    \frac{1}{p^1} = \int_0^\infty \mathrm{d}u \, e^{- up^1} \,  .
\end{equation}

The Lorentz-scalar part of \eqref{eq:J1def} and the momentum integrals are easily dealt with using the above techniques so we momentarily omit them by defining a ``reduced'' integral
\begin{equation}
    \tilde{J}_1 = \int_0^\infty \mathrm{d}u \, e^{- up^1} k^1 (p^0)^2 \,  .
\end{equation}

We now perform a series expansion of this exponential which we understand will give a series of loop integrals we can evaluate using some combinatorics and formulas in Appendix \ref{app:Integrals}.  Even powers of $u$  drop out  since the numerator of the corresponding momentum integral will contain an odd number of $\sigma$ components of momentum - this implies that any scalar contraction will come with a factor of $\eta_{01}$ and hence will drop.  So we need retain only  
\begin{equation} \label{eq:j11}
    \tilde{J}_1 = - \int_0^\infty \mathrm{d}u u \, \sum_{n=0}^{\infty}  \frac{u^{2n}}{(2n+1)!} k^1 (p^0)^2 (p^1)^{2n+1}\,  .
 \end{equation}

Now, the precise replacement stems from the usual combinatorics in loop integrals
\begin{equation}
    k^{\mu} p^{\nu_1} \dots p^{\nu_{2n+3}} \rightarrow  \frac{1}{P^{(2n+4)}} P^{\mu \nu_1 \dots \nu_{2n+3}}(k \cdot p) (p^2)^{n+1} \,  .
\end{equation}
 
 In a scheme where $\eta^{01}=0$, there are $(2n+1)!!$ non vanishing contributions in $P^{\mu \nu_1 \dots \nu_{2n+3}}$ with a choice of indices as in \eqref{eq:j11}. We then end up with
 \begin{equation}
     \tilde{J}_1 =  \varphi(\epsilon) \psi(\epsilon) (k \cdot p) p^2 \int_0^{\infty} \mathrm{d}u u \sum_{n=0}^\infty \frac{(2n+1)!!}{(2n+1)! P^{(2n+4)}} (- u p^2 \psi(\epsilon))^n \, ,
 \end{equation}
 where we have adopted the parametrisations $\eta^{11} = - \psi(\epsilon)$ and $\eta^{00} = \varphi(\epsilon)$, with $\psi = 1+ \mathfrak{g} \epsilon$ and $\varphi = 1+ \mathfrak{g} \epsilon$ are both positive definite. Introducing a new integration variable $z = up^2 \psi/4$, one can prove that the series can be resummed to yield the regularised hypergeometric function ${}_0\tilde{F}_1$
 \begin{equation}
     \tilde{J}_1 = \frac{\varphi(\epsilon)}{2} \Gamma\left(\frac{d}{2}\right) (k \cdot p) \int_0^\infty \mathrm{d}z \, {}_0 \tilde{F}_1\left(2+ \frac{d}{2},-z\right) \,  .
 \end{equation}

The remaining integral is part of a family of parametric integrals whose precise evaluation is known and equals
\begin{equation} \label{eq:zint1}
    \int_0^\infty \mathrm{d}z \, z^{\alpha -1} \, {}_0\tilde{F}_1 (b; -z) = \frac{ \Gamma(\alpha)}{\Gamma(b-\alpha)} \,  .
\end{equation}

Adapting this formula to the case at hand we find
\begin{equation}
    \tilde{J}_1 = \frac{\varphi(\epsilon)}{d} (k \cdot p) \, , \qquad     J_1 = -\frac{\varphi(\epsilon)}{2d} I_1^2 \,  .   
\end{equation}

Other than $J_1$, we have a family of relevant non-invariant integrals depending on a integer parameter $\alpha \in \mathbb{N}$ reading
\begin{equation} \label{eq:J2a}
    J^{(\alpha)}_2 =  \int \frac{\mathrm{d}^d k}{(2 \pi)^d} \frac{\mathrm{d}^d p}{(2 \pi)^d} \frac{1}{p^1} \frac{k^0 k^1 (p^0)^{2 \alpha+1}}{(k^2-m^2) (p^2-m^2)^{\alpha + 1} [(k+p)^2-m^2]} \,  .
\end{equation}

Using the same technique the relevant cases evaluate to
\begin{align}
    J_2^{(0)} &= \frac{3d-4}{4d (d-1)} \varphi(\epsilon) I_1^2 \,  , \\
    J_2^{(1)} &= \frac{3(3d-4)}{4(d-1)d(d+2)} \varphi(\epsilon)^2 I_1^2 +\frac{3 (d-2)  \varphi(\epsilon)^2}{2(d-1)d(d+2)}\varphi(\epsilon)^2 m^2 I_1 I_2 \, , \\
    J_2^{(2)} &= \frac{15 (3 d-4)}{4 (d-1)d(d+2)(d+4)} \varphi(\epsilon)^3 I_1^2 +\frac{15 (d-2) m^2 }{ (d-1)d(d+2)(d+4)} \varphi(\epsilon)^3 m^2 I_1 \left(I_2 + \frac{m^2}{2}I_3 \right) \,  .
\end{align}

\section{Example}\label{app:ex}

We complement the main presentation with an explicit toy example.  This allows for an independent computerised cross-check of the calculations done elsewhere in the project. In this toy model we are also able to explicitly examine the implication of removing the base manifold counter-term at one-loop via a field redefinition/addition of total derivative.  

We begin with a model consisting of a single $S^1$ direction of radius $r = r(y)$ for the fibre with Lagrangian
\begin{equation}
    \mathcal{L} = \frac{1}{2} (\partial_1 x \partial_0 \tilde{x} + \partial_0 x \partial_1 \tilde{x}) - \frac{1}{2} \left( r^2 \partial_1 x \partial_1 x + r^{-2} \partial_1 \tilde{x} \partial_1 \tilde{x}\right) + \frac{1}{2} \partial_\mu y \partial^\mu y \, .
\end{equation}

The one-loop renormalisation yields
\begin{equation}
    \mathcal{L}_{\mathrm{CT}} = \frac{1}{4 \pi \epsilon} \frac{r \ddot{r} - \dot{r}^2}{ r^4} (\partial_1 \tilde{x} \partial_1 \tilde{x} - r^4 \partial_1 x \partial_1 x) + \frac{1}{4 \pi \epsilon} \frac{\dot{r}^2}{r^2} \partial_\mu y \partial^\mu y \, ,
\end{equation}
where over-dots denote, as usual, derivatives with respect to $y$. Using  Method 2 followed by Method 1 for final evaluation of integrals, we find the two-loop contributions 
\begin{align}
     \widetilde{T}_2^{(2)}|_{11} &= \frac{1}{8 \pi^2 \epsilon^2} \frac{5 (r^{(1)})^4 - 8 r (r^{(1)})^2 r^{(2)} +r^2 (r^{(2)})^2 + 2 r^2 r^{(1)} r^{(3)}}{r^4} =  \widetilde{T}_2^{(2)}|_{00} \, , \\ 
    \widetilde{T}_2^{(2)}|_{01} &= 0 \, , \\
    T_2^{(2)}|_{xx} &= - \frac{1}{16 \pi^2 \epsilon^2} \frac{2 (r^{(1)})^4 - 6 r (r^{(1)})^2 r^{(2)} + 4 r^2 r^{(1)}r^{(3)} + r^2 ((r^{(2)})^2 -  r^{(4)})}{r^2} \, , \\ 
    T_2^{(2)}|_{\tilde{x}\tilde{x}} &=  \frac{1}{16 \pi^2 \epsilon^2} \frac{6 (r^{(1)})^4 -14 r (r^{(1)})^2 r^{(2)} + 4 r^2 r^{(1)}r^{(3)} + r^2 ( 5(r^{(2)})^2 - r r^{(4)})}{r^6} \, , \\
    \widetilde{T}_1^{(2)}|_{00} &= \frac{\mathfrak{g}}{8 \pi^2 \epsilon} \frac{((r^{(1)})^2 - r r^{(2)})^2}{r^4} \, , \\
    \widetilde{T}_1^{(2)}|_{11} &= \frac{1}{8 \pi^2 \epsilon} \frac{(\mathfrak{g}-1) (r^{(1)})^4 + 2 (3 \mathfrak{g}-1) r (r^{(1)})^2 r^{(2)}  - (3 \mathfrak{g}-1) r^2 (r^{(2)})^2}{r^4} \, , \\
    \widetilde{T}_1^{(2)}|_{01} &= 0 \, , \\
    T_1^{(2)}|_{xx} &=  \frac{1}{16 \pi^2 \epsilon} \frac{2(1- \mathfrak{g}) (r^{(1)})^4 +  (2 \mathfrak{g}-1) r(r^{(1)})^2 r^{(2)} +2 (2 \mathfrak{g}-1)r^2 (r^{(2)})^2}{r^2} \, , \\ 
    T_1^{(2)}|_{\tilde{x}\tilde{x}} &=  \frac{1}{16 \pi^2 \epsilon} \frac{2(-4+9 \mathfrak{g}) (r^{(1)})^4 - 9(2 \mathfrak{g}-1) r(r^{(1)})^2 r^{(2)} +2 (2 \mathfrak{g}-1)r^2 (r^{(2)})^2}{r^6}\, ,
    \end{align}
where $r^{(n)}$ is an alternative notation for the $n$-th derivative of $r$ with respect to $y$. These  agree with the results of the main text  upon the substitution of the generalised metric $\H=\mathrm{diag} (r^2 , r^{-2})$. We remark that the vanishing of $\widetilde{T}^{(2)}|_{01}$ is due to the triviality of the tensorial combinations  within this example.  

Let us note that we can amend the one-loop counter-term through the inclusion of any piece that vanishes upon integration by parts and application of the equations of motion
\begin{equation}
 \mathcal{L}_{\mathrm{on-shell}}=  \dot{f}(y) \partial_\mu y \partial^\mu y - \frac{1}{2} f(y) \H^{(1)}_{\bullet \bullet} \approx 0    \,  ,
\end{equation}
for some function $f(y)$. With the choice that $\dot{f} = - \frac{1}{2} \widetilde{T}^{(1)}$   the base divergence of the one-loop counter can be removed, considering instead 
\begin{equation}
\mathcal{L}_{\mathrm{CT}} + \mathcal{L}_{\mathrm{on-shell}} =
\frac{f(y)}{4 \pi \epsilon} \left( r \dot{r} \,  \partial_1 x \partial_1 x - \frac{\dot{r}}{r^3} \partial_1 \tilde{x} \partial_1 \tilde{x} \right) + \frac{1}{4 \pi \epsilon} \frac{r \ddot{r} - \dot{r}^2}{ r^4} (\partial_1 \tilde{x} \partial_1 \tilde{x} - r^4 \partial_1 x \partial_1 x) \,  .
\end{equation}

We may now proceed to re-calculate the two-loop divergences using this modified one-loop counter term.  One might anticipate that this resolves some of the discrepancies between $\widetilde{T}^{(2)}_1|_{00}$ and $\widetilde{T}^{(2)}_1|_{11}$ seen in the above. However, an explicit calculation yields
\begin{align}
    \widetilde{T}_1^{(2)}|_{00} &= \frac{1}{16 \pi^2 \epsilon} \frac{(1- \mathfrak{g}) (r^{(1)})^4  - 2 \mathfrak{g} r (r^{(1)})^2 r^{(2)} + \mathfrak{g} r^2 (r^{(2)})^2}{r^4} \nonumber \\ &+ \frac{1}{16 \pi^2 \epsilon} \frac{(-1 + 2 \mathfrak{g}) f r r^{(1)} ((r^{(1)})^2 - r r^{(2)})}{r^4}  + \dots  \, , \\
     \widetilde{T}_1^{(2)}|_{11} &= \frac{1}{16 \pi^2 \epsilon} \frac{(5- 14\mathfrak{g}) (r^{(1)})^4  +4 (1-3 \mathfrak{g}) r (r^{(1)})^2 r^{(2)} + 2 (1-3\mathfrak{g}) r^2 (r^{(2)})^2}{r^4} \nonumber \\ 
     &+ \frac{1}{8 \pi^2 \epsilon}\frac{3(-1 + 2 \mathfrak{g}) f r r^{(1)} ((r^{(1)})^2 - r r^{(2)})}{r^4}  + \dots  \, , 
\end{align}
where ellipsis indicate terms proportional to  $\frac{\bar{\gamma}}{\epsilon}$ (which do \textit{not} now vanish). Far from ameliorating the situation, we  still have Lorentz violation from the base counter terms and, moreover, un-cancelled $\bar{\gamma}$ terms.

\end{appendix}

\end{document}